\documentclass[twocolumn]{aastex63}
\usepackage{xcolor}
\usepackage{gensymb}
\usepackage[nointegrals]{wasysym}
\usepackage{amsmath}

\LTcapwidth=\textwidth

\submitjournal{AJ}

\shorttitle{New \textit{Kepler} Candidates}
\shortauthors{Kunimoto et al.}

\begin{document}

\title{Searching the Entirety of \textit{Kepler} Data. I. 17 New Planet Candidates \\ Including 1 Habitable Zone World}

\correspondingauthor{Michelle Kunimoto}
\email{mkunimoto@phas.ubc.ca}

\author[0000-0001-9269-8060]{Michelle Kunimoto}
\affiliation{Department of Physics and Astronomy, University of British Columbia, 6224 Agricultural Road, Vancouver, BC V6T 1Z1, Canada}

\author[0000-0002-4461-080X]{Jaymie M. Matthews}
\affiliation{Department of Physics and Astronomy, University of British Columbia, 6224 Agricultural Road, Vancouver, BC V6T 1Z1, Canada}

\author[0000-0001-5172-4859]{Henry Ngo}
\affiliation{NRC Herzberg Astronomy and Astrophysics, 5071 W Saanich Road, Victoria, BC V9E 2E7, Canada}

\begin{abstract}
We present the results of an independent search of all $\sim$200,000 stars observed over the four year \textit{Kepler} mission (Q1$-$Q17) for multiplanet systems, using a three-transit minimum detection criteria to search orbital periods up to hundreds of days. We incorporate both automated and manual triage, and provide estimates of the completeness and reliability of our vetting pipeline. Our search returned 17 planet candidates (PCs) in addition to thousands of known \textit{Kepler} Objects of Interest (KOIs), with a 98.8$\%$ recovery rate of already confirmed planets. We highlight the discovery of one candidate, KIC-7340288 b, that is both rocky (radius $ \leq 1.6 R_{\bigoplus}$) and in the Habitable Zone (insolation between $0.25 $ and $ 2.2$ times the Earth's insolation). Another candidate is an addition to the already known KOI-4509 system. We also present adaptive optics imaging follow-up for six of our new PCs, two of which reveal a line-of-sight stellar companion within $4^{\prime\prime}$.
\end{abstract}

\section{Introduction}\label{sec:intro}

Ultraprecise photometry from space satellites like NASA's \textit{Kepler} mission has lead to a revolution in the discovery and characterization of planets beyond the solar system. Launched in 2009, \textit{Kepler} alone has confirmed 2345 planets out of 4765 announced candidates\footnote{Based off of confirmed and candidate \textit{Kepler} planets listed on the NASA Exoplanet Archive at https://exoplanetarchive.ipac.caltech.edu/docs/counts\textunderscore detail.html, accessed 2019 October 27.} from its original four year mission, accounting for more than half of all planets known today \citep{bor10, bor11, bat13, bur14, mul15, row15, cou16, tho18}.

\textit{Kepler} archival data, which is publicly available on the Mikulski Archive for Space Telescopes\footnote{http://archive.stsci.edu/} (MAST), has been a popular focus of independent searches for exoplanets over the years. The citizen science initiative known as ``Planet Hunters,'' for instance, was launched with the goal of involving the general public in \textit{Kepler} data analysis and planet detection. Since their first discoveries in \citet{fis12}, the Planet Hunters project has uncovered over 100 planet candidates (PCs) with \textit{Kepler} data \citep{lin13, sch13, wan13, sch14A, sch14B, wan15}. \citet{ofi13} is one of the earliest systematic searches in the literature, restricting the subset of stars searched to \textit{Kepler} Objects of Interest (KOIs), which are already known to contain transit-like signals. In Q0$-$Q6 data, they found 84 new candidates. \citet{hua13} performed a search of 124,840 stars observed in Q1$-$Q6, finding 150 new candidates. \citet{jac13} and \citet{san14} searched all $\sim$200,000 stars for ultrashort period planets, finding 4 new candidates in Q0$-$Q11 data and 16 new candidates in Q0$-$Q16 data respectively.

Even following the release of \textit{Kepler} catalogues based on all four years of \textit{Kepler} data \citep{cou16, tho18}, independent authors have shown that there are still scientifically interesting candidates to be found.  \citet{sha18} represents the first application of machine learning to searching \textit{Kepler} data for exoplanets, finding two new planets among a subset of Q1$-$Q17 KOIs including an eighth planet in a planetary system. \citet{kun18} also searched a subset of KOIs, finding four new candidates including a Neptune-sized candidate in the Habitable Zone. More recently, \citet{cac19} searched 156,717 stars for planets with orbital periods between 0.2 and 100 days, finding 97 new candidates.

Independent searches have also contributed to exoplanet occurrence rate studies. \citet{pet13} and \citet{dre15} are two such studies that used searches of \textit{Kepler} data as a precursor for occurrence rate statistics, focused on GK and M dwarfs respectively. Since all other estimates so far have been based on \textit{Kepler} catalogues, independent searches are unique contributions to the field, provided they are performed systematically and with measured detection efficiency. We use these papers as inspiration for our future plans to apply our search to occurrence rate estimates.

\subsection{Paper Outline}

So far, independent \textit{Kepler} searches have focused on only a subset of light-curves \citep[e.g.,][]{kun18, sha18} or a more limited range of orbital periods than examined by the \textit{Kepler} team \citep[e.g.,][]{wan15, cac19}. In this paper, we present an independent systematic search of the entirety of \textit{Kepler} data ($\sim200,000$ stars) for planets, using the same three-transit minimum detection criteria as the \textit{Kepler} team.

We describe our planet detection and vetting pipelines in Sections \ref{sec:search} and \ref{sec:vet}, which we make available for public use on Github\footnote{https://github.com/mkunimoto/Transit-Search-and-Vetting} under the BSD 3-Clause License. The main difference between our search and \textit{Kepler}'s is our use of a Box-Least Squares (BLS) algorithm \citep{kov02} and the associated effective BLS signal-to-noise ratio (S/N) for measuring the significance of a detection, instead of a wavelet-based algorithm \citep{jen10} and the associated Multiple Event Statistic (MES). Furthermore, while our vetting is largely inspired by the automated DR25 Robovetter \citep{tho18}, we introduce different tests and follow our automated vetting with a final manual vetting stage, similar to \citet{pet13}.

Our ability to differentiate between planets and noise-like false positives (FPs) is discussed in Section \ref{sec:performance}. We compare our results to the findings across all \textit{Kepler} catalogues in Section \ref{sec:kepler}, and detail 17 new PCs in Section \ref{sec:newcands}. These candidates are processed through further analysis, including centroid vetting, adaptive optics (AO) imaging follow-up, and astrophysical FP calculation to increase confidence in their planet status.

\section{Planet Detection Pipeline}\label{sec:search}

\subsection{Preparing the light-curves}

Q1$-$Q17 DR25 long-cadence light-curve files were downloaded from MAST. This photometry includes systematic corrections for instrumental trends and estimates of dilution due to other stars that may contaminate the photometric aperture \citep{stu14}. To initially set up the data, we used the \texttt{kfitsread} routine from the \textit{Kepler} Transit Model Codebase \citep{row16}, which includes code previously used by the \textit{Kepler} team for transit detection and characterization. \texttt{kfitsread} reads in each FITS file, removes data flagged as low quality, stitches all quarters of data together to create one continuous light-curve, and subtracts the median flux from each data point.

We then detrended each light-curve to filter out astrophysical and instrumental signatures using the \texttt{detrend5} routine. Each observation is corrected by fitting a cubic polynomial to a segment $W$ days wide centred on the time of measurement. A good choice of $W$ is longer than the duration of a typical transit (several hours) to prevent significant transit shape distortion, while short enough to adequately filter out astrophysical signatures (several days). The ideal choice of $W$ is star-dependent, as stars having varying levels of noise and intrinsic stellar variability. In an effort to reflect this, we detrended each light-curve using $W = 1$, 1.5, and 2 days, and measured the corresponding standard deviations $\sigma_{1}$, $\sigma_{1.5}$, and $\sigma_{2}$. We used $W = 1$ if $\sigma_{1}/\sigma_{1.5} < 0.8$ or $\sigma_{1}/\sigma_{2} < 0.8$, $W = 1.5$ if $\sigma_{1.5}/\sigma_{2} < 0.8$, and $W = 2$ otherwise, similar to the process used by \citet{dre15} to flag stars with greater levels of noise. In other words, a more aggressive detrend would only be favoured if it resulted in a significant decrease in overall light-curve variability.

We then 5$\sigma$-clipped outliers in the data. Only outliers in the positive flux direction were removed so as to leave any deep transits untouched. Lastly, we searched for data gaps within the detrended light-curves. Data gaps are frequently accompanied by sharp increases or decreases in flux that can interfere with the search for planets. For example, the \textit{Kepler} telescope would execute a 90$\degree$ roll every 90 days to reorient its solar panels, resulting in a break in observations of approximately one day. We defined a data gap as 0.75 or more days of missing photometry. Gaps are often accompanied by sharp increases or decreases in flux, which can interfere with the search for transits. Thus, we removed all data points within one day of the start and end of each gap.

\subsection{Searching for Transiting Planets}

We searched the light-curves using a box-fitting least squares (BLS) routine, based on the original algorithm by \citet{kov02} which was designed to identify periodic transit signals in time-series photometry. In the \textit{Kepler} Transit Model Codebase, this is available as \texttt{transitfind2}.

Once a possible transit was identified, its period $P$, epoch $T_{0}$, depth $T_{\text{dep}}$, and duration $T_{\text{dur}}$ were estimated. We calculated the S/N of an event to be the ratio of the mean transit depth integrated over the transit duration relative to the standard deviation of out-of-transit observations $\sigma_{\text{OT}}$,

\begin{equation}
\text{S/N} = \frac{\sqrt{N}}{\sigma_{\text{OT}}}T_{\text{dep}},\label{eqn:SN}
\end{equation}

\noindent where $N$ is the number of in-transit data points. To be more robust to outliers, $\sigma_{\text{OT}}$ was calculated using the Median Absolute deviation (MAD) with $\sigma_{\text{OT}} = 1.48\text{MAD}$ \citep{hoa83}. Eqn. \ref{eqn:SN} is comparable to the ``effective'' S/N mentioned in \citet{kov02}, and assumes that the depth of the transit is uniform. While this is a good approximation for small Earth-sized planets with central transits, relatively large planet-to-star radius ratios and/or large impact parameters can have significant ingress and egress durations. In these cases, the S/N will be overestimated, but this is expected to have minimal impact on the full assessment of PC events \citep{row14}.

For each light-curve, we searched for transit signals with S/N $>$ 6. After identifying a transit signal, we removed its associated events from the data and searched the residuals. This enabled sensitivity to multiplanet systems. We capped this multipassthrough search at five consecutive searches and set aside light-curves that reached this maximum for manual inspection. Usually this meant a particularly noisy light-curve was causing the BLS algorithm to return many obviously poor signals. If this was not the case, we would continue searching until no more S/N $> 6$ signals were detected.

\subsubsection{Choice of S/N Threshold}

The significance of a detection depends primarily on its associated S/N. The \textit{Kepler} team defined the so-called MES as their S/N, establishing an MES $>$ 7.1 threshold by a Monte Carlo approach to confine the false alarm rate due to statistical fluctuations to $<$ 1 for the \textit{Kepler} campaign \citep{jen02}. We differ in our definition of S/N, so our thresholds are not directly comparable. Instead, we follow the suggestion of \citet{kov02} that the threshold for a significant detection with the BLS algorithm is S/N = 6.

While it is true that the rate of false alarms increases rapidly toward lower signal-to-noise, the true floor depends on characteristics of the host star, and on behaviour of the instrument on timescales related to the properties of the transit candidates. Digging deeper into the noise increases the probabilities of discovering and characterizing transiting planets that are small and/or have long periods. These both represent regimes of great interest in the exploration of exoplanetary parameter space. Furthermore, searching to lower S/N can increase sensitivity of Transit Candidates (TCs) above more conservative cutoffs. S/N can often be underestimated, such as through the assumption of a box-shaped transit or the distortion of the transit shape from the detrending algorithm, causing a planet TC to be erroneously rejected. For these reasons, recent searches have begun to relax the noise floor, such as \citet{sha18} in which a BLS algorithm was used to search as low as S/N $=$ 5, and \citet{kun18}, in which members of our team searched down to S/N $=$ 6.

\subsection{Identification of TCs}

A total of 130,312 signals with S/N $>$ 6 were detected around the 198,640 stars searched. An overwhelming number of these are false alarms due to instrumental or astrophysical systematics in the time series. This is a weakness of using S/N as the only detection criterion. Thus, before following up each signal with the full suite of candidacy tests, we ran a first stage of vetting to discard likely false alarms. Signals that passed these tests were designated TCs.

Since the detrending process is destructive to the shape of a planetary transit and often results in a loss in S/N, we produced a re-detrended version of each light-curve before running the tests. We masked out each transit by excluding all observations within one transit duration of the central time of each transit. Then, we ran the detrending algorithm to determine the cubic polynomial fit to each segment of data as before, essentially only fitting to all out-of-transit observations. Finally, we unmasked the transit in the light-curve and used an extrapolation of the fit to estimate corrections during transit. These re-detrended light-curves are used for the remainder of the analysis in this paper unless otherwise specified.

\subsubsection{S/N Recalculation Test}
The signal of a planet in the re-detrended light-curve should still be strong enough compared to the noise to warrant transit candidacy. Thus, we require both the original and recalculated S/N to remain above 6.

\subsubsection{Robust Statistic Test}
A weakness of using the S/N to indicate the strength of a signal is that it is unable to discriminate between a consistent set of transit events of uniform depths and durations, and a chance combination of dissimilar events. This test calculates a new S/N, RS, using the median depth instead of the mean to reduce the influence of outliers on S/N. A signal passes this test with RS $>$ 6 in both the original and re-detrended light-curve.

\subsubsection{S/N Consistency Test}
This test examines the signal-to-noise ratios of each transit individually and compares them to what is expected based off the full transit S/N. Since the depths of individual transits of PCs should be equal to each other, the $i$th transit comprised of $n_{i}$ observations has an expected expected S/N of

\begin{equation}
\langle\text{S/N}_{i}\rangle = \frac{\sqrt{n_{i}}}{\sigma_{\text{OT}}}T_{\text{dep}}.
\end{equation}

\noindent This is compared to the actual S/N of the $i$th transit, S/N$_{i}$, using a $\chi^{2}$ statistic with $N_{T}$ degrees of freedom

\begin{equation}
\chi^{2} = \sum_{i=1}^{N_{T}} (\text{S/N}_{i} - \langle \text{S/N}_{i} \rangle)^{2}
\end{equation}

\noindent where $N_{T}$ is the number of transits. We define

\begin{equation}
\text{CHI} = \text{S/N}\bigg(\frac{\chi^{2}}{N_{T}}\bigg)^{-1/2}
\end{equation}

\noindent for use as the false alarm discriminator, requiring a candidate to pass with CHI $> 6$ in both the original and re-detrended light-curves.

\subsubsection{Number of Transits}
Each signal must have at least three transits. A minimum of two transits is required to determine orbital period, while requiring a third improves reliability of the period estimate, reduces false detections, and increases the overall S/N. Simply dividing the total length of observations by the orbital period to get an estimate of the number of transits is insufficient as some transits may lie in gaps in the data. To avoid counting transits in gaps, we only count the number of transits that occur at epochs where data exist within 0.5 transit durations of the midpoint.

After applying all of the above cuts, we were left with 33,322 TCs out of the 130,312 signals with S/N $>$ 6.

\section{Vetting Pipeline}\label{sec:vet}
While the first stage of vetting significantly reduces the rate of false detections, some of the 33,322 TCs could still be due to noise, systematics, or astrophysical FPs. Thus, a suite of diagnostic tests must be performed to confirm (or refute) the candidacy of each signal as a bona fide transiting planet.

A transit model fit, followed by each candidacy test, was run for each TC. Several of the candidacy tests require a transit model fit, and fitting better characterizes the candidate parameters. TCs that pass each of the automated candidacy tests, as well as a round of manual inspection, are upgraded to PCs.

Our vetting pipeline was largely inspired by \textit{Kepler}'s Robovetter, an automated vetting tool first used for \textit{Kepler}'s DR24 catalogue \citep{cou16} and again for DR25 \citep{tho18} (hereafter KDR25). Prior to the Robovetter, KOI catalogues were primarily based on manual inspection. While we hope to make our vetting pipeline completely automated in the future, manual inspection still plays an integral role in our vetting process. Thus, we note that the main goal of our automated candidacy tests is to reduce the number of FPs sufficiently enough that the number of transit candidates requiring manual review is feasible. This mentality influenced our candidacy test thresholds. When possible, we used the same or similar cutoffs as the \textit{Kepler}-equivalent tests. Otherwise, our cutoffs were empirically chosen with this goal in mind.

\subsection{Transit Model Fitting}\label{sec:modelfit}

We used a \citet{man02} quadratic limb darkening transit model assuming circular orbits,\footnote{Adapted from Ian's Astro-Python Codes at http://www.lpl.arizona.edu/\textasciitilde ianc/python/} fit to each transit with least-squares. Limb darkening parameters were taken from \citet{cla11} based on the known $T_{\text{eff}}$, $\text{log} g$, and [Fe/H] from the \citet{mat17} stellar properties catalogue. To speed up the fit process, data more than two transit durations from the centre of each transit were ignored.

\subsubsection{Fitted Parameters}
The model is parameterized by orbital period, transit epoch, ratio of planet and star radii ($R_{p}/R_{s}$), distance between planet and star at midtransit in units of stellar radius ($a/R_{s}$), impact parameter ($b$), and zero-point flux ($z$).

For initial guesses, $P$ and $T_{0}$ were taken from the BLS search results. $R_{p}/R_{s}$ was estimated as the square root of the BLS transit depth,

\begin{equation}
\frac{R_{p}}{R_{s}} = \sqrt{T_{\text{dep}}},
\end{equation}

\noindent while the initial guess for $z$ was 0. $a/R_{s}$ was estimated using Eqn. 8 in \citet{sea03}:

\begin{equation}
\frac{a}{R_{s}} = \bigg[\frac{(1+\sqrt{T_{\text{dep}}})^{2} - b^{2}(1 - \sin^{2}{\frac{\pi T_{\text{dur}}}{P}})}{\sin^{2}{\frac{\pi T_{\text{dur}}}{P}}}\bigg]^{1/2}
\end{equation}

\noindent with $b$ set to its initial guess and transit duration $T_{\text{dur}}$ set to the BLS estimate. Since the model is sensitive to $b$, the fit was run once for each $b = 0, 0.1, 0.2, ..., 0.9$. The fit with reduced $\chi^{2}$ closest to 1 was chosen to determine the best-fit parameters.

In case the transit model fit would fail to converge, a trapezoid fit parameterized by $T_{0}$, $R_{p}/R_{s}$, $z$, width of the flat part of transit, and slope of the sides of the trapezoid was used instead. $P$ was set fixed to the BLS-detected value.

\subsection{Candidacy Tests against Non-transit-like FPs}

The first candidacy tests aim to identify non-transit-like (NTL) FPs --- signals that do not resemble transiting or eclipsing objects. Frequently, candidates with low S/N and/or few transits are simply due to noise or instrumental artifacts. Candidates may also be due to quasi-sinusoidal signals such as pulsating stars or star spots.

\subsubsection{Transit Model Fit Test}
The transit model should fit the data better than a straight line, parameterized by the zero-point flux $z$. We compare each model fit's reduced chi-squared values

\begin{equation}
\chi_{\text{red}}^{2} = \frac{1}{\nu}\sum_{i=1}^{N}\frac{(y_{i} - m_{i})^{2}}{\sigma_{i}^{2}}
\end{equation}

\noindent where $y_{i}$, $m_{i}$, and $\sigma_{i}$ represent the flux, modeled flux, and error of the $i$th data point, and $\nu$ are the total degrees of freedom. Given $N$ points fitted, the transit model with 6 parameters will have $N - 6$ degrees of freedom while the straight line model with 1 parameter has $N - 1$. A signal passes this test if the reduced chi-squared of the transit model is less than the reduced chi-squared of the straight line model.

\subsubsection{Transit Model S/N Test}
The S/N of the model fit, MOD, should be slightly larger than the S/N of the signal due to a variety of reasons: namely, the model should match the shape of the TC better than the BLS square pulse, and the ephemerides are more refined. A significantly lower MOD than S/N calls into question the planetary origin of the signal. This test requires both MOD $>$ 6 and MOD/S/N $>$ 0.75.

\subsubsection{Depth Mean-to-Median Ratio Test}

The mean of all measured transit depths should be consistent with the median of all transit depths. Thus, the depth mean-to-median (DMM) ratio can be used to identify potential scenarios when a candidate is due to a systematic error. If the DMM value is significantly different from 1.0, it indicates that some transits have significantly different depths from the rest, and thus the candidate is unlikely to be astrophysical in origin. A candidate fails this test if DMM $>$ 1.5.

\subsubsection{Chases Test}

As described in Section \ref{sec:chases}, the \textit{Kepler} team developed an individual transit metric called Chases to assess the detection strength of transit events relative to nearby signals \citep{tho18}. Chases is only calculated for candidates with five or fewer transits. As in Appendix A.3.3 of KDR25, this test takes the median of the individual Chases metrics and fails candidates with a value less than 0.8.

\subsubsection{Uniqueness Tests}\label{sec:uni}

For a transit to be considered ``unique,'' there should not be any other transit-like events in the folded light-curve with a depth, duration, and period similar to the primary signal, in either the positive or negative flux directions. 

Two uniqueness statistics are calculated for each TC:

\begin{equation}
\sigma_{\text{U1}} = \frac{|d_{\text{pri}} - d_{\text{sec}}|}{\sqrt{\sigma_{\text{pri}}^{2} + \sigma_{\text{sec}}^{2}}}
\end{equation}

and

\begin{equation}
\sigma_{\text{U2}} = \frac{|d_{\text{pri}} - d_{\text{ter}}|}{\sqrt{\sigma_{\text{pri}}^{2} + \sigma_{\text{ter}}^{2}}}
\end{equation}

\noindent where $d_{\text{pri}}$, $d_{\text{sec}}$, and $d_{\text{ter}}$ are the depths of the TC event, second-largest event, and third-largest event, respectively, and $\sigma_{\text{pri}}$, $\sigma_{\text{sec}}$, and $\sigma_{\text{ter}}$ are their uncertainties. Secondary and tertiary events may be in either the positive or negative flux direction. A TC must have both $\sigma_{\text{U1}} > 3.0$ and $\sigma_{\text{U2}} > 3.0$ (i.e., at least 3$\sigma$ significance). 

Running this analysis on the re-detrended light-curve is a good choice when the transit is due to a planet, as it ensures the planet transit is not distorted by detrending and the test correctly indicates strong uniqueness. However, a noise TC is essentially the only noise in the light-curve that is not detrended in this version of the light-curve, making an indication of uniqueness against the rest of the noise misleading. Thus, we also perform this analysis on the original light-curve, requiring a slightly lower 2$\sigma$ significance to pass.

The \textit{Kepler} team also developed their own ``model-shift uniqueness test,'' which is publicly available on GitHub\footnote{https://github.com/JeffLCoughlin/Model-Shift} \citep{cou17a} and described in Appendix A.3.4 of KDR25. While the statistics described above take into account the uniqueness of the TC in the form of a square pulse, this test takes into account the full transit shape as follows. After removing outliers, the best-fit model of the primary transit is used to measure the best-fit depth at all other phases. The two deepest events aside from the primary event (called the secondary and tertiary events) and the most positive flux event are all identified. The significances of these events ($\sigma_{\text{pri}}$, $\sigma_{\text{sec}}$, $\sigma_{\text{ter}}$, and $\sigma_{\text{pos}}$) are computed by dividing their depths by the standard deviation of the light-curve residuals outside of the primary and secondary events, assuming white noise. The amount of systematic red noise in the light-curve on the timescale of the transit is also computed, as the standard deviation of the best-fit depths at phases outside of the primary and secondary events. Taking the ratio of the red noise to the white noise gives the value $F_{\text{red}}$. $F_{\text{red}} = 1$ means there is no red noise in the light-curve.

The threshold at which an event is considered statistically significant is given by 

\begin{equation}
F\!A_{1} = \sqrt{2}\text{ erfcinv}\bigg(\frac{T_{\text{dur}}}{P\cdot N_{\text{TCs}}}\bigg).
\end{equation}

\noindent Here $N_{\text{TCs}}$ is the number of transit candidates examined, the quantity $P/T_{\text{dur}}$ represents the number of independent statistical tests for a single target, and erfcinv is the inverse complementary error function. Similarly, the threshold at which the difference in significance between two events is considered to be significant is given by

\begin{equation}
F\!A_{2} = \sqrt{2}\text{ erfcinv}\bigg(\frac{T_{\text{dur}}}{P}\bigg).
\end{equation}

The following quantities are used as decision metrics:

\begin{equation}
M\!S_{1} = F\!A_{1} - \sigma_{\text{pri}}/F_{\text{red}},
\end{equation}

\begin{equation}
M\!S_{2} = F\!A_{2} - (\sigma_{\text{pri}} - \sigma_{\text{ter}}),
\end{equation}

and

\begin{equation}
M\!S_{3} = F\!A_{2} - (\sigma_{\text{pri}} - \sigma_{\text{pos}}).
\end{equation}

\noindent A candidate fails the test if either $M\!S_{1} > -3$, $M\!S_{2} > 1$, or $M\!S_{3} > 1$. These criteria ensure that the primary event is statistically significant when compared to the systematic noise level of the light-curve, the tertiary event, and the positive event, respectively.

\subsubsection{Transit Shape Test}

The transit shape test determines if the measured depth deviates from the mean value more in the positive flux direction, negative flux direction, or are symmetrically distributed in both directions. The SHP metric, provided alongside the model-shift uniqueness test from \citet{cou17a}, is defined by 

\begin{equation}
\text{SHP} = \frac{F_{\text{max}}}{F_{\text{max}} - F_{\text{min}}}
\end{equation}

\noindent where $F_{\text{max}}$ and $F_{\text{min}}$ are the maximum and minimum measured flux amplitudes, respectively. Since the light-curve is normalized, $F_{\text{max}}$ is always a positive value and $F_{\text{min}}$ is always negative. SHP lies between 0 and 1, where 0 indicates the light-curve only decreases in flux, consistent with a planet transit, and a value near 1 indicates the light-curve only increases in flux, such as for a lensing event or systematic outlier. A candidate passes with SHP $< 0.5$.

\subsubsection{Single Event Domination Test}

Assuming all individual transits have equal S/Ns, S/N$_{\text{I}}$, the full transit S/N given in Eqn. \ref{eqn:SN} can be rewritten as

\begin{equation}
\text{S/N} = \sqrt{N_{T}} \text{ S/N}_{\text{I}}
\end{equation}

\noindent where $N_{T}$ is the number of individual events. It follows that if the largest individual transit's S/N value, S/N$_{i,\text{max}}$, divided by the S/N is much larger than $\sqrt{N_{T}}$, the calculation of the candidate's S/N is likely dominated by one of the individual events.

A candidate fails this test if S/N$_{i,\text{max}}$/S/N$ > $ 0.8, as in the \textit{Kepler} team's own signal event domination test (Appendix A.3.5 of KDR25). Only candidates with $P > 90$ days are tested, as short-period candidates often have a large number of individual transit events, increasing the chance of one event coinciding with a large systematic feature.

\subsubsection{Individual Transit Metrics}\label{sec:chases}

This series of metrics examines individual transits and flags those that fail. After removing flagged events, the resulting signal must still have at least three transits and S/N $>$ 6.

\textit{Rubble Metric}. As per the \textit{Kepler} team's ``Rubble'' metric described in Appendix A.3.7.1 of KDR25, transit events may be missing a significant amount of data, either during transit or before and/or after. For each event we count the number of data points within one transit duration of the centre of the transit and divide this by the number of cadences expected given 29.42 minutes per cadence. An event is flagged if this value is less than 0.75 as in KDR25.

\textit{Chases Metric}. The \textit{Kepler} team developed the ``Chases'' metric to identify NTL events in long period, low S/N candidates by mimicking the tendency of human vetters to classify transits that ``stand out'' as PCs (see Appendix A.3.7.3 of KDR25). Chases uses the Single Event Statistic (SES) time series generated by the Transit Pipeline Search (TPS) module of the \textit{Kepler} Pipeline \citep{jen17}, which measures the significance of a signal centred on every cadence. A transit produces a peak in the SES time series. 

We created an analogous time series of S/N values centred on every cadence for the purpose of this test. The Chases metric is determined by first identifying the maximum S/N value for cadences in transit, S/N$_{\text{max}}$. The S/N time series is searched for $\Delta_{t}$, the time of the closest signal with $|\text{S/N}| > $ 0.6 S/N$_{\text{max}}$. As in KDR25, the search range starts at 1.5 $T_{\text{dur}}$ from midtransit, up to a maximum $\Delta_{\text{tmax}} = P/10$, on either side of the transit candidate signal. The final Chase metric is determined as $C_{i} = \text{min}(\Delta_{t},\Delta_{tmax})/\Delta_{tmax}$.

A value of $C_{i} \approx 0$ indicates an event of comparable strength to the transit is close to the transit event, while a value of $C_{i} = 1$ indicates there is no comparable peak or trough, and the transit is unique.

Chases metrics are only computed for TCs with five or fewer transit events, as these events are expected to be especially significant in order to combine to have S/N $>$ 6. Events with $C_{i} < 0.01$ are flagged.

\textit{Negative Significance}. A valid transit should only be comprised of events corresponding to decreases in the flux. Any individual event with S/N $<$ 0, indicating a flux increase, is flagged. 

\subsection{Candidacy Tests against Eclipsing Binary FPs}

TCs that pass the previous tests are designated transit-like. However, some may still be nonplanetary in origin. One of the most common types of astrophysical FPs are eclipsing binary stars (EBs), which could just graze the target star enough for the eclipse depth to be consistent with a planet transit.

Transit-like FPs may also be due to off-target signals, such as background eclipsing binaries or planet transit signals coming from off-target sources. These scenarios can typically be indicated by identifying significant centroid offsets. Our vetting pipeline does not currently incorporate automated tests to identify these FPs. However, we later perform centroid analysis as part of a more in-depth analysis of new PCs.

\subsubsection{Significant Secondary Test}
A secondary eclipse could manifest as the secondary event in the phased light-curve. This test follows the same procedure as the uniqueness test, but assesses the uniqueness of the secondary event rather than the primary using a new set of metrics (see Appendix A.4.1.2 of KDR25):

\begin{equation}
M\!S_{4} = F\!A_{1} - \sigma_{\text{sec}}/F_{\text{red}},
\end{equation}

\begin{equation}
M\!S_{5} = F\!A_{2} - (\sigma_{\text{sec}} - \sigma_{\text{ter}}),
\end{equation}

and 

\begin{equation}
M\!S_{6} = F\!A_{2} - (\sigma_{\text{sec}} - \sigma_{\text{pos}}).
\end{equation}

\noindent If either $M\!S_{4} > 2$, $M\!S_{5} > 1$, or $M\!S_{6} > 1$, the candidate fails due to having a significant secondary event.

\subsubsection{Planet Candidates with Significant Secondaries}

Significant secondary events are not necessarily confirmation of an eclipsing binary FP.

Following Appendix A.4.1.3 of KDR25, if the primary and secondary events have statistically indistinguishable depths and the secondary is at phase 0.5, a PC may have been detected at twice its actual orbital period. Thus, a TC is allowed to pass the Significant Secondary test if $\sigma_{\text{pri}} - \sigma_{\text{sec}} < F\!A_{2}$ and the phase of the secondary is within $T_{\text{dur}}/4$ of 0.5.

Additionally, some giant planets close to their stars such as hot Jupiters, can have eclipses due to planetary occultations via reflected light and thermal emission. The depths of these eclipses are typically much smaller than those due to eclipsing binaries, while the properties of the primary events themselves should still be consistent with a planetary origin. A TC is allowed to pass the Significant Secondary test if the depth of the secondary is less than 10$\%$ of the primary, the impact parameter is less than 0.95, and the planet's radius as derived using the fitted parameter $R_{p}/R_{s}$ is $R_{p} < 30 R_{\bigoplus}$.

\subsubsection{Odd-Even Depth Tests}

Secondary eclipses could also be erroneously marked as half of the primary events if the eclipsing binary is detected at half its actual period and its eclipses would otherwise occur at phase 0.5. These eclipsing binaries can be identified as candidates with significantly different odd and even transit depths. As with the S/N Consistency Test, the Odd-Even Depth Tests are only used for candidates with $P < 90$ days.

An odd-even depth statistic is calculated for each TC:

\begin{equation}
\sigma_{\text{OE1}} = \frac{|d_{\text{odd}} - d_{\text{even}}|}{\sqrt{\sigma_{\text{odd}}^{2} + \sigma_{\text{even}}^{2}}} 
\end{equation}

\noindent where $d_{\text{odd}}$ and $d_{\text{even}}$ are the median of all points within 30 minutes of the centre of odd and even transits, respectively, and $\sigma_{\text{odd}}$ and $\sigma_{\text{even}}$ are the standard deviations of those points. For the case of trapezoidal model fits, all points making up the flat part in transit are also included. A TC fails if $\sigma_{\text{OE1}} > 1.0$.

A second odd-even depth statistic is also calculated as part of the model-shift uniqueness test. This method takes into account the full transit shape as well as the noise level of the full light-curve. However, it is more susceptible to outliers and systematics compared to the first statistic. Thus, we use a lenient requirement of $\sigma_{\text{OE}} - FA_{1} < 10$.

\subsubsection{V-shape Test}

Candidates where the ingress and egress times are a significant fraction of the total transit duration are most likely FPs. Planetary transits typically have a U-shape, while V-shaped transits are often created by EBs. The V-shape metric is defined as $V = b + R_{p}/R_{s}$, in order to identify eclipsing binaries both due to grazing eclipses (large impact parameter, $b$) and being too deep (large $R_{p}/R_{s}$). A candidate fails with $V > 1.05$.

\subsection{Manual Inspection}

The final round of vetting involves a visual inspection of each of the TCs that passed the automated vetting stage. We look at the full light-curve, the light-curve phase-folded to the transit's period, a close-up of the transit in the phase diagram, and a side-by-side comparison of odd and even transits. The latter two images include the data averaged into 30 minute bins as well as the model fit to the light-curve to assess the fit. While the previous tests are able to remove the majority of FPs and attempt to mimic decisions made by human vetters, manual inspection still serves as an important ``reality check'' that each passing TC is convincing enough to be promoted to PC.

A total of 5608 of the 33,322 TCs survived the automated vetting stage. Of those, 3972 passed manual vetting to become PCs.

\section{Assessing Vetting Performance}\label{sec:performance}

Ideally, the vetting pipeline is accurate when classifying planets as planets and FPs as FPs. Realistically, no pipeline is perfect, and sacrifices must be made to achieve balance. For example, lenient candidacy test thresholds will cause more real planets to be accepted, at the cost of more FPs incorrectly passed as PCs. This will call the validity of any new PCs coming out of the pipeline into question.

Two useful metrics used to assess vetting performance are the completeness (the fraction of true transiting planets passed as PCs) and reliability (the fraction of PCs that are actually planets). These numbers are unknown. However, we can estimate them using simulated data. Injecting fake planet transits into real \textit{Kepler} data and vetting the resulting detections gives an estimate of the vetting completeness. Likewise, we can simulate FPs to estimate how often the vetting process mistakenly labels FPs as planets.

We note that this work only attempts to measure reliability against noise FPs. These are the largest concerns for low-S/N TCs, among which we expect most of our new PCs to lie.

\subsection{Simulated Data}

We injected 120,642 planet transits into the light-curves and prepared, searched, and vetted the data using the same process as for the actual observed data. The only exception was that we tested the manual component on a small subset of the injected detections. The overall process is consistent with injection and recovery tests performed for completeness measurements of other independent pipelines in the literature \citep[e.g.][]{pet13, dre15}. We injected signals log-uniformly distributed over the ranges $0.5 < P < 500$ days and $0.5 < R_{p} < 16.0$ $R_{\bigoplus}$. Each transit was created using a quadratic limb darkening \citet{man02} model, with impact parameters uniformly distributed between 0 and 1 and assuming circular orbits.

For testing against noise FPs, the simulated data should allow realistic signals with noise properties similar to the real data, while ensuring no possibility of detecting true exoplanets still in the light-curve. To achieve this, we took the 198,640 light-curves originally searched and inverted them. Essentially, this recreated the Inverted (INV) set of simulations described in \citet{chr17} for their own vetting tests. Any ``transit'' would actually be a positive flux increase in the observed data, and thus not a planet. 

The \textit{Kepler} team also created a Scrambled (SCR) data set for testing against noise FPs, corresponding to reordering of the \textit{Kepler} quarters by yearly chunks. Three orders were created, as described in \citet{cou17b}. We tested our pipeline on Scrambled Group 1 (SCR1). 

\subsection{Vetting Completeness}

\begin{figure}[h]
\centering
\includegraphics[width=0.5\textwidth]{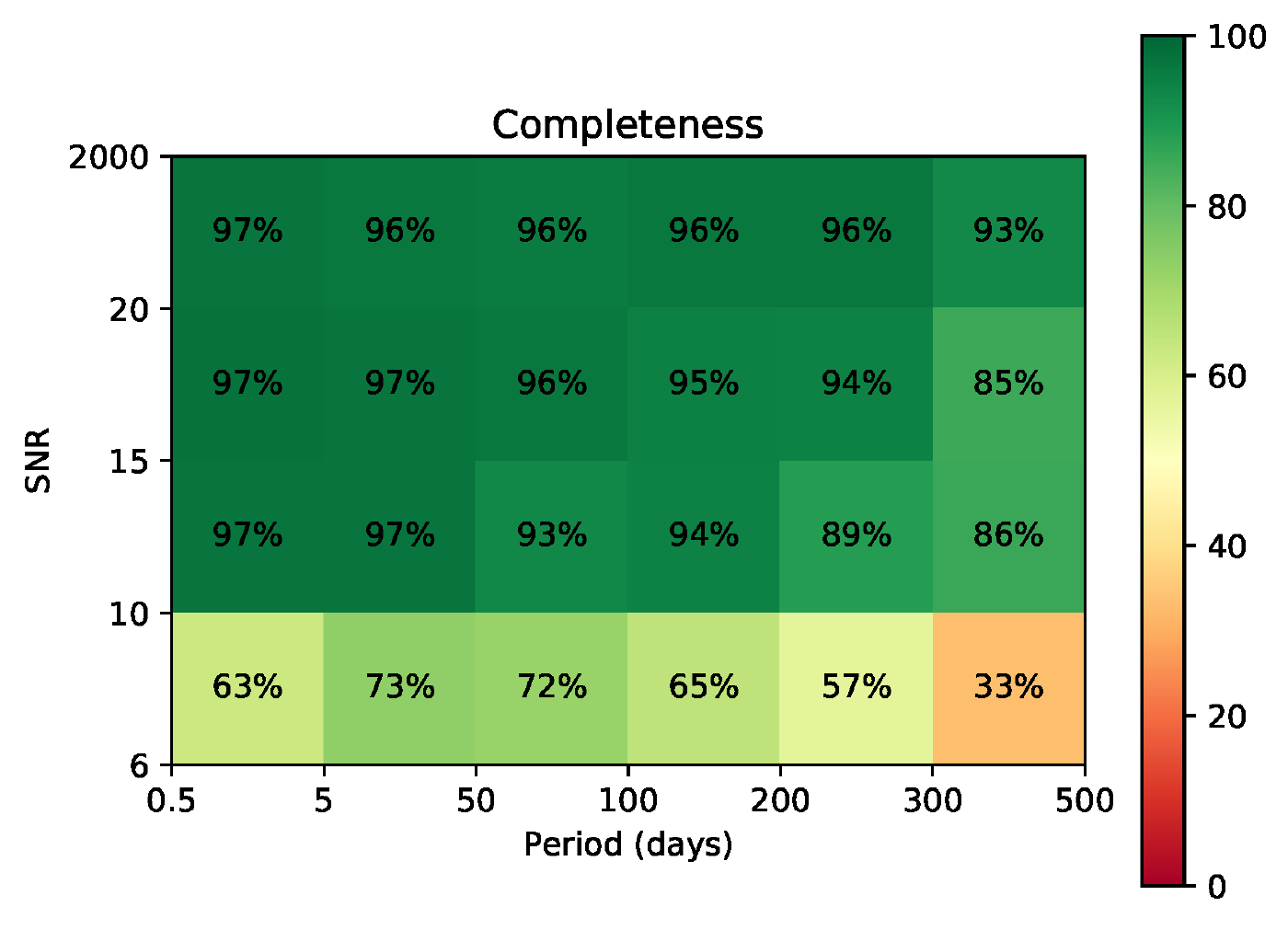}
\caption{Completeness of the vetting pipeline based on running the automated tests on simulated planet TCs.}\label{fig:comp}
\end{figure}

Of the 48,610 simulated TCs detected by our search pipeline, 45,676 (94.0$\%$) passed the automated candidacy tests. More usefully, completeness is binned over period and S/N in Fig. \ref{fig:comp}. As expected, completeness decreases with lower S/N and larger period, where most noise TCs would be expected to lie. 

\begin{figure}[h]
\centering
\includegraphics[width=0.5\textwidth]{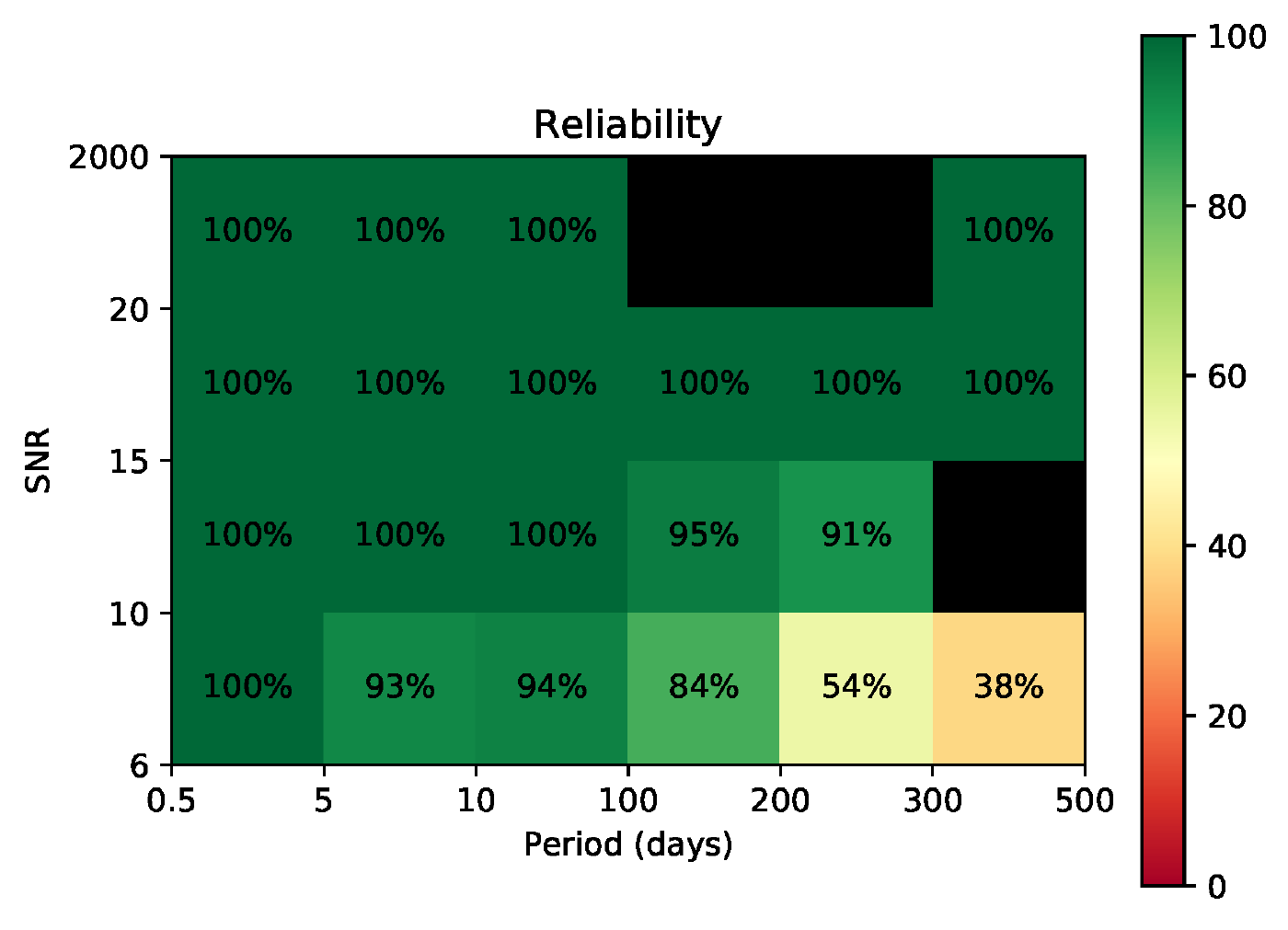}
\caption{Reliability of the vetting pipeline against noise false positives based on running the automated and manual tests on simulated noise TCs (inverted + scrambled). Bins with fewer than three candidates or fewer than 20 simulated noise FPs are not shown.}\label{fig:rel}
\vspace{0.1in}
\end{figure}

The vetting process also involves a manual component in the form of the final visual inspection. Performing a full completeness measurement that takes this into account is difficult due to the presence of human bias. Additionally, it is infeasible to manually review each of the simulated TCs that passed the automated tests. Thus, we chose a random subset of 1,000 of the passing TCs to review. In an attempt to remove human bias, we combined these TCs with all passing TCs from the simulated false positive set, and removed any labeling that would indicate the origin set of each TC. We failed 15 of the 1000 planet TCs (1.5$\%$). Overall, we expect the manual vetting to reduce our overall vetting completeness by 1-2$\%$ from the 94$\%$ success rate of the automated component.

\subsection{Vetting Reliability}

Our pipeline identified 15,283 TCs in the INV set, and 12,103 in SCR set. The automated tests failed 14,494 (94.8$\%$) and 11,222 (92.7$\%$) of these, respectively. We then manually reviewed all surviving TCs, combined with the simulated planet TCs as described above. We found that the majority of the FP TCs were high-S/N events that had obviously asymmetric transits, making them easy to distinguish from bona fide planets. Overall, we failed all but 8 TCs in the INV set and 28 TCs in the SCR set, giving a total success rate of 99.8$\%$.

The fraction of FPs successfully classified as FPs is also known as the effectiveness of the pipeline. This can be combined with the final vetting results to estimate the reliability. Letting $E$ denote the effectiveness, KDR25 define reliability $R$ as

\begin{equation}
    R = 1 - \frac{N_{\text{FP}}}{N_{\text{PC}}} \bigg(\frac{1-E}{E}\bigg).
\end{equation}

\noindent where $N_{\text{PC}}$ and $N_{\text{FP}}$ are the numbers of observed PCs and FPs identified by the vetting pipeline, respectively. 

Considering we identified 3971 PCs and 29,348 FPs out of all TCs, our effectiveness of 99.8$\%$ gives an overall reliability of 98.3$\%$. However, plotting reliability as in Fig. \ref{fig:rel} reveals areas in period-S/N space where the pipeline is particularly unreliable, namely S/N $<$ 10 and $P > 200$ days. While our effectiveness in this regime was $99.6\%$, we only identified 10 PCs compared to 1214 FPs. 

\section{Results Compared to \textit{Kepler}}\label{sec:kepler}

We used the federation process described in \citet{mul15} to match 3915 of our 3972 PCs with known KOIs identified by the \textit{Kepler} team, accumulated over all \textit{Kepler} catalogues. The NASA Exoplanet Archive\footnote{https://exoplanetarchive.ipac.caltech.edu/} was access on 2019 May 09.

\subsection{Confirmed Planets}

We successfully detected and passed 2268 of the 2295 (98.8$\%$) planets confirmed by \textit{Kepler}, defined as having an Exoplanet Archive Disposition of CONFIRMED and a Disposition Using \textit{Kepler} Data of CANDIDATE on the NASA Exoplanet Archive.

Another 20 were marked as TCs, but failed our candidacy tests. Upon manual inspection, it appears that significant Transit Timing Variations (TTVs) were to blame for the failing of five confirmed planets (KOI-142.01, 227.01, 377.01, 377.02, and 884.02). The other 15 planets were either very close to passing or only failed a single test (KOI-46.02, 172.02, 701.04, 1236.03, 1574.02, 2038.03, 2298.02, 2365.02, 2533.01, 3458.01, 4034.01, 4384.01, 5416.01, 5706.01, and 7016.01). We found that these planets often had much lower calculated S/N than what was listed on the NASA Exoplanet Archive (for example, KOI-4384.01 had an S/N of only 6.3 according to our pipeline, but 12.2 from \textit{Kepler}). Thus, it is likely that the lack of whitening in our search pipeline can explain these discrepancies, rather than these signals being intrinsically poor candidates.

Six were detected but failed to meet the requirements to be a TC (KOI-179.02, 245.03, 490.02, 1274.01, 1718.02, and 3234.01). KOI-179.02, KOI-490.02, and KOI-1274.01 had only one or two detected transits, lower than the required three. KOI-1718.02 had a barely failing RS (5.8), KOI-3234.01 had too low of a CHI value (4.6), and KOI-245.03 failed both. Only a single confirmed planet, KOI-4846.01, was missed entirely.

\subsection{Candidate Planets}

We successfully detected and passed 1447 of the 2421 ($59.8\%$) known \textit{Kepler} candidate planets, defined as having both dispositions listed as CANDIDATE.

The lower rate of recovery among PCs is to be expected given that confirmed planets typically have higher S/N and transit shapes more clearly consistent with a planetary origin. Furthermore, we note that 576 (around 60$\%$) of the 974 candidates missed or failed by our pipeline were also not detected by \textit{Kepler}'s DR25 pipeline.

\subsection{False Positives}

Of our PCs, 193 have both dispositions listed as FALSE POSITIVE. Of these, 109 were flagged by \textit{Kepler} as FPs solely due to having a significant centroid offset, while another 71 had an ephemeris match indicating contamination. Given that we did not incorporate centroid tests or ephemeris matching between KOIs into our vetting pipeline, it is unsurprising that we would pass these as candidates. However, we address both of these issues for our new candidates.

Six of our PCs have a Disposition Using \textit{Kepler} Data of FALSE POSITIVE, but an Exoplanet Archive Disposition of CONFIRMED (KOI-125.01, 129.01, 631.01, 1416.01, 1450.01, and 3032.01). Furthermore, one of our PCs is the sole KOI on the NASA Exoplanet Archive with a Disposition Using \textit{Kepler} Data of CANDIDATE, but an Exoplanet Archive Disposition of FALSE POSITIVE (KOI-242.01). 

\section{New Planet Candidates}\label{sec:newcands}

After removing all federated \textit{Kepler} confirmed planets, candidate planets, and FPs from our PC list, we were left with 57 new PCs. All of our new PCs have low S/N, ranging from S/N $=$ 7.1 to 10.7, which are the kind of candidate most susceptible to being missed by detection pipelines. For the remainder of this section, we list candidates according to their Kepler Input Catalogue (KIC) number \citep{bro11}.

We performed additional follow-up analysis on each of the candidates to more rigorously assess their candidacy. This involved ephemeris matching, centroid analysis, AO imaging follow-up (in select cases), and false positive probability (FPP) calculation. We also performed a Markov Chain Monte Carlo (MCMC) refit to each transit, taking into account dilution effects of companions detected in the AO imaging. These fits produced our final reported planet parameters.

As discussed in Section \ref{sec:performance}, we found that our pipeline has significantly lower reliability for S/N $<$ 10 and $P > 200$ days than other regimes. Our reliability estimate would indicate that $\sim$5 of the 10 PCs detected with these properties are likely FPs. Considering that our pipeline contributed five of these PCs while the other five are known KOIs, we made the conservative decision to downgrade the new candidates with these properties to FP status and continue the analysis with the remaining 52 PCs.

\subsection{Ephemeris Matching}

Light that contributes to the target's light-curve may not necessarily originate from the target. If this contamination is caused by a star with a variable signal, then the same signal will be observed in the target with reduced amplitude due to dilution. Thus, if two signals have the same ephemeris, then at least one of them is an FP due to contamination.

We compared the periods and epochs of each new PC to all KOIs, searching for cases where

\begin{equation}
    |P - P_{\text{match}}| \leq \text{min}(\text{2 hours}, 0.001P)
\end{equation}
\noindent and 
\begin{equation}
    |T_{0} - T_{0, \text{match}}| \leq \text{min}(\text{4 hours}, 0.001P)
\end{equation}

\noindent as in \citet{dre15}. We did not find any matches.

\subsection{Stellar Variability}\label{sec:variability}

False positives may also be due to stellar variability that was not fully removed during the detrending process. In particular, failing to remove the rotation signal of the star can create a periodic, transit-like signal in the light-curve. Finding a match between the orbital period of the PC and the rotation period of the host star would indicate an FP due to stellar variability.

We ran each un-detrended light-curve through the Lomb-Scargle periodogram in Astropy \citep{ast13, ast18} and searched for cases where the rotation period (or a multiple thereof) matched the detected orbital period of the PC. We determined rotation periods from the period corresponding to the highest peak in the periodogram. Fourteen of our host stars also had rotation periods listed in \citet{mcq14}. Furthermore, we manually inspected each periodogram to search for smaller peaks or excess noise at the orbital periods, which could confound the search for planets. For periods that corresponded to a peak, we determined its false alarm probability (the probability of measuring a given peak height under the assumption that the noise is Gaussian with no periodic component), and flagged cases where the power had a probability less than 0.05. Following this analysis, we identified 30 of our 52 PCs as likely FPs due to stellar variability.

We also investigated whether or not transits would change or disappear depending on different stellar variability removal methods. We used the \texttt{biweight} time-windowed slider implemented in the W$\overline{\text{o}}$tan Python package,\footnote{https://github.com/hippke/wotan} which was identified by \citet{hip19} as the ideal method for recovering transits from light-curve data based on comprehensive comparison of common detrending routines. Using window lengths of 0.5, 1, and 2 days, we detrended each raw light-curve, examined the data phased at the planet period, and calculated the S/N by measuring the depth of the transit and assuming the same duration, period, and epoch as the PC. The only exception was that we did not use a 0.5 day width for transits with durations greater than 0.2 days, so as to avoid significantly distorting the transit itself. Four of the PCs had either S/N $<$ 6 (KIC-6937870 b) or the transit itself was inconsistent in shape and duration with the original PC (KIC-2985262 b, KIC-6380164 b, and KIC-10419787 b) using one of these alternate detrends. Then, we re-detrended the remaining 18 light-curves after masking out the transits, re-examined the phase diagram, recalculated the S/N, and fit a least-squares transit model. We found that regardless of window length used, the S/N remained above 6 for all 18 PCs, and the model best-fit parameters were within 1$\sigma$ of the results using our original detrending algorithm, giving further confidence that these signals were not an artifact of stellar variability.

\subsection{Centroid Analysis}
We used the difference imaging method described in \citet{bry13} to identify background FPs for the remaining 18 PCs, which is summarized here. We downloaded all necessary target pixel files from MAST. For each quarter, we combined all in-transit cadences to produce an average in-transit pixel image. We took an equal number of cadences on either side of the transit to produce an average out-of-transit pixel image. Subtracting the in-transit from the out-of-transit image gives the difference image. We fit the \textit{Kepler} Pixel Response Function (PRF) to each of the out-of-transit and difference images. The PRF is defined as the composite of \textit{Kepler}’s optical point-spread function, integrated spacecraft pointing jitter during a nominal cadence, and other systematic effects, and is represented as a piece-wise continuous polynomial on a subpixel mesh \citep{bry10}. We used \texttt{PyKE} \citep{sti12}, which provides fitting of the \textit{Kepler} PRF as a function of flux, centre positions, width, and rotation angle to a given target pixel file. Respectively, the centre positions of the out-of-transit and difference images give the location of the target star and transit source, providing a direct measurement of the centroid offset for that quarter.

\citet{bry13} discuss that the difference images for low-S/N transits are typically noise dominated. The difference image can appear significantly different from the out-of-transit image in one quarter, and may show the transit at other locations or on the target star in others. Thus, we attain a more reliable estimate of the centroid offset and its uncertainty by robustly averaging all quarterly offsets. We also use the bootstrapping technique described in \citet{bry13} to estimate the uncertainty in the result, taking the larger of the two values. Given the $Q$ measured offsets (where $Q$ is the number of quarters analyzed), we produce $Q^{2}$ different sets, randomly selecting from the list of offsets to fill each set. We then find the average of each set. The standard deviation of the $Q^{2}$ averages provides the bootstrap uncertainty estimate.

Following \citet{bry13}, we classify candidates as FPs if they have a 3$\sigma$ significant offset larger than 2$^{\prime\prime}$, or 4$\sigma$ offset larger than 1$^{\prime\prime}$. One of our PCs (KIC-3336146 b) met these thresholds and was reclassified as an FP. We complete the rest of our analysis with the remaining 17 new PCs.

\subsection{AO Observations}\label{sec:AO}

We obtained AO follow-up imaging for six of our host stars, prioritizing our potentially rocky candidates. The uses of AO data are twofold: first, nearby stars dilute the observed transit depth, resulting in an underestimated planet radius. This is especially of concern for small, rocky planets due to their relative rarity, and those just under the proposed 1.6$R_{\bigoplus}$ ``rocky limit,'' past which most planets are not rocky \citep{rog15}. Second, a contaminant star could be the source of an FP signal, whether as a background or foreground eclipsing binary. Contrast curves derived from the AO images serve as effective constraints for unseen companions in our FPP calculations. We found that our AO images reduced FPPs by a factor of $\sim16$ on average (see Section \ref{sec:FPP}), emphasizing the usefulness of AO follow-up for planet validation.

We collected observations of three stars on the Gemini North 8.1 m telescope in the $K_{s}$ band with the Natural Guide Star (NGS) AO assisted Near InfraRed Imager and spectrograph (NIRI, \citealt{hod03}). Data were taken between 2018 July and 2019 June (Program ID GN-2018B-Q-134). Another three stars were observed with the Laser Guide Star (LGS) AO system and NIRI in 2019 July as part of a Fast Turnaround program (Program ID GN-2019A-FT-213). Total exposure time for each target was between 5 and 6 minutes. We used the $f/32$ NIRI camera, providing a plate scale of $0.022^{\prime\prime}$ $\text{px}^{-1}$ and a $22^{\prime\prime}\times22^{\prime\prime}$ field of view.

\begin{figure}[ht!]
\centering
\includegraphics[width=\linewidth]{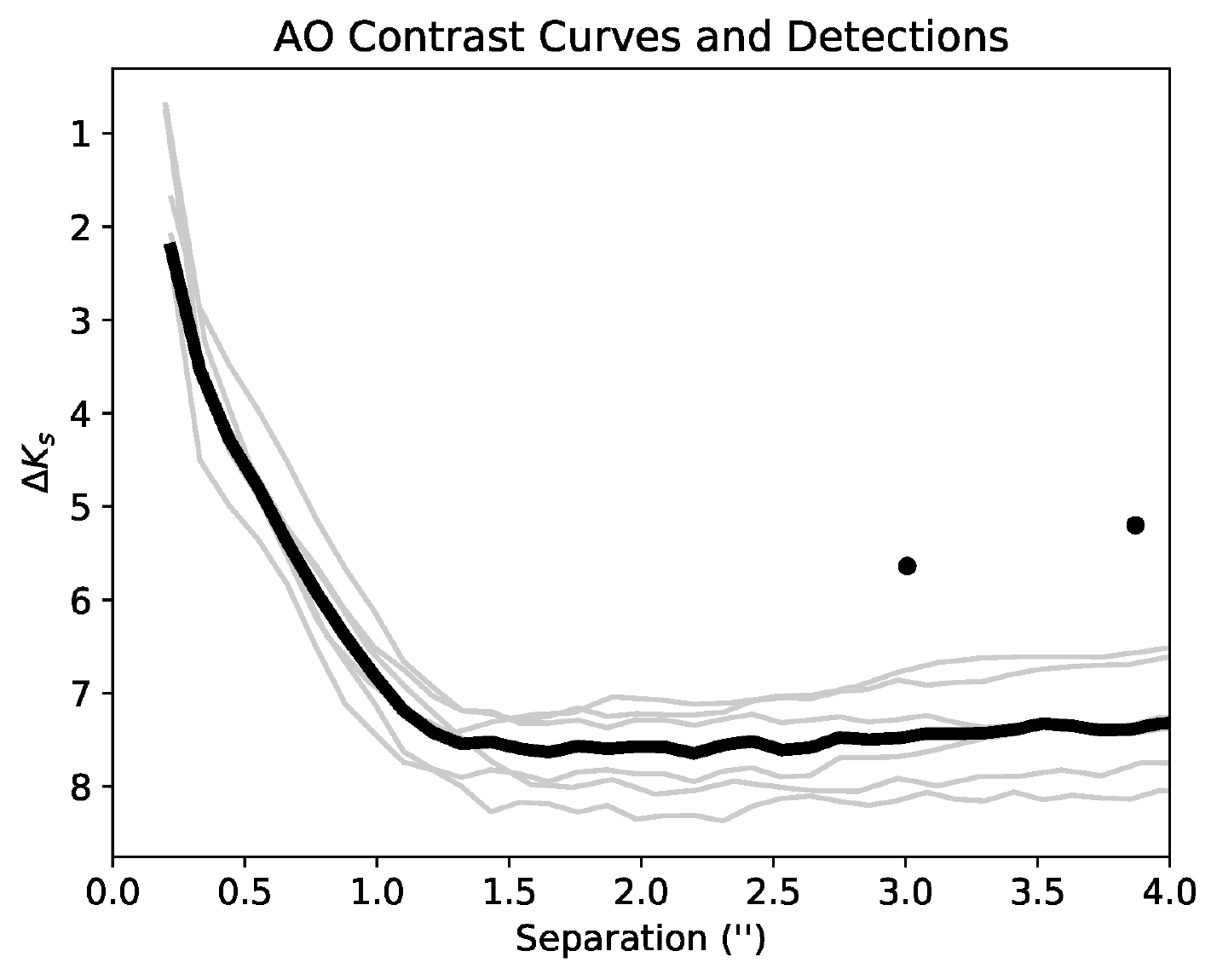}
\caption{In grey are the 5$\sigma$ contrast curves for all Gemini NGS-AO- and LGS-AO-observed targets. The black curve indicates the median. The black points indicate the best-fit locations of detected companions, determined from the AO images.} \label{fig:medcontrast}
\end{figure}

Data were reduced by median-stacking each dark-subtracted and flat-divided image into a single AO image per star. We manually inspected each image for artifacts and potential companions in order to mask them out before computing 5$\sigma$ contrast curves. To calculate each curve, we used the procedure outlined in \citet{ngo15}, computing the standard deviation of flux values in a series of annuli with widths equal to twice the FWHM of the central star's point-spread function. Fig. \ref{fig:medcontrast} shows all 5$\sigma$ contrast curves along with the median to indicate our typical sensitivity. We provide the data for all contrast curves in Table \ref{tbl:ccurves}, out to a maximum separation of $\sim$8$^{\prime\prime}$ (typical). We chose our maximum separation on a per-target basis based on the limit at which separations were no longer covered by all median-stacked images, due to the dither pattern used to take our observations.

\begin{table}[h!]
\centering
    \caption{Contrast curve data for all six targets observed with Gemini NGS-AO and LGS-AO in the $K_{s}$ band. Only a portion of this table is shown here. A machine-readable version of the full table is available.}\label{tbl:ccurves}
    \begin{tabular}{c|c|c|c|c}
        \hline\hline
        KIC & Guide Star & UT Obs. Date & Sep. ($^{\prime\prime}$) & $\Delta K_{s}$\\
        & System & & & \\
        \hline
        6126245 & NGS-AO & 31 May 2019 & 0.20 & 0.76265 \\
        & & & 0.35 & 3.80569 \\
        & & & 0.51 & 4.62049 \\
        & & & 0.66 & 5.22778 \\
        & & & ... & ... \\
        6224562 & LGS-AO & 30 June 2019 & 0.20 & 0.69001 \\
        & & & 0.35 & 3.25464 \\
        & & & 0.51 & 4.47640 \\
        & & & 0.66 & 5.49310 \\
        & & & ... & ... \\
    \end{tabular}
\end{table}

\begin{figure*}[ht!]
\centering
\includegraphics[width=0.4\linewidth]{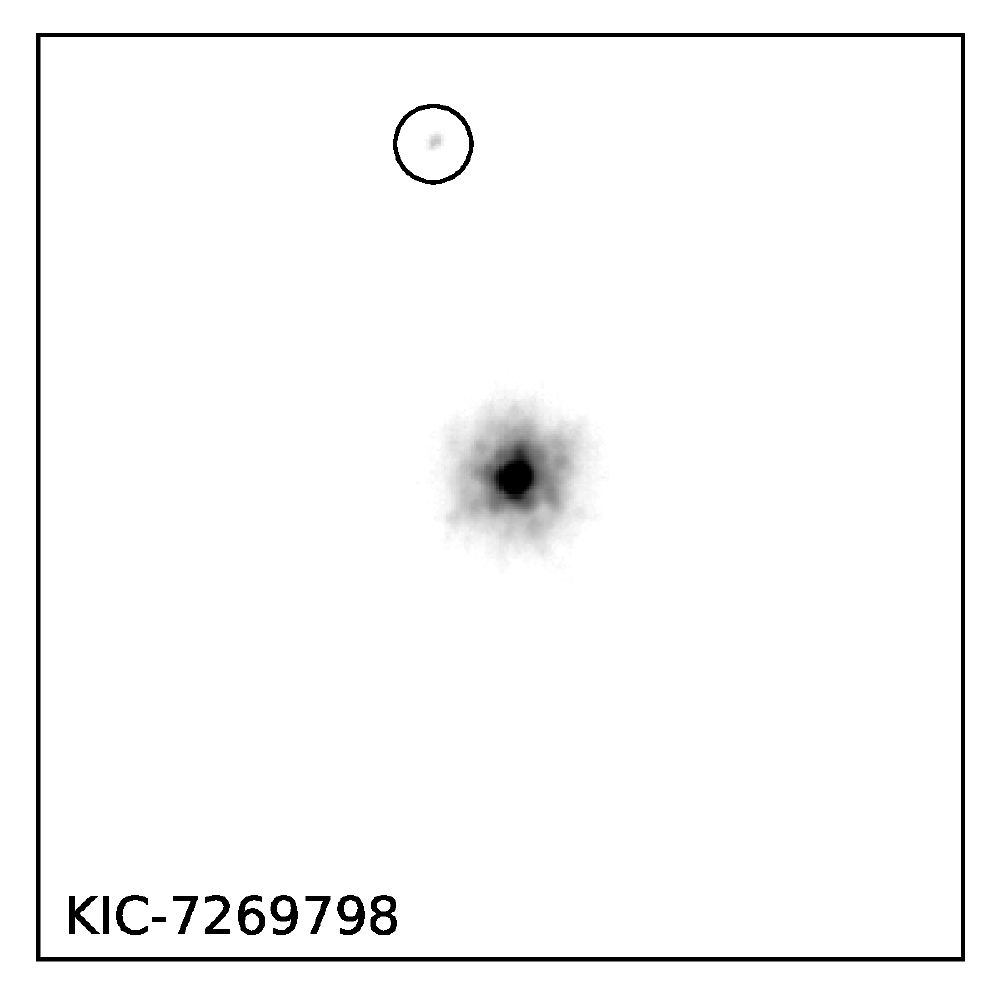}
\includegraphics[width=0.4\linewidth]{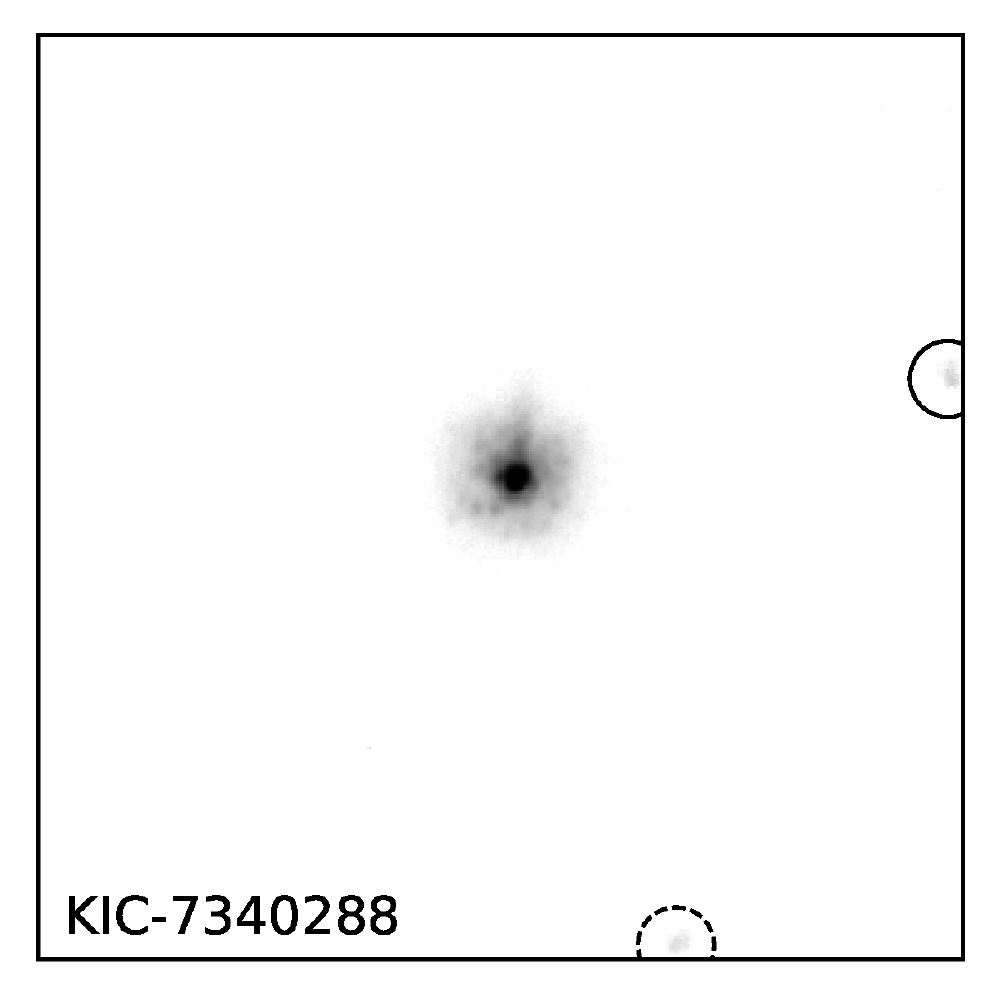}
\caption{AO images $4^{\prime\prime}\times4^{\prime\prime}$ in size and plotted in log scale, centred on each target with resolved companions within $4^{\prime\prime}$ indicated by a black circle. KIC-7340288 has second companion just outside of the $4^{\prime\prime}$ threshold, indicated by a dotted black circle. For all images, north points up and east points left.}\label{fig:aocomps}
\end{figure*}

We manually examined each image for contaminant stars within 4$^{\prime\prime}$, the size of a \textit{Kepler} pixel. We fit a two-Gaussian model to the two targets with detected companions in order to derive the angular separation, position angle (PA), and $\Delta K_{s}$. Results are shown in Table \ref{tbl:AOtable}, and the AO images of targets with companions are shown in Fig. \ref{fig:aocomps}. One of our targets, KIC-7340288, has a potential companion just outside of 4$^{\prime\prime}$ (at $4.2^{\prime\prime}$) that is thus excluded from our analysis but indicated in the plots by dotted circles. 

\begin{table*}[ht!]
    \centering
    \caption{Gemini NIRI and Robo-AO imaging searches for companions within 4$^{\prime\prime}$ of our target stars. Zie17 refers to \citet{zie17}.}\label{tbl:AOtable}
    \begin{tabular}{c|c|c|c|c|c|c|c|c}
        \hline\hline
        KIC & Telescope & Filter & UT Obs. Date & Comp? & Sep. ($^{\prime\prime}$) & PA ($\degree$)  & $\Delta m$ & Ref.\\
        \hline
        6126245 & Gemini NGS-AO &$K_{s}$ & 31 May 2019 &  N & - & - & - & This work \\
        6224562 & Gemini LGS-AO &$K_{s}$ & 30 June 2019 & N &   - & - & - & This work\\
        6782399 & Gemini NGS-AO &$K_{s}$ & 14 June 2019 & N &  - & - & - &  this work\\
        7269798 & Gemini LGS-AO &$K_{s}$ & 30 June 2019 & Y & 3.006$\pm$0.002 & $13.034\pm0.001$ & 5.64$\pm$0.04  & This work\\
        7340288 & Gemini LGS-AO &$K_{s}$ & 01 July 2019 & Y & 3.870$\pm$0.002 & 283.380$\pm$0.001 & 5.20$\pm$0.03 & This work\\
        7747788 & Gemini NGS-AO &$K_{s}$ & 11 June 2019 & N &  - & - & - & This work\\
        11350118 & Robo-AO & LP600 & 01 Sept 2014 &   N & - & - & - & Zie17 \\
    \end{tabular}
    \end{table*}

One of our new PCs (KIC-11350118 c) corresponds to a known KOI already observed in the LP600 band as part of the Robo-AO KOI surveys \citep{law14}. Robo-AO did not detect any nearby stars.

We observed an additional 56 targets across both programs. However, during the execution of the observation program, the vetting pipeline described in Section \ref{sec:vet} and follow-up analysis described earlier in this section was modified and these targets are no longer PCs. We provide contrast curves and companions for these targets in the Appendix. Of note, 32 of our targets are KOIs, 12 of which do not have Robo-AO observations. Our observations can be used in future follow-up analysis of all the confirmed and candidate planets associated with these KOIs.

\subsection{Astrophysical FPPs}\label{sec:FPP}

We tested each of our candidates against astrophysical FP hypotheses using \texttt{vespa}, a Python package built to enable astrophysical FPP analysis of transiting signals \citep{mor12,mor15b}. 

\texttt{vespa} uses stellar posteriors calculated with \texttt{isochrones}, a Python package that provides MCMC fitting of single-, binary-, and triple-star model stellar properties to MIST stellar model grids \citep{mor15a}, as an input. We provided \texttt{isochrones} with R.A./Decl coordinates and $grizJHK$ photometry from the KIC, with $griz$ bands corrected to the Sloan Digital Sky Survey (SDSS) according to \citet{pin12}. $T_{\text{eff}}$, log$g$, and [Fe/H] from \citet{mat17} were used if the provenance of these values is from spectroscopy or asteroseismology. We provided parallaxes from \textit{Gaia} DR2 \citep{gai16,gai18} when available.

As constraints, we followed the convention of \citet{mor16} by setting the allowed ``exclusion'' radius for a blend scenario as three times the uncertainty in the fitted centroid position, floored at 0.5$^{\prime\prime}$. We also set the maximum secondary eclipse depth allowed by the \textit{Kepler} photometry as 

\begin{equation}
    \delta_{\text{max}} = \delta_{\text{sec}} + 3\sigma_{\text{sec}},
\end{equation}

\noindent where $\delta_{\text{sec}}$ and $\sigma_{\text{sec}}$ are the fitted depth and uncertainty of the secondary event in the light-curve, as calculated by the model-shift uniqueness test in the vetting pipeline. Lastly, we inputted contrast curves derived from the AO imaging. In \texttt{vespa}, these eliminate the possibility of bound or background stars above a certain brightness at a given projected distance.

Considering all of these inputs, \texttt{vespa} assigns probabilities to different hypotheses that might describe a transiting PC signal: unblended eclipsing binary, hierarchical-triple eclipsing binary, chance-aligned background/foreground eclipsing binary, and transiting planet. We consider candidates with total non-transiting-planet probabilities FPP $>$ 0.9 as FPs. All other candidates, including those that have \texttt{vespa} fail to return an FPP (typically due to a nonconverging MCMC fit), remain planet candidates. None of our 17 remaining PCs were classified as FPs based on our \texttt{vespa} results.

Twelve of the candidates have FPP $<$ 0.01 (confidence at the 99\% level). \texttt{vespa} has been used to validate over a thousand KOIs \citep{mor16} as confirmed planets using this threshold. However, given that all of these PCs have low signal-to-noise ratios (S/N $<$ 10), a noise or systematic explanation for the signals cannot be ignored. \citet{bur19} indicated that statistical validation methods based only on astrophysical scenarios, such as \texttt{vespa}, are insufficient for such a low-S/N regime. Thus, we chose to retain their candidate disposition.

\subsection{MCMC Fit}

\begin{figure*}[h!]
\centering
\includegraphics[width=0.3\linewidth]{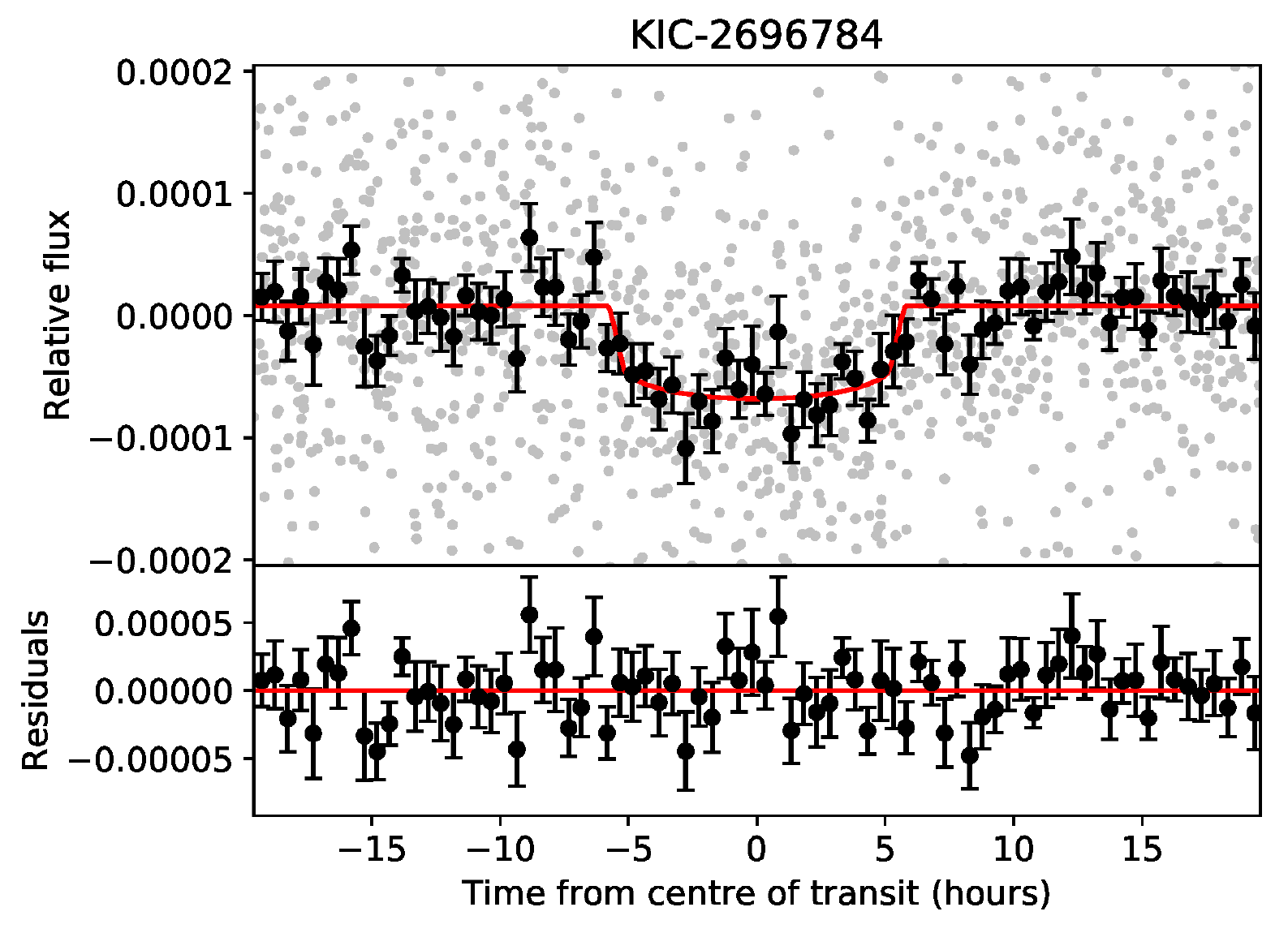}
\includegraphics[width=0.3\linewidth]{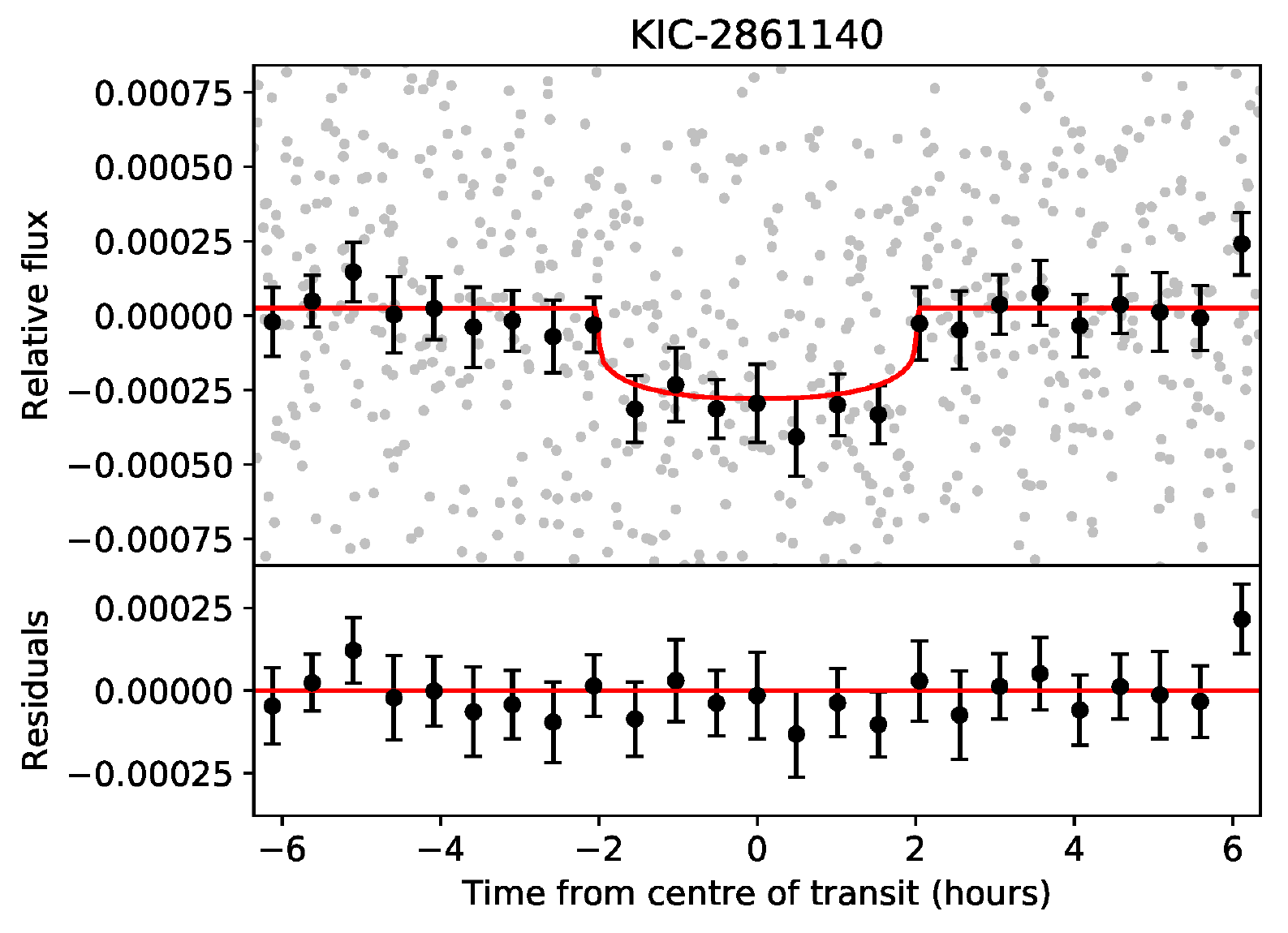}
\includegraphics[width=0.3\linewidth]{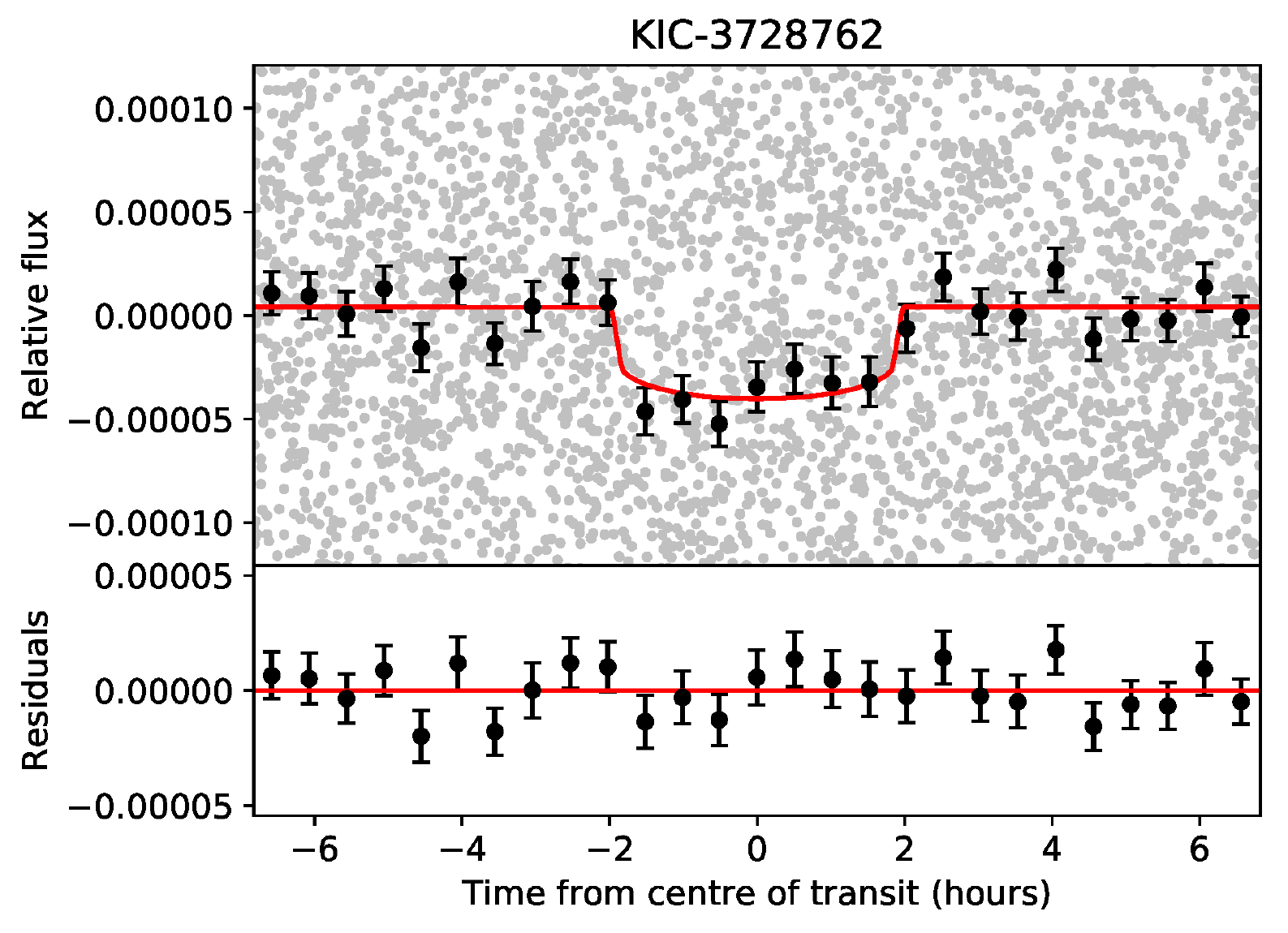}
\includegraphics[width=0.3\linewidth]{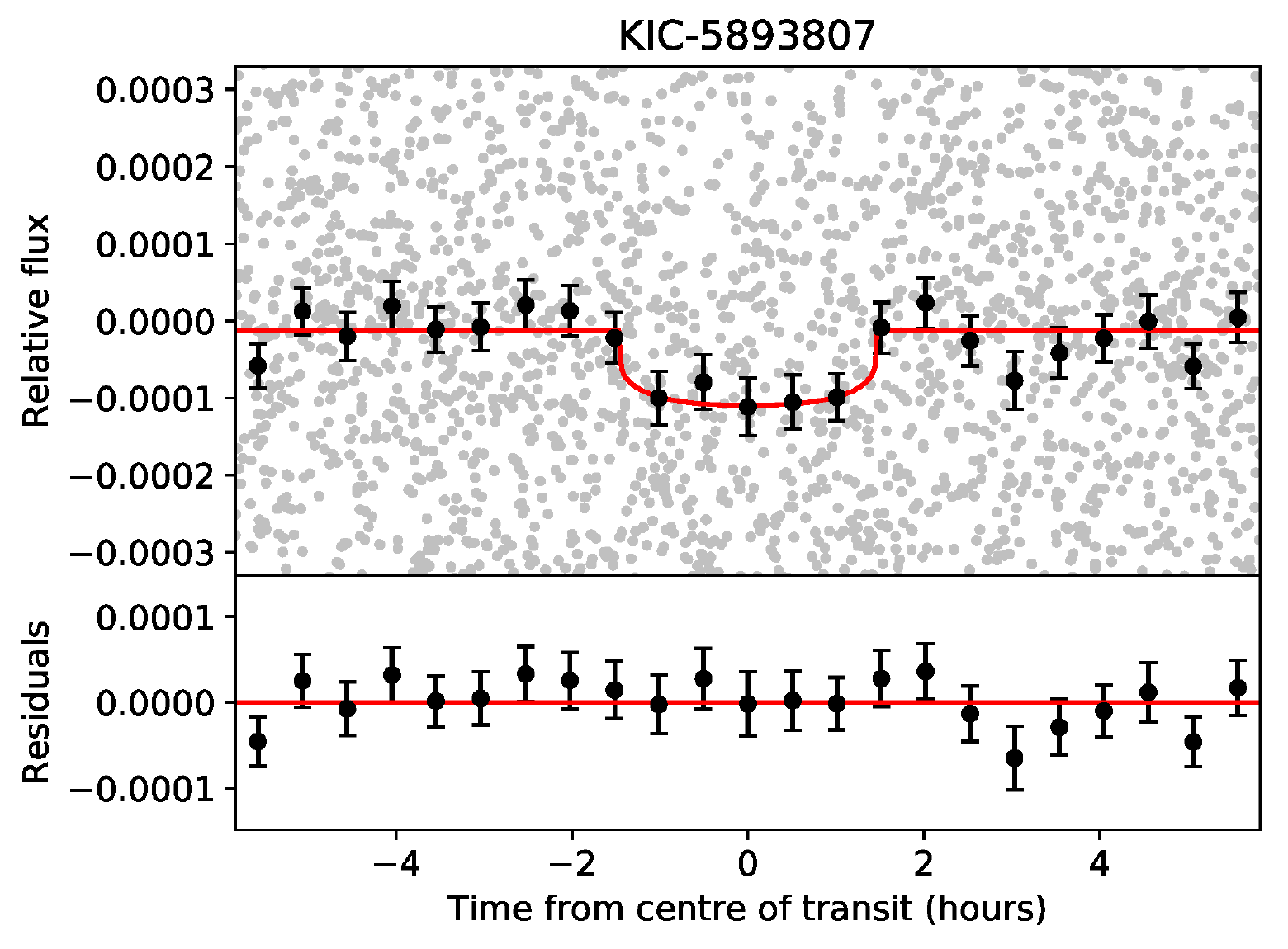}
\includegraphics[width=0.3\linewidth]{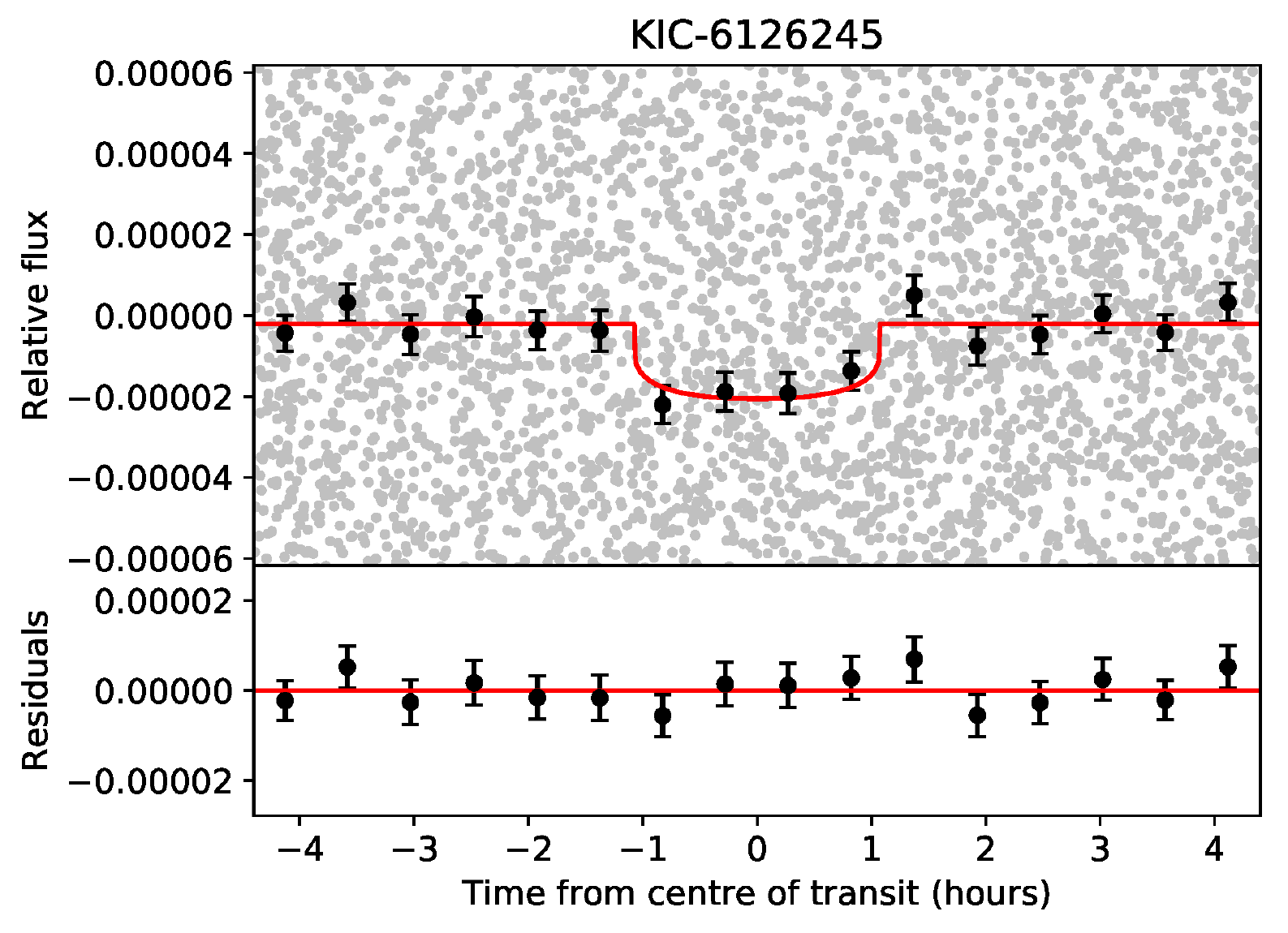}
\includegraphics[width=0.3\linewidth]{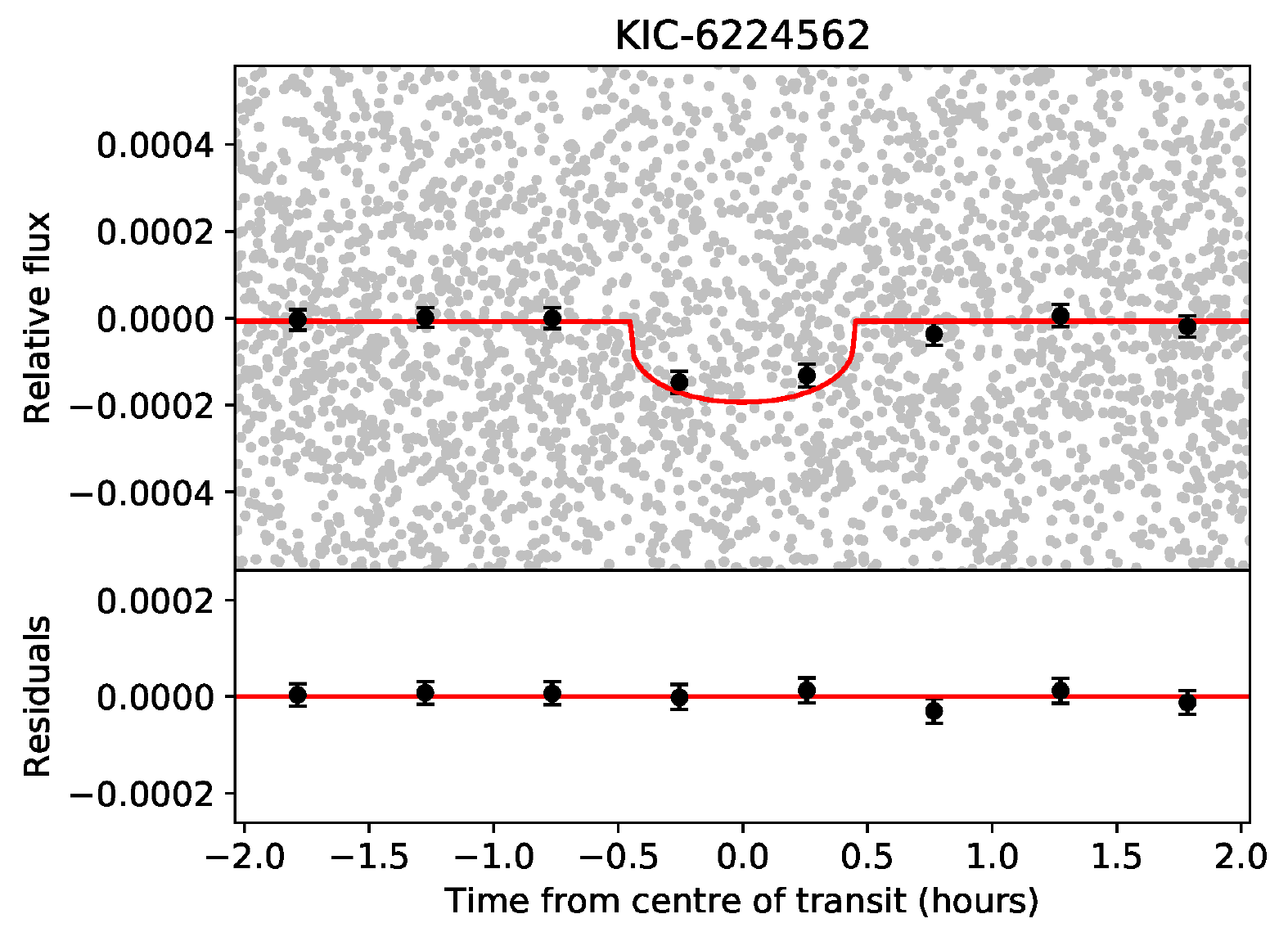}
\includegraphics[width=0.3\linewidth]{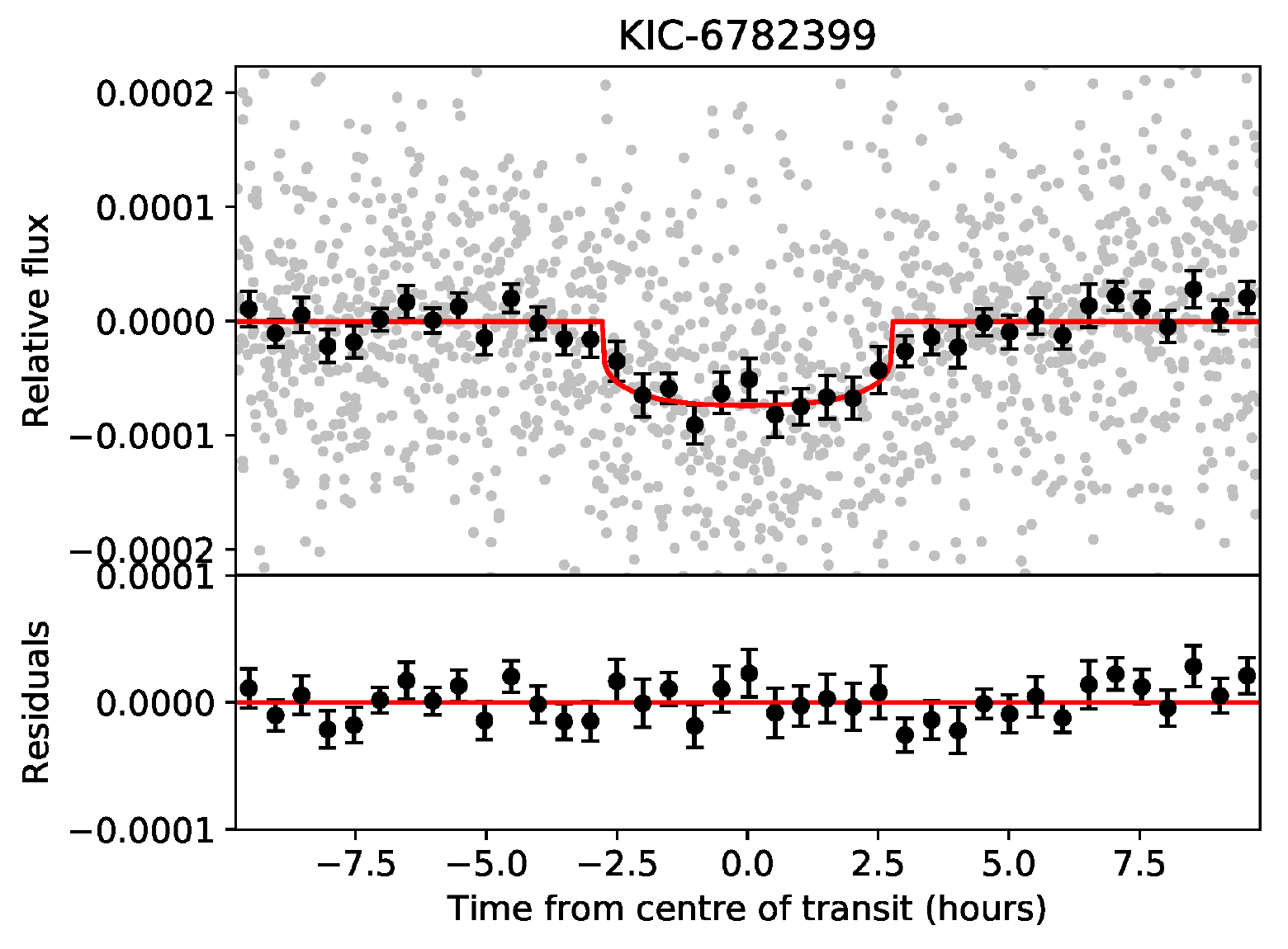}
\includegraphics[width=0.3\linewidth]{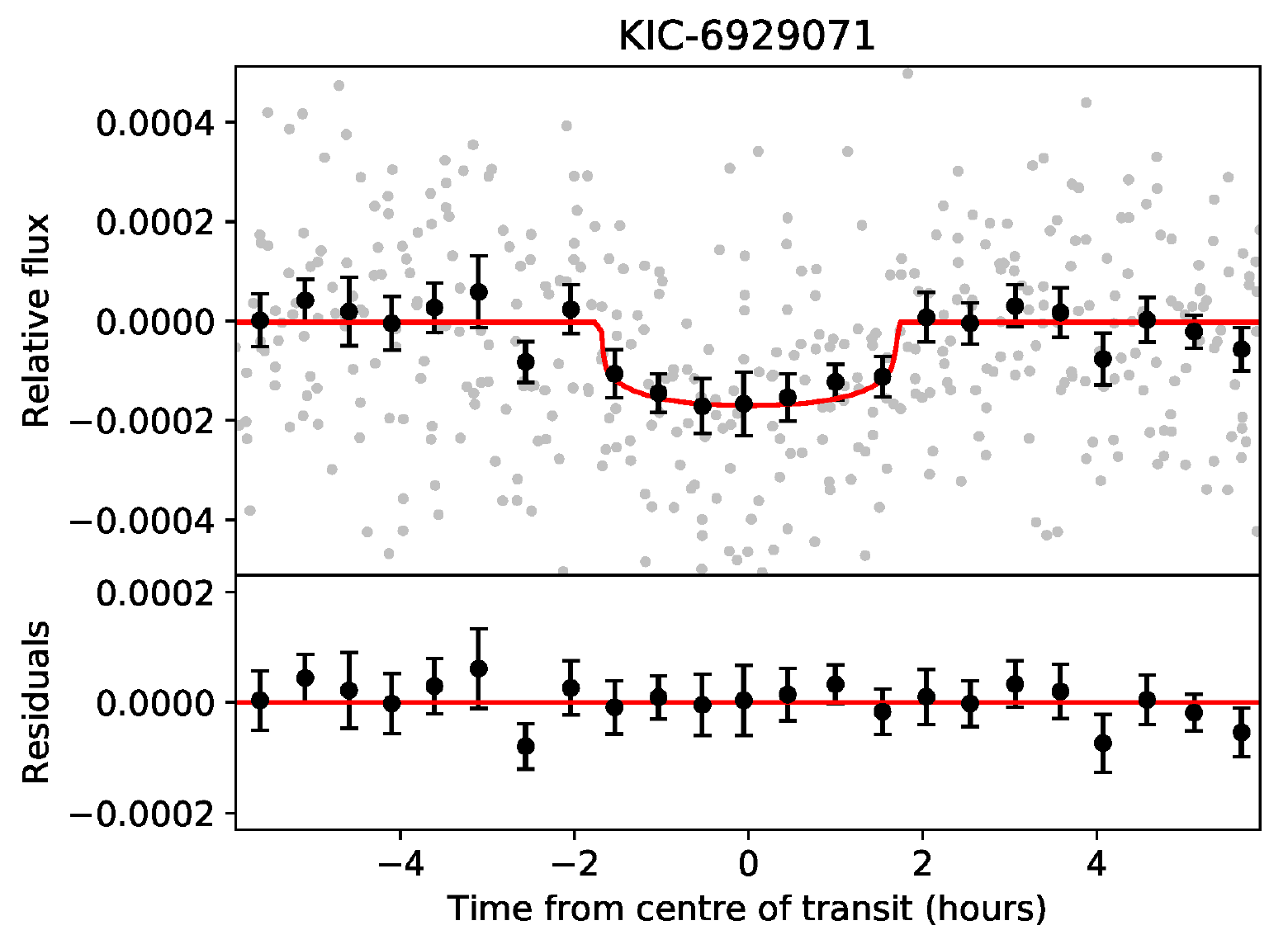}
\includegraphics[width=0.3\linewidth]{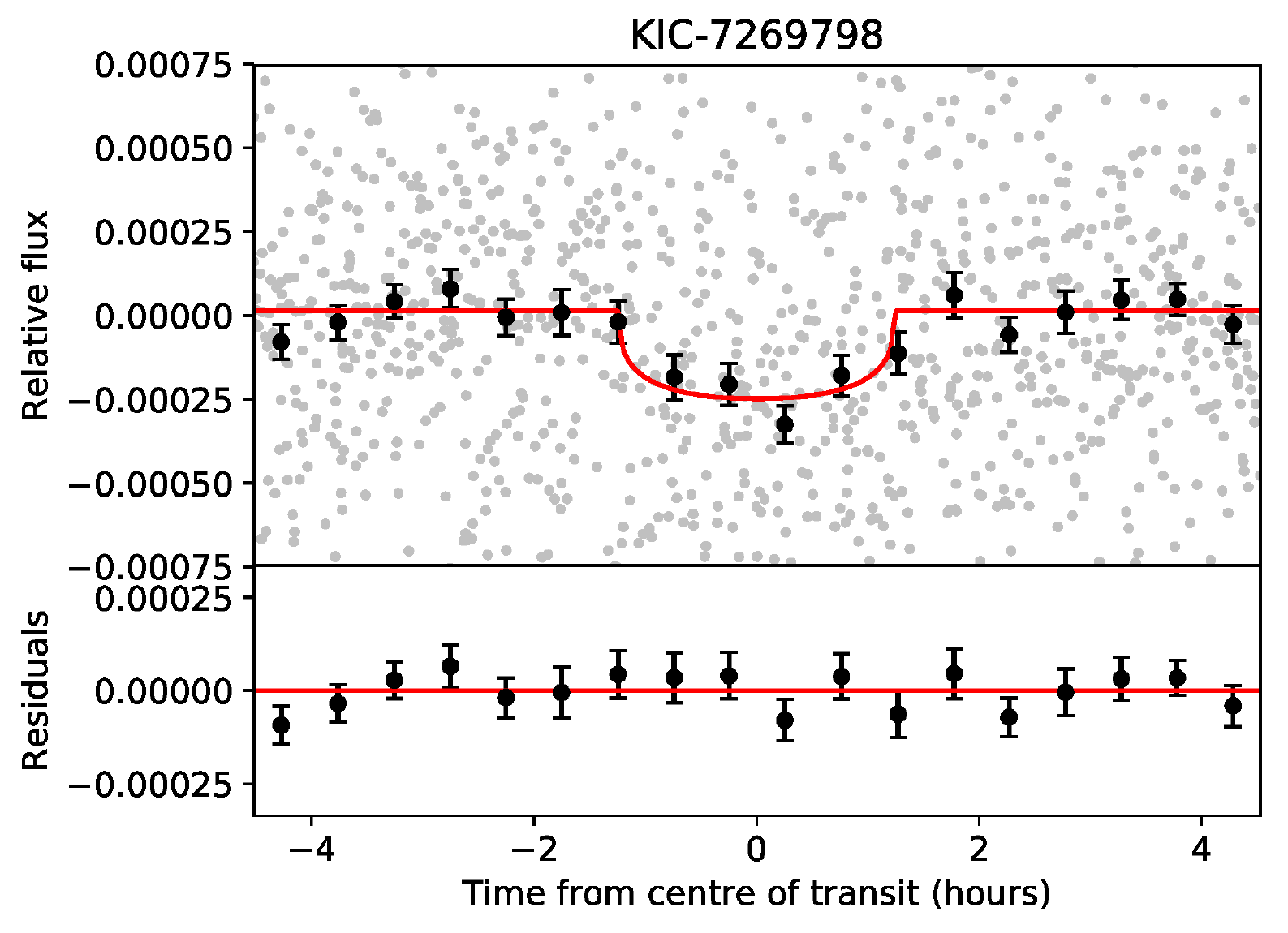}
\includegraphics[width=0.3\linewidth]{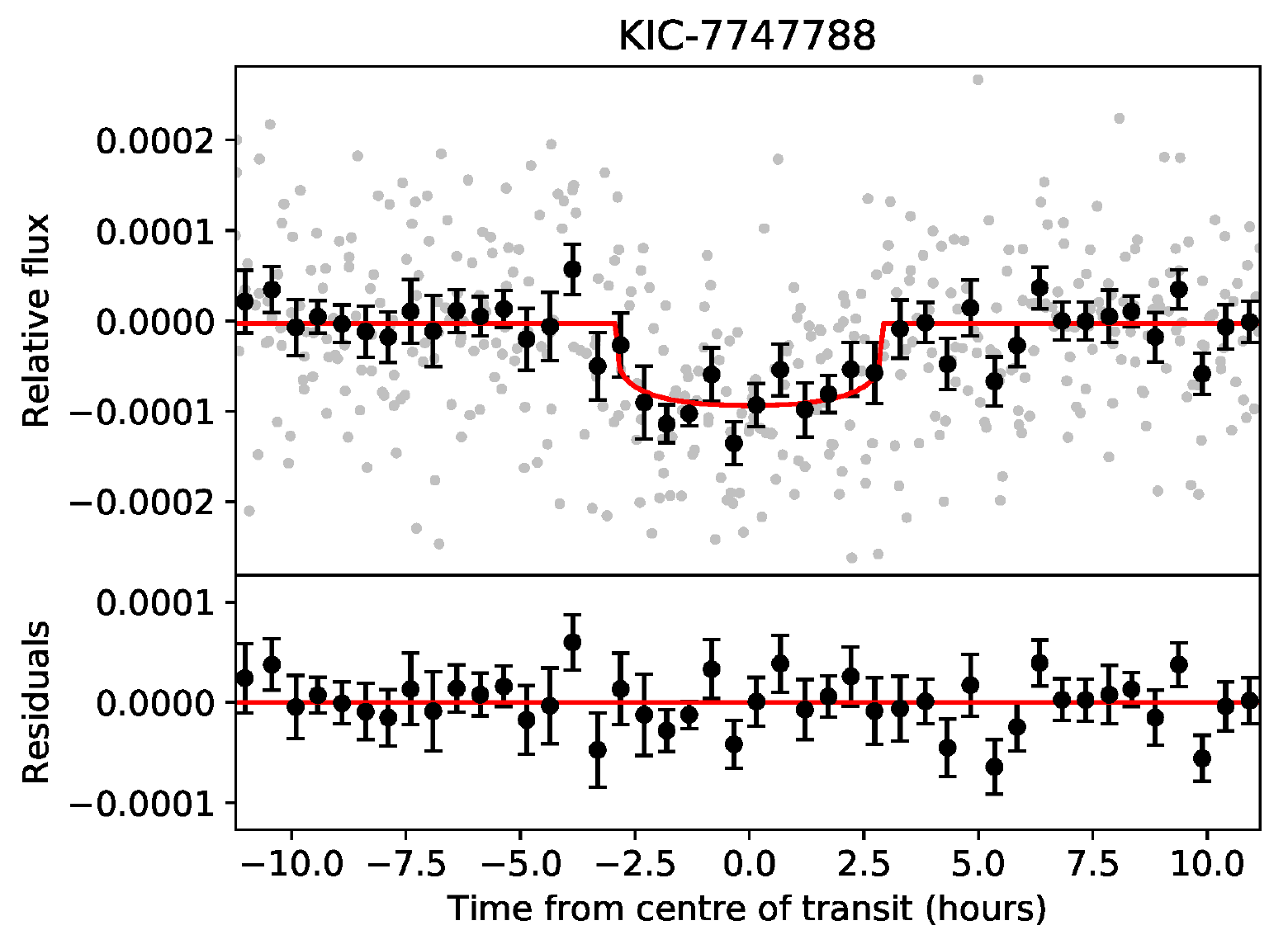}
\includegraphics[width=0.3\linewidth]{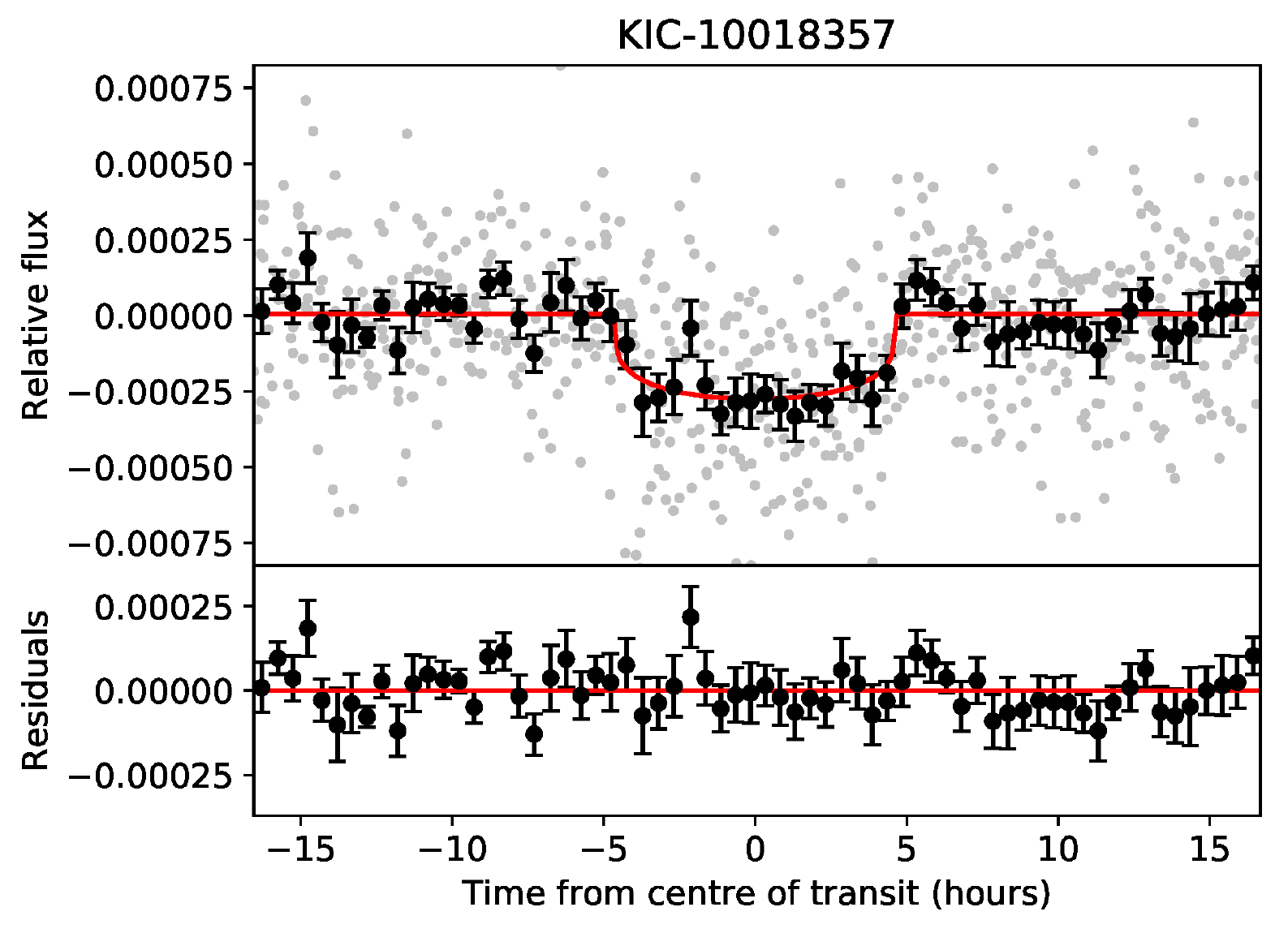}
\includegraphics[width=0.3\linewidth]{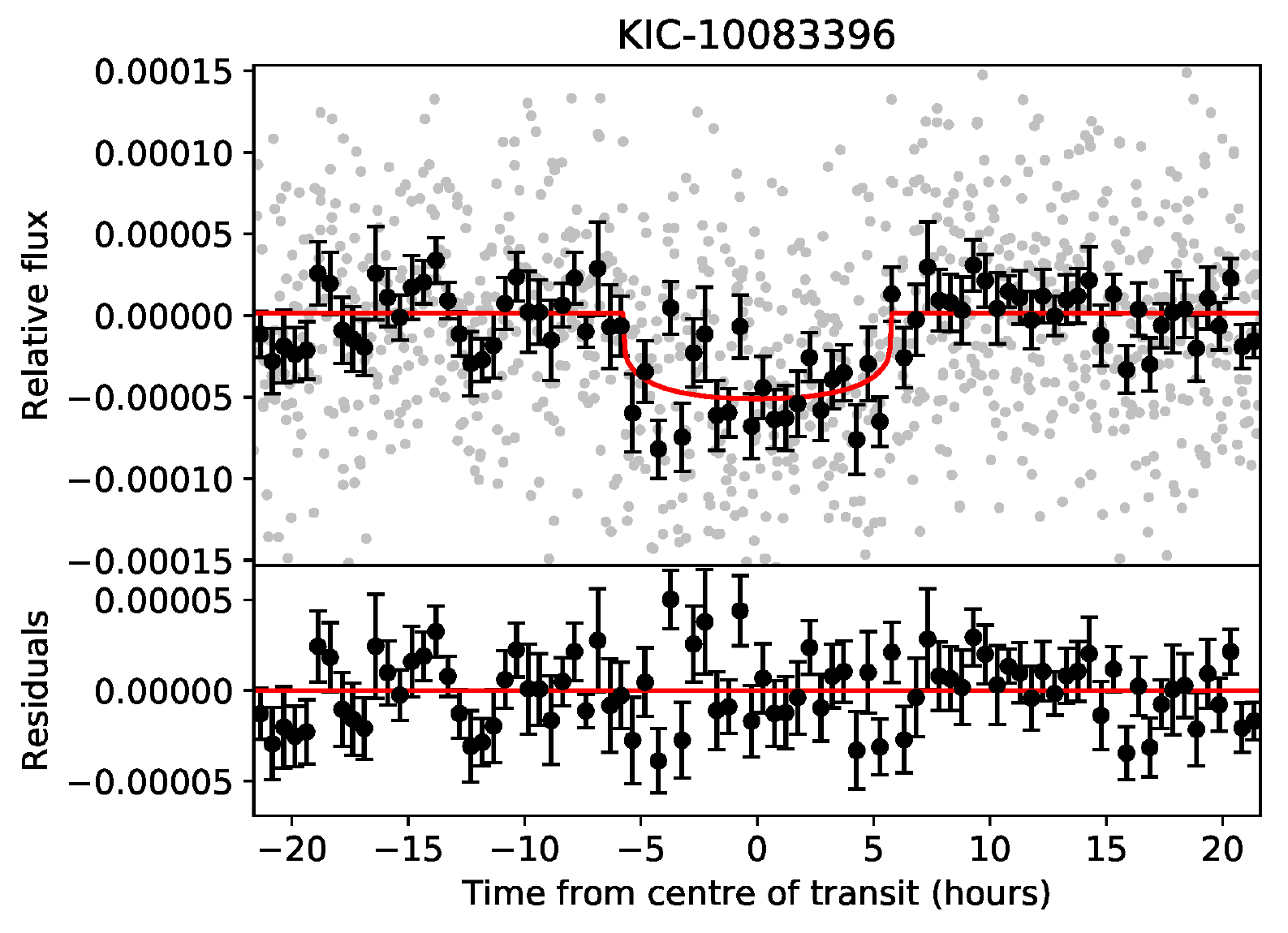}
\includegraphics[width=0.3\linewidth]{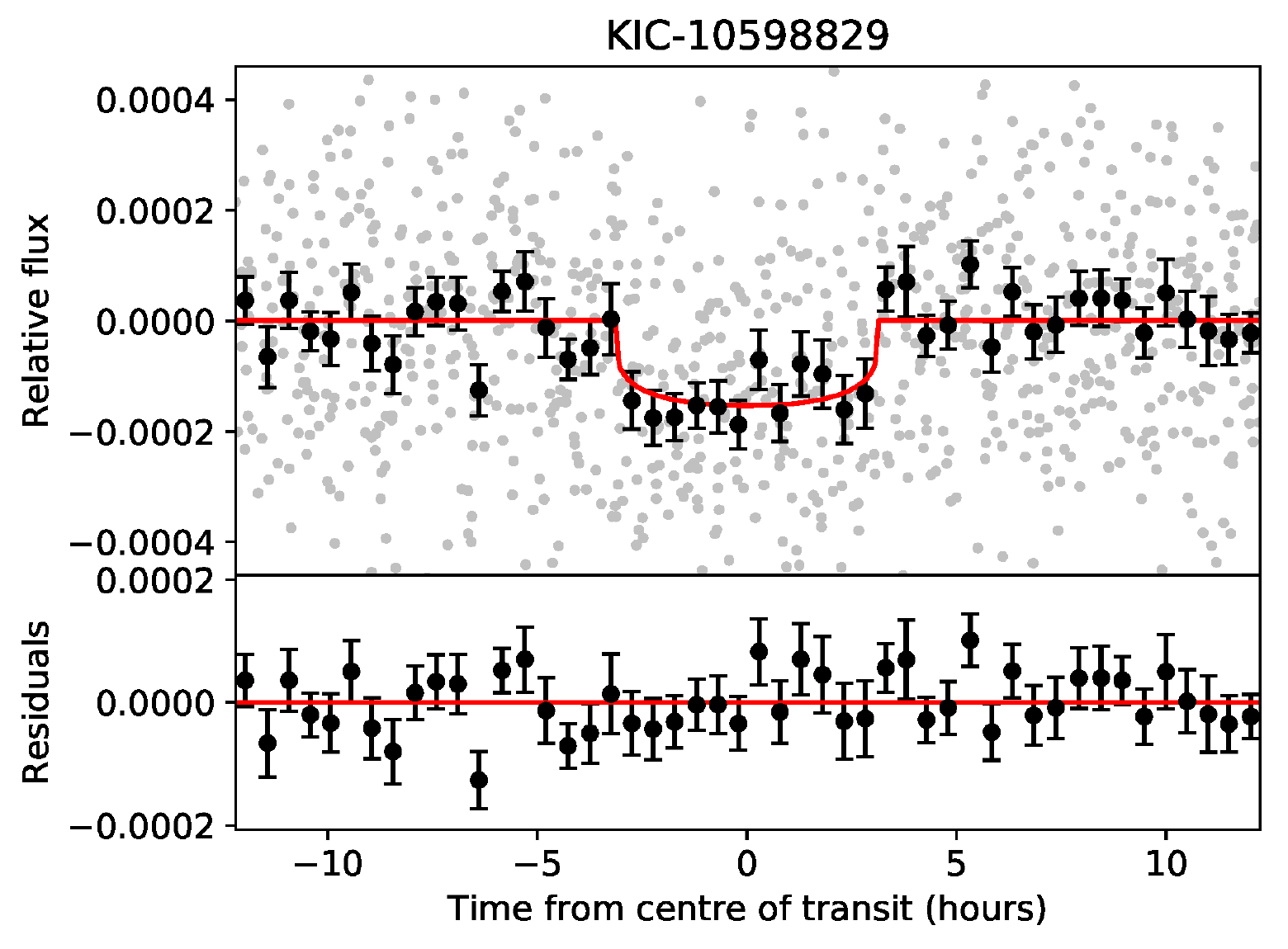}
\includegraphics[width=0.3\linewidth]{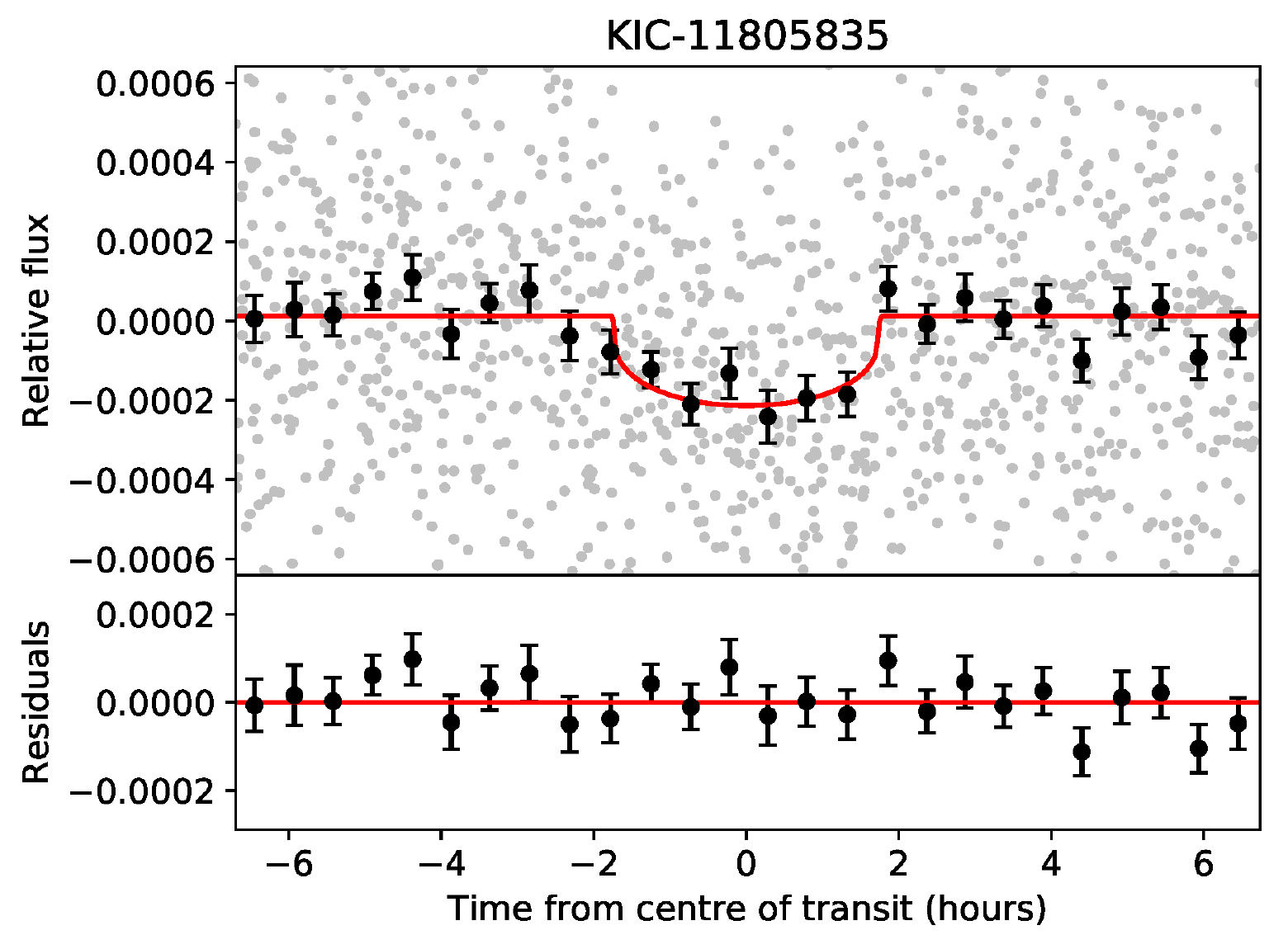}
\includegraphics[width=0.3\linewidth]{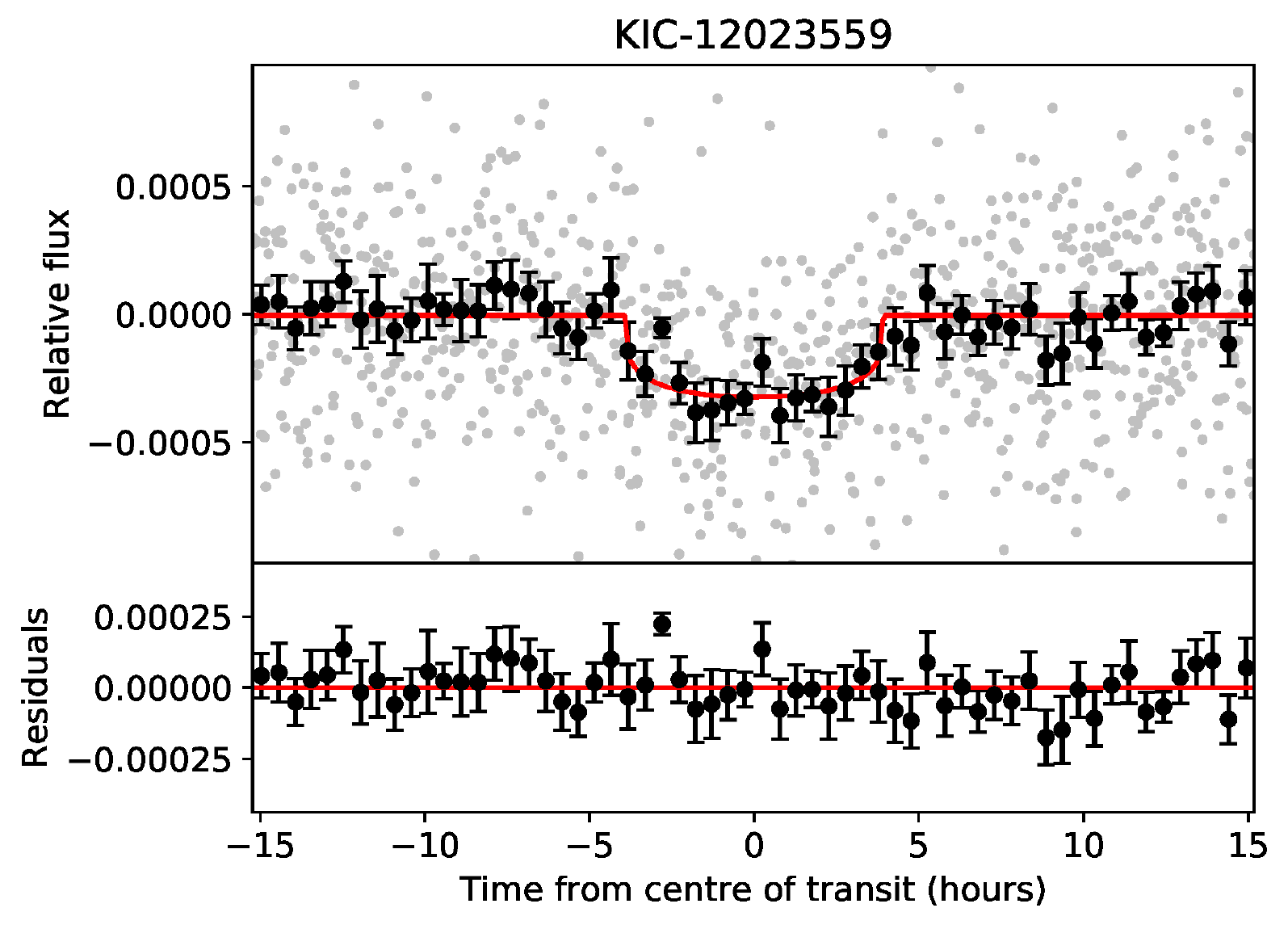}
\caption{Binned phase diagrams of 15 of the 17 new PCs, showing data and model fit with residuals. Original data points are plotted in grey, while data binned into 30 minute bins is in black. Error bars represent the standard error of each bin. The transit model MCMC fit to the data is plotted in red. KIC-7340288 b and KIC-11350118 c are plotted in Section \ref{sec:highlight}.}\label{fig:plots}
\end{figure*}

We refit each transit using \texttt{emcee}, a Python implementation of an affine invariant Markov Chain Monte Carlo (MCMC) ensemble sampler \citep{for13}, seeded by the best-fit parameters from the least-squares fit discussed in Section \ref{sec:modelfit}. We set $P$ and $T_{0}$ fixed to their least-squares values. Fit results are shown in Table \ref{tbl:MCMC}. These fit results are mixed with \texttt{isochrones} stellar parameter posteriors, given in Table \ref{tbl:stellar}, to produce derived planet parameters shown in Table \ref{tbl:results}. The reported values use the median, with uncertainties given by 15.9$\%$ and 84.1$\%$ percentiles, corresponding to a 68.2$\%$ confidence region. Plots of the phase diagrams of each transit with the MCMC fits and residuals are shown in Fig. \ref{fig:plots}.

\subsubsection{Derived Parameters from MCMC}
The planet radius $R_{p}$ is determined from the fitted parameter $R_{p}/R_{s}$ using

\begin{equation}\label{eqn:radius}
R_{p} = \bigg(\frac{R_{p}}{R_{s}}\bigg)R_{s},
\end{equation}

\noindent where $R_{s}$ is the known stellar radius.

The semimajor axis of the planet's orbit $a$ is determined from the fitted $P$ and known stellar parameters, rather than the fitted $a/R_{s}$,

\begin{equation}
    a = \frac{GM_{s}P^{2}}{4\pi^{2}},
\end{equation}

\noindent where $G$ is the gravitational constant and $M_{s}$ is the stellar mass.

The planet equilibrium temperature $T_{\text{eq}}$ is calculated assuming thermodynamic equilibrium between the incident stellar flux and the radiated heat from the planet,

\begin{equation}
T_{\text{eq}} = T_{\text{eff}} (1 - A)^{1/4}\sqrt{\frac{R_{s}}{2a}},
\end{equation}

\noindent where $A$ is the albedo of the planet. We assume Earth's albedo, $A = 0.3$, for all cases.

The planet's stellar insolation $S$, defined as the ratio of the flux of the host star at the planet to the solar flux at Earth, is determined by

\begin{equation}
S = \bigg(\frac{R_{s}/R_{\astrosun}}{a}\bigg)^{2}\bigg(\frac{T_{\text{eff}}}{T_{\text{eff},\astrosun}}\bigg)^{4}.
\end{equation}

\begin{longtable*}[h!]{c|c|c|c|c|c}
\caption{MCMC fit results for select fitted planet parameters ($R_{p}/R_{s}$, $a/R_{s}$, and $b$). $P$ and $T_{0}$ were set to their least-squares best-fit values.}\label{tbl:MCMC}\\
        \hline\hline
        KIC & $P$ (days) & $T_{0}$ (BKJD) & $R_{p}/R_{s}$ & $a/R_{s}$ & $b$ \\
        \hline
        \endfirsthead
        KIC & $P$ (days) & $T_{0}$ (BKJD) & $R_{p}/R_{s}$ & $a/R_{s}$ & $b$ \\
        \hline
        \endhead
        1570311 b & $23.44253108\pm0.00046135$ & $148.98683\pm0.01301$ & $0.017478_{-0.002075}^{+0.005396}$ & $8.487_{-2.065}^{+4.907}$ & $0.977_{-0.036}^{+0.016}$ \\
        2696784 b & $82.30223397\pm0.00196956$ & $130.31551\pm0.01966$ & $0.009613_{-0.000561}^{+0.000652}$ & $22.104_{-6.402}^{+5.043}$ & $0.923_{-0.042}^{+0.038}$ \\
        2861140 b & $36.87848594\pm0.00033789$ & $364.07192\pm0.00612$ & $0.016318_{-0.001597}^{+0.00173}$ & $63.344_{-19.783}^{+10.242}$ & $0.43_{-0.296}^{+0.351}$\\
        2985262 b & $13.0351506\pm5.443\times10^{-5}$ & $140.06886\pm0.00392$ & $0.009084_{-0.000449}^{+0.000517}$ & $29.901_{-6.193}^{+2.269}$ & $0.383_{-0.266}^{+0.3}$ \\
        3336146 b & $3.27626622\pm1.652\times10^{-5}$ & $134.62799\pm0.00298$ & $0.007098_{-0.000508}^{+0.000596}$ & $14.447_{-3.703}^{+1.248}$ & $0.41_{-0.279}^{+0.33}$ \\
        3345775 b & $6.22112577\pm2.924\times10^{-5}$ & $122.33384\pm0.00401$ & $0.004294_{-0.000189}^{+0.000232}$ & $13.329_{-2.986}^{+1.97}$ & $0.847_{-0.055}^{+0.064}$ \\
        3347135 b & $226.52674578\pm0.00125028$ & $160.17927\pm0.0051$ & $0.017495_{-0.000459}^{+0.000529}$ & $242.158_{-23.965}^{+8.295}$ & $0.271_{-0.185}^{+0.229}$ \\
        3662290 b & $288.23951462\pm0.01314029$ & $322.24425\pm0.04069$ & $0.011739_{-0.000658}^{+0.000882}$ & $101.68_{-29.933}^{+17.014}$ & $0.815_{-0.081}^{+0.101}$\\
        3728762 b & $6.73928932\pm6.924\times10^{-5}$ & $122.20544\pm0.00867$ & $0.007275_{-0.000501}^{+0.000553}$ & $5.557_{-1.476}^{+1.023}$ & $0.914_{-0.04}^{+0.042}$\\
        3967744 b & $57.88535219\pm0.00040897$ & $143.26298\pm0.0065$ & $0.011696_{-0.000693}^{+0.000888}$ & $47.663_{-15.352}^{+6.252}$ & $0.439_{-0.304}^{+0.358}$ \\
        4346258 b & $4.90776291\pm1.936\times10^{-5}$ & $354.03734\pm0.00291$ & $0.013645_{-0.00114}^{+0.001287}$ & $19.513_{-5.171}^{+1.919}$ & $0.418_{-0.291}^{+0.329}$ \\
        4551429 b & $35.37625461\pm0.00023684$ & $147.08621\pm0.00434$ & $0.012458_{-0.000774}^{+0.00096}$ & $75.028_{-17.846}^{+6.129}$ & $0.386_{-0.262}^{+0.325}$ \\
        4556565 b & $5.54559321\pm3.186\times10^{-5}$ & $353.80389\pm0.004$ & $0.011497_{-0.000861}^{+0.001035}$ & $16.36_{-4.241}^{+1.875}$ & $0.427_{-0.289}^{+0.32}$\\
        5095499 b & $4.29501271\pm1.861\times10^{-5}$ & $134.0872\pm0.00411$ & $0.008775_{-0.000546}^{+0.000575}$ & $9.371_{-1.944}^{+0.698}$ & $0.379_{-0.274}^{+0.305}$ \\
        5184017 b & $6.25985244\pm2.664\times10^{-5}$ & $135.91422\pm0.00367$ & $0.006305_{-0.000386}^{+0.000557}$ & $11.235_{-3.042}^{+1.007}$ & $0.409_{-0.279}^{+0.345}$\\
       5342061 c & $11.49044213\pm7.721\times10^{-5}$ & $137.65928\pm0.00558$ & $0.013311_{-0.000975}^{+0.001104}$ & $23.445_{-5.977}^{+2.578}$ & $0.413_{-0.29}^{+0.327}$ \\
        5628770 b  & $11.42952942\pm6.504\times10^{-5}$ & $132.35509\pm0.00511$ & $0.008349_{-0.000623}^{+0.000662}$ & $39.42_{-8.723}^{+3.527}$ & $0.4_{-0.278}^{+0.303}$ \\
        5649129 b & $2.82857392\pm1.138\times10^{-5}$ & $133.09613\pm0.00321$ & $0.007052_{-0.000446}^{+0.000701}$ & $8.999_{-2.605}^{+1.073}$ & $0.469_{-0.324}^{+0.314}$\\
        5794479 b & $5.92912541\pm3.014\times10^{-5}$ & $125.34438\pm0.00419$ & $0.007263_{-0.000324}^{+0.000445}$ & $14.551_{-2.751}^{+1.261}$ & $0.412_{-0.277}^{+0.263}$ \\
        5893807 b & $7.66465733\pm5.173\times10^{-5}$ & $133.13869\pm0.00444$ & $0.009196_{-0.000853}^{+0.000885}$ & $18.378_{-4.85}^{+1.785}$ & $0.41_{-0.287}^{+0.338}$ \\
        6021193 e & $26.48588826\pm0.0001843$ & $126.89598\pm0.00597$ & $0.009068_{-0.000482}^{+0.000736}$ & $27.39_{-7.436}^{+3.054}$ & $0.401_{-0.275}^{+0.351}$ \\
        6126245 b & $3.48546885\pm1.049\times10^{-5}$ & $134.83373\pm0.0022$ & $0.004019_{-0.000334}^{+0.000346}$ & $11.424_{-2.809}^{+1.283}$ & $0.4_{-0.276}^{+0.332}$\\
        6139884 b & $4.80084532\pm2.32\times10^{-5}$ & $122.04373\pm0.00401$ & $0.005274_{-0.000372}^{+0.000447}$ & $10.93_{-2.984}^{+1.501}$ & $0.611_{-0.184}^{+0.21}$\\
        6224562 b & $2.32907482\pm4.46\times10^{-6}$ & $133.31436\pm0.00164$ & $0.012356_{-0.000948}^{+0.001302}$ & $18.11_{-4.558}^{+1.936}$ & $0.42_{-0.287}^{+0.316}$ \\
        6347299 b & $38.64138416\pm0.00025459$ & $148.42833\pm0.006$ & $0.009747_{-0.000596}^{+0.000678}$ & $43.08_{-9.797}^{+3.242}$ & $0.374_{-0.26}^{+0.326}$ \\
        6380164 d & $167.78839179\pm0.00333756$ & $206.04216\pm0.01562$ & $0.014345_{-0.000478}^{+0.000522}$ & $189.128_{-26.707}^{+9.523}$ & $0.326_{-0.225}^{+0.26}$ \\
        6440915 b & $365.41156475\pm0.01311087$ & $317.94313\pm0.01688$ & $0.023918_{-0.001569}^{+0.00191}$ & $105.273_{-30.741}^{+35.13}$ & $0.829_{-0.163}^{+0.087}$ \\
        6782399 b & $34.20150223\pm0.00018815$ & $134.63965\pm0.00479$ & $0.008004_{-0.000335}^{+0.000517}$ & $43.655_{-11.713}^{+5.013}$ & $0.387_{-0.274}^{+0.359}$ \\
        6837899 b & $8.99702134\pm5.882\times10^{-5}$ & $137.73126\pm0.00441$ & $0.010686_{-0.000889}^{+0.00098}$ & $22.957_{-5.947}^{+2.935}$ & $0.428_{-0.293}^{+0.32}$ \\
        6888194 b & $46.04031439\pm0.00077561$ & $163.79766\pm0.01181$ & $0.009233_{-0.000632}^{+0.000687}$ & $96.009_{-21.171}^{+8.866}$ & $0.382_{-0.267}^{+0.317}$\\
        6929071 b & $61.85364904\pm0.00031287$ & $183.29319\pm0.00792$ & $0.012057_{-0.000935}^{+0.000992}$ & $126.476_{-33.009}^{+13.322}$ & $0.433_{-0.295}^{+0.315}$ \\
        6937870 b & $27.46007046\pm0.00027629$ & $140.54046\pm0.00769$ & $0.010037_{-0.000705}^{+0.001033}$ & $45.134_{-12.159}^{+4.254}$ & $0.423_{-0.299}^{+0.33}$ \\
        7020834 b & $369.4781786\pm0.01213654$ & $187.4524\pm0.01568$ & $0.018399_{-0.00126}^{+0.003931}$ & $43.816_{-4.313}^{+4.82}$ & $0.985_{-0.006}^{+0.009}$ \\
        7119412 b & $10.54126725\pm0.00012629$ & $136.69996\pm0.01003$ & $0.011053_{-0.00088}^{+0.001328}$ & $8.323_{-2.616}^{+1.625}$ & $0.919_{-0.04}^{+0.047}$ \\
        7186892 b & $17.23935628\pm7.05\times10^{-5}$ & $131.71803\pm0.00421$ & $0.006875_{-0.000397}^{+0.000666}$ & $38.689_{-9.623}^{+5.189}$ & $0.416_{-0.284}^{+0.329}$  \\
        7187389 b & $23.77032399\pm0.00027154$ & $359.57155\pm0.00879$ & $0.01145_{-0.00086}^{+0.000984}$ & $27.557_{-6.466}^{+2.576}$ & $0.411_{-0.272}^{+0.308}$\\
        7269798 b & $21.44308742\pm0.00011838$ & $152.55136\pm0.00472$ & $0.014886_{-0.001118}^{+0.00144}$ & $59.932_{-16.839}^{+7.183}$ & $0.44_{-0.296}^{+0.326}$ \\
        7340288 b & $142.53244069\pm0.00335958$ & $204.71041\pm0.01799$ & $0.025258_{-0.001766}^{+0.00201}$ & $156.563_{-32.38}^{+12.21}$ & $0.369_{-0.255}^{+0.311}$ \\
        7747788 b & $133.09439782\pm0.00221722$ & $215.34785\pm0.01357$ & $0.00893_{-0.000531}^{+0.000627}$ & $161.086_{-42.123}^{+15.814}$ & $0.409_{-0.286}^{+0.328}$ \\
        7974496 b & $3.96943045\pm2.652\times10^{-5}$ & $133.95697\pm0.0056$ & $0.008433_{-0.000747}^{+0.000843}$ & $8.43_{-2.343}^{+1.201}$ & $0.632_{-0.188}^{+0.203}$\\
        8172679 b & $194.05437841\pm0.00399057$ & $197.40431\pm0.00878$ & $0.017765_{-0.000449}^{+0.000949}$ & $182.339_{-32.662}^{+11.466}$ & $0.359_{-0.244}^{+0.287}$\\
        9274173 b & $4.43040627\pm1.401\times10^{-5}$ & $134.02566\pm0.00261$ & $0.011594_{-0.000835}^{+0.000968}$ & $17.616_{-4.748}^{+3.045}$ & $0.804_{-0.089}^{+0.097}$\\
        9716483 b & $209.40859648\pm0.00225065$ & $166.46013\pm0.00829$ & $0.012104_{-0.000497}^{+0.000696}$ & $128.227_{-32.405}^{+10.877}$ & $0.4_{-0.281}^{+0.33}$ \\
        9777962 b & $367.20909928\pm0.00969414$ & $359.4191\pm0.02022$ & $0.02824_{-0.00282}^{+0.004365}$ & $80.972_{-19.615}^{+40.269}$ & $0.949_{-0.08}^{+0.027}$ \\
        10018357 b & $133.78748404\pm0.00230495$ & $253.99492\pm0.01191$ & $0.01555_{-0.000618}^{+0.001482}$ & $97.298_{-31.568}^{+12.202}$ & $0.481_{-0.341}^{+0.328}$ \\
        10083396 b & $113.46453674\pm0.0020158$ & $210.81344\pm0.01041$ & $0.006687_{-0.000304}^{+0.000366}$ & $70.969_{-13.633}^{+4.505}$ & $0.347_{-0.243}^{+0.309}$ \\
        10419787 b & $122.71394705\pm0.00096645$ & $208.46146\pm0.00552$ & $0.015731_{-0.001016}^{+0.001187}$ & $140.29_{-34.704}^{+12.715}$ & $0.413_{-0.283}^{+0.319}$  \\
        10598829 b & $67.52966257\pm0.00045439$ & $191.28971\pm0.00578$ & $0.011558_{-0.000819}^{+0.000907}$ & $76.452_{-18.646}^{+6.627}$ & $0.396_{-0.275}^{+0.329}$\\
        10879314 b & $49.19380825\pm0.00037771$ & $164.61498\pm0.0069$ & $0.014659_{-0.001153}^{+0.001378}$ & $70.276_{-18.504}^{+9.683}$ & $0.418_{-0.289}^{+0.338}$\\
        11092463 b & $6.87343628\pm8.866\times10^{-5}$ & $135.83495\pm0.01066$ & $0.013895_{-0.001421}^{+0.002293}$ & $7.309_{-2.8}^{+1.561}$ & $0.923_{-0.043}^{+0.054}$ \\
        11139863 b & $7.22517263\pm3.616\times10^{-5}$ & $120.75217\pm0.00397$ & $0.004123_{-0.000177}^{+0.00029}$ & $19.012_{-5.91}^{+3.395}$ & $0.558_{-0.356}^{+0.264}$ \\
        11350118 c & $2.65550668\pm1.532\times10^{-5}$ & $133.27041\pm0.00483$ & $0.009019_{-0.0007}^{+0.000877}$ & $6.396_{-1.649}^{+0.612}$ & $0.419_{-0.293}^{+0.33}$\\
        11565976 b & $24.24399123\pm0.00016113$ & $143.36071\pm0.0042$ & $0.006635_{-0.00048}^{+0.000506}$ & $52.545_{-12.444}^{+4.43}$ & $0.397_{-0.272}^{+0.318}$\\
        11805835 b & $23.52676998\pm0.00024171$ & $142.63053\pm0.00886$ & $0.013617_{-0.001029}^{+0.001429}$ & $46.944_{-12.626}^{+5.132}$ & $0.427_{-0.296}^{+0.33}$ \\
        12023559 b & $84.55709677\pm0.00127186$ & $206.60834\pm0.01139$ & $0.016528_{-0.000877}^{+0.001056}$ & $77.011_{-16.798}^{+6.172}$ & $0.39_{-0.271}^{+0.305}$ \\
        12216301 b & $116.53116276\pm0.00220776$ & $160.14551\pm0.01344$ & $0.012935_{-0.000673}^{+0.000896}$ & $83.959_{-23.552}^{+11.893}$ & $0.676_{-0.148}^{+0.173}$\\
        12505309 b & $2.89755848\pm3.03\times10^{-6}$ & $122.99708\pm0.00109$ & $0.003888_{-0.000306}^{+0.00036}$ & $16.478_{-4.924}^{+2.215}$ & $0.447_{-0.305}^{+0.337}$\\
    \end{longtable*}

\begin{longtable*}[h!]{c|c|c|c|c|c|c}
        \caption{\texttt{isochrones} fit results for select fitted stellar parameters ($R_{s}$, $M_{s}$, $T_{\text{eff}}$, $\log{g}$, [Fe/H], and distance $d$).}\label{tbl:stellar}\\
        \hline\hline
        KIC & $R_{s}$ ($R_{\astrosun}$) & $M_{s}$ ($M_{\astrosun}$) & $T_{\text{eff}}$ ($K$) & $\log{g}$ (cm/s$^{2}$) & [Fe/H] (dex) & $d$ (kpc) \\
        \hline
        \endfirsthead
        KIC & $R_{s}$ ($R_{\astrosun}$) & $M_{s}$ $M_{\astrosun}$ & $T_{\text{eff}}$ ($K$) & $\log{g}$ (cm/s$^{2}$) & [Fe/H] (dex) & $d$ (kpc) \\
        \hline
        \endhead
        1570311 & $5.26_{-0.34}^{+0.26}$ & $2.11_{-0.2}^{+0.27}$ & $5062_{-45}^{+39}$ & $3.347_{-0.072}^{+0.035}$ & $-0.181_{-0.171}^{+0.086}$ & $2.17_{-0.13}^{+0.11} $ \\
        2696784 & $1.42_{-0.03}^{+0.04}$ & $1.45_{-0.06}^{+0.04}$ & $7106_{-291}^{+419}$ & $4.292_{-0.026}^{+0.028}$ & $-0.021_{-0.126}^{+0.133}$ & $0.62_{-0.01}^{+0.01}$\\
2861140 & $1.28_{-0.09}^{+0.1}$ & $1.2_{-0.08}^{+0.09}$ & $6391_{-203}^{+221}$ & $4.302_{-0.06}^{+0.053}$ & $-0.026_{-0.139}^{+0.149}$ & $1.78_{-0.13}^{+0.14} $ \\
2985262 & $0.95_{-0.01}^{+0.01}$ & $1.02_{-0.04}^{+0.03}$ & $5902_{-167}^{+158}$ & $4.494_{-0.017}^{+0.011}$ & $-0.059_{-0.176}^{+0.153}$ & $0.53_{-0.0}^{+0.0} $ \\
3336146 & $1.11_{-0.02}^{+0.02}$ & $1.13_{-0.05}^{+0.04}$ & $6281_{-195}^{+176}$ & $4.407_{-0.029}^{+0.017}$ & $-0.08_{-0.157}^{+0.172}$ & $0.69_{-0.01}^{+0.01} $ \\
3345775 & $1.87_{-0.04}^{+0.03}$ & $2.58_{-0.13}^{+0.07}$ & $11580_{-193}^{+214}$ & $4.306_{-0.022}^{+0.018}$ & $-0.215_{-0.199}^{+0.114}$ & $0.28_{-0.0}^{+0.0} $ \\
3347135 & $1.13_{-0.04}^{+0.05}$ & $1.16_{-0.05}^{+0.05}$ & $5932_{-89}^{+65}$ & $4.395_{-0.029}^{+0.029}$ & $0.272_{-0.095}^{+0.091}$ & $0.38_{-0.02}^{+0.02}$\\
3662290 & $1.56_{-0.04}^{+0.05}$ & $1.48_{-0.1}^{+0.09}$ & $6986_{-322}^{+265}$ & $4.221_{-0.05}^{+0.039}$ & $0.078_{-0.133}^{+0.132}$ & $0.58_{-0.01}^{+0.01}$ \\
3728762 & $1.7_{-0.07}^{+0.06}$ & $1.43_{-0.07}^{+0.16}$ & $6706_{-234}^{+269}$ & $4.129_{-0.042}^{+0.084}$ & $0.111_{-0.196}^{+0.151}$ & $1.02_{-0.03}^{+0.03}$ \\
3967744 & $2.43_{-0.1}^{+0.12}$ & $1.85_{-0.07}^{+0.08}$ & $7004_{-260}^{+266}$ & $3.932_{-0.039}^{+0.041}$ & $0.273_{-0.121}^{+0.106}$ & $1.86_{-0.08}^{+0.08}$ \\
4346258 & $0.8_{-0.03}^{+0.03}$ & $0.86_{-0.04}^{+0.04}$ & $5279_{-146}^{+163}$ & $4.567_{-0.025}^{+0.021}$ & $-0.038_{-0.125}^{+0.145}$ & $0.97_{-0.03}^{+0.04}$ \\
4551429 & $0.55_{-0.0}^{+0.0}$ & $0.58_{-0.01}^{+0.01}$ & $3786_{-32}^{+52}$ & $4.727_{-0.01}^{+0.007}$ & $0.334_{-0.115}^{+0.086}$ & $0.15_{-0.0}^{+0.0}$ \\
4556565 & $1.37_{-0.1}^{+0.19}$ & $1.29_{-0.04}^{+0.05}$ & $6063_{-139}^{+120}$ & $4.272_{-0.113}^{+0.078}$ & $0.345_{-0.049}^{+0.096}$ & $1.51_{-0.09}^{+0.19}$ \\
5095499 & $1.22_{-0.04}^{+0.05}$ & $1.14_{-0.06}^{+0.07}$ & $6286_{-185}^{+194}$ & $4.323_{-0.041}^{+0.039}$ & $-0.044_{-0.167}^{+0.153}$ & $1.51_{-0.05}^{+0.07}$ \\
5184017 & $2.61_{-0.5}^{+0.16}$ & $1.73_{-0.1}^{+0.04}$ & $6187_{-81}^{+293}$ & $3.841_{-0.041}^{+0.159}$ & $0.388_{-0.082}^{+0.05}$ & $1.27_{-0.14}^{+0.05}$ \\
5342061 & $1.09_{-0.04}^{+0.05}$ & $1.09_{-0.08}^{+0.06}$ & $6120_{-193}^{+214}$ & $4.395_{-0.038}^{+0.037}$ & $-0.047_{-0.154}^{+0.147}$ & $1.29_{-0.05}^{+0.06}$ \\
5628770 & $1.2_{-0.02}^{+0.03}$ & $1.22_{-0.05}^{+0.04}$ & $6510_{-193}^{+195}$ & $4.363_{-0.025}^{+0.017}$ & $-0.057_{-0.169}^{+0.148}$ & $0.84_{-0.01}^{+0.01}$ \\
5649129 & $4.31_{-0.35}^{+0.19}$ & $1.6_{-0.26}^{+0.26}$ & $4839_{-108}^{+173}$ & $3.37_{-0.058}^{+0.072}$ & $0.004_{-0.438}^{+0.111}$ & $1.47_{-0.08}^{+0.06}$ \\
5794479 & $2.56_{-0.12}^{+0.16}$ & $3.65_{-0.85}^{+0.19}$ & $12631_{-2685}^{+961}$ & $4.189_{-0.14}^{+0.055}$ & $0.257_{-0.174}^{+0.113}$ & $1.44_{-0.06}^{+0.04}$ \\
5893807 & $1.71_{-0.11}^{+0.16}$ & $1.51_{-0.11}^{+0.09}$ & $6692_{-283}^{+260}$ & $4.147_{-0.083}^{+0.076}$ & $0.212_{-0.156}^{+0.119}$ & $2.1_{-0.13}^{+0.16}$ \\
6021193 & $1.62_{-0.02}^{+0.03}$ & $1.29_{-0.03}^{+0.03}$ & $5919_{-85}^{+131}$ & $4.128_{-0.015}^{+0.019}$ & $0.302_{-0.079}^{+0.077}$ & $0.78_{-0.01}^{+0.01}$ \\
6126245 & $1.54_{-0.04}^{+0.04}$ & $1.5_{-0.1}^{+0.07}$ & $7190_{-330}^{+346}$ & $4.235_{-0.044}^{+0.035}$ & $0.013_{-0.172}^{+0.137}$ & $0.76_{-0.02}^{+0.02}$ \\
6139884 & $0.89_{-0.01}^{+0.01}$ & $0.96_{-0.04}^{+0.03}$ & $5931_{-166}^{+194}$ & $4.523_{-0.018}^{+0.012}$ & $-0.263_{-0.196}^{+0.178}$ & $0.37_{-0.0}^{+0.0}$ \\
6224562 & $0.8_{-0.02}^{+0.03}$ & $0.86_{-0.03}^{+0.04}$ & $4977_{-121}^{+106}$ & $4.571_{-0.024}^{+0.018}$ & $0.234_{-0.125}^{+0.108}$ & $0.68_{-0.03}^{+0.03}$ \\
6347299 & $1.01_{-0.01}^{+0.01}$ & $1.07_{-0.04}^{+0.03}$ & $5965_{-79}^{+80}$ & $4.455_{-0.024}^{+0.014}$ & $0.01_{-0.091}^{+0.085}$ & $0.7_{-0.01}^{+0.01}$ \\
6380164 & $2.19_{-0.13}^{+0.1}$ & $1.67_{-0.06}^{+0.05}$ & $6717_{-120}^{+135}$ & $3.977_{-0.037}^{+0.054}$ & $0.21_{-0.119}^{+0.113}$ & $1.03_{-0.05}^{+0.04}$ \\
6440915 & $2.14_{-0.15}^{+0.13}$ & $1.64_{-0.08}^{+0.07}$ & $6577_{-184}^{+220}$ & $3.986_{-0.05}^{+0.07}$ & $0.265_{-0.145}^{+0.12}$ & $1.89_{-0.11}^{+0.11}$ \\
6782399 & $1.89_{-0.09}^{+0.06}$ & $1.51_{-0.08}^{+0.08}$ & $6659_{-110}^{+101}$ & $4.06_{0.031}^{+0.062}$ & $0.133_{-0.154}^{+0.176}$ & $0.84_{-0.03}^{+0.02}$ \\
6837899 & $1.09_{-0.02}^{+0.03}$ & $1.12_{-0.05}^{+0.04}$ & $6228_{-188}^{+191}$ & $4.415_{-0.031}^{+0.018}$ & $-0.078_{-0.153}^{+0.168}$ & $1.1_{-0.02}^{+0.02}$ \\
6888194 & $2.74_{-0.35}^{+0.25}$ & $1.82_{-0.07}^{+0.08}$ & $6482_{-149}^{+282}$ & $3.812_{-0.058}^{+0.118}$ & $0.33_{-0.128}^{+0.077}$ & $1.36_{-0.12}^{+0.1}$ \\
6929071 & $1.98_{-0.07}^{+0.08}$ & $1.54_{-0.07}^{+0.09}$ & $6633_{-251}^{+254}$ & $4.03_{-0.035}^{+0.041}$ & $0.148_{-0.153}^{+0.153}$ & $1.51_{-0.04}^{+0.05}$ \\
6937870 & $0.6_{-0.01}^{+0.01}$ & $0.62_{-0.02}^{+0.02}$ & $4209_{-89}^{+89}$ & $4.681_{-0.014}^{+0.008}$ & $-0.089_{-0.144}^{+0.146}$ & $0.21_{-0.0}^{+0.0}$ \\
7020834 & $2.14_{-0.08}^{+0.09}$ & $1.76_{-0.11}^{+0.17}$ & $7166_{-356}^{+589}$ & $4.018_{-0.049}^{+0.077}$ & $0.203_{-0.164}^{+0.146}$ & $0.76_{-0.02}^{+0.02}$ \\
7119412 & $0.74_{-0.01}^{+0.01}$ & $0.79_{-0.03}^{+0.02}$ & $4880_{-83}^{+102}$ & $4.6_{-0.017}^{+0.011}$ & $0.021_{-0.1}^{+0.094}$ & $0.4_{-0.0}^{+0.0}$ \\
7186892 & $0.74_{-0.02}^{+0.03}$ & $0.81_{-0.03}^{+0.03}$ & $4979_{-72}^{+77}$ & $4.603_{-0.016}^{+0.017}$ & $-0.027_{-0.095}^{+0.117}$ & $0.19_{-0.01}^{+0.01}$ \\
7187389 & $0.88_{-0.02}^{+0.02}$ & $0.95_{-0.04}^{+0.03}$ & $5658_{-165}^{+176}$ & $4.529_{-0.02}^{+0.015}$ & $-0.044_{-0.182}^{+0.148}$ & $0.86_{-0.02}^{+0.02}$ \\
7269798 & $0.54_{-0.01}^{+0.01}$ & $0.58_{-0.01}^{+0.01}$ & $3758_{-21}^{+28}$ & $4.73_{-0.008}^{+0.009}$ & $0.377_{-0.075}^{+0.055}$ & $0.22_{-0.0}^{+0.0}$ \\
7340288 & $0.55_{-0.01}^{+0.01}$ & $0.57_{-0.01}^{+0.02}$ & $3949_{-52}^{+79}$ & $4.722_{-0.012}^{+0.008}$ & $0.029_{-0.149}^{+0.114}$ & $0.33_{-0.0}^{+0.0}$ \\
7747788 & $1.71_{-0.05}^{+0.05}$ & $1.63_{-0.1}^{+0.07}$ & $7146_{-314}^{+389}$ & $4.186_{-0.042}^{+0.031}$ & $0.174_{-0.124}^{+0.115}$ & $0.75_{-0.01}^{+0.02}$ \\
7974496 & $1.62_{-0.1}^{+0.09}$ & $1.33_{-0.07}^{+0.08}$ & $6448_{-284}^{+218}$ & $4.14_{-0.043}^{+0.065}$ & $0.079_{-0.126}^{+0.156}$ & $1.91_{-0.1}^{+0.1}$ \\
8172679 & $4.89_{-0.48}^{+0.55}$ & $1.7_{-0.17}^{+0.19}$ & $5040_{-72}^{+64}$ & $3.284_{-0.078}^{+0.084}$ & $-0.452_{-0.18}^{+0.181}$ & $1.37_{-0.12}^{+0.17}$ \\
9274173 & $1.12_{-0.03}^{+0.04}$ & $1.12_{-0.06}^{+0.05}$ & $6006_{-106}^{+97}$ & $4.385_{-0.038}^{+0.035}$ & $0.122_{-0.112}^{+0.122}$ & $1.17_{-0.03}^{+0.04}$ \\
9716483 & $1.57_{-0.05}^{+0.05}$ & $1.54_{-0.1}^{+0.09}$ & $7327_{-417}^{+410}$ & $4.229_{-0.043}^{+0.041}$ & $0.008_{-0.128}^{+0.143}$ & $0.98_{-0.02}^{+0.02}$ \\
9777962 & $2.42_{-0.15}^{+0.14}$ & $1.77_{-0.08}^{+0.08}$ & $6764_{-241}^{+282}$ & $3.916_{-0.045}^{+0.05}$ & $0.24_{-0.124}^{+0.126}$ & $2.6_{-0.14}^{+0.15}$ \\
10018357 & $4.69_{-1.02}^{+0.14}$ & $1.99_{-0.21}^{+0.78}$ & $5251_{-178}^{+667}$ & $3.406_{-0.056}^{+0.252}$ & $-0.435_{-0.491}^{+0.382}$ & $1.83_{-0.11}^{+0.04}$ \\
10083396 & $1.56_{-0.03}^{+0.03}$ & $1.33_{-0.06}^{+0.05}$ & $6387_{-137}^{+116}$ & $4.176_{-0.026}^{+0.026}$ & $0.109_{-0.087}^{+0.14}$ & $0.46_{-0.01}^{+0.01}$ \\
10419787 & $1.2_{-0.03}^{+0.03}$ & $1.21_{-0.06}^{+0.05}$ & $6361_{-195}^{+208}$ & $4.364_{-0.032}^{+0.022}$ & $0.012_{-0.163}^{+0.141}$ & $1.01_{-0.02}^{+0.03}$ \\
10598829 & $1.55_{-0.13}^{+0.2}$ & $1.47_{-0.08}^{+0.1}$ & $6743_{-235}^{+166}$ & $4.235_{-0.109}^{+0.054}$ & $0.206_{-0.15}^{+0.137}$ & $1.35_{-0.11}^{+0.17}$ \\
10879314 & $2.47_{-0.3}^{+0.21}$ & $1.69_{-0.08}^{+0.08}$ & $6235_{-120}^{+152}$ & $3.877_{-0.054}^{+0.092}$ & $0.378_{-0.087}^{+0.062}$ & $2.18_{-0.24}^{+0.19}$ \\
11092463 & $0.88_{-0.03}^{+0.03}$ & $0.95_{-0.04}^{+0.03}$ & $5601_{-168}^{+148}$ & $4.533_{-0.024}^{+0.02}$ & $-0.004_{-0.144}^{+0.154}$ & $1.1_{-0.04}^{+0.04}$ \\
11139863 & $1.8_{-0.06}^{+0.05}$ & $1.63_{-0.13}^{+0.14}$ & $7185_{-353}^{+330}$ & $4.14_{-0.058}^{+0.057}$ & $0.119_{-0.277}^{+0.196}$ & $0.26_{-0.0}^{+0.0}$ \\
11350118 & $0.67_{-0.01}^{+0.01}$ & $0.72_{-0.02}^{+0.02}$ & $4751_{-153}^{+154}$ & $4.646_{-0.014}^{+0.011}$ & $-0.171_{-0.153}^{+0.15}$ & $0.52_{-0.01}^{+0.01}$ \\
11565976 & $1.93_{-0.05}^{+0.05}$ & $1.51_{-0.07}^{+0.1}$ & $6668_{-211}^{+239}$ & $4.044_{-0.031}^{+0.04}$ & $0.102_{-0.132}^{+0.167}$ & $0.81_{-0.02}^{+0.02}$ \\
11805835 & $0.63_{-0.01}^{+0.01}$ & $0.67_{-0.02}^{+0.02}$ & $4723_{-141}^{+184}$ & $4.663_{-0.014}^{+0.012}$ & $-0.364_{-0.191}^{+0.151}$ & $0.38_{-0.01}^{+0.0}$ \\
12023559 & $1.03_{-0.02}^{+0.02}$ & $1.06_{-0.05}^{+0.04}$ & $6136_{-201}^{+172}$ & $4.438_{-0.026}^{+0.019}$ & $-0.126_{-0.175}^{+0.194}$ & $1.02_{-0.02}^{+0.02}$ \\
12216301 & $2.57_{-0.1}^{+0.18}$ & $3.54_{-0.33}^{+0.16}$ & $12170_{-1230}^{+930}$ & $4.165_{-0.087}^{+0.054}$ & $0.293_{-0.159}^{+0.092}$ & $1.63_{-0.04}^{+0.04}$ \\
12505309 & $1.63_{-0.03}^{+0.04}$ & $1.52_{-0.09}^{+0.09}$ & $6931_{-270}^{+289}$ & $4.196_{-0.041}^{+0.034}$ & $0.138_{-0.134}^{+0.13}$ & $0.48_{-0.01}^{+0.01}$ \\
\end{longtable*}

\begin{longtable*}[h!]{c|c|c|c|c|c|c|c|c|c}
    \caption{Summary of results for all new candidate planets (CAND; FPP $<$ 0.9) and false positives (FP; due to low reliability, stellar variability, centroid offset, or FPP $>$ 0.9). Planetary radii do not take into account dilution; refer to Table \ref{tbl:dilution}.}\label{tbl:results} \\
    \hline\hline
    KIC & $P$ (days) & $R_{p}$ ($R_{\bigoplus}$) & $a$ (au) &  $T_{\text{eq}}$ ($K$) & $S$ $(S_{\bigoplus})$ & S/N & FPP & Status & Notes \\
        \hline
        \endfirsthead
        KIC & $P$ (days) & $R_{p}$ ($R_{\bigoplus}$) & $a$ (au) &  $T_{\text{eq}}$ ($K$) & $S$ $(S_{\bigoplus})$ & S/N & FPP & Status & Notes \\
        \hline
        \endhead
        1570311 b & 23.4 & $8.40_{-1.77}^{+2.49}$ & $0.197_{-0.008}^{+0.005}$ & $1033_{-88}^{+88}$ & $270.45_{-81.37}^{+104.82}$ & 8.9 & - & FP & Stellar variability \\
        2696784 b & 82.3 & $1.50_{-0.10}^{+0.11}$ & $0.418_{-0.007}^{+0.005}$ & $579_{-25}^{+30}$ & $26.68_{-4.27}^{+6.06}$ & 7.2 & - & CAND & \texttt{vespa} failed \\
        2861140 b & 36.9 & $2.28_{-0.27}^{+0.32}$ & $0.230_{-0.005}^{+0.006}$ & $666_{-33}^{+37}$ & $46.52_{-8.50}^{+11.15}$ & 7.2 & 0.0526 & CAND & \\
        2985262 b & 13.0 & $0.94_{-0.05}^{+0.06}$ & $0.109_{-0.001}^{+0.001}$ & $767_{-22}^{+23}$ & $81.92_{-9.15}^{+10.26}$ & 10.7 & - & FP & Stellar variability \\
        3336146 b & 3.3 & $0.86_{-0.07}^{+0.08}$ & $0.045_{-0.001}^{+0.001}$ & $1374_{-44}^{+45}$ & $845.59_{-102.32}^{+115.09}$ & 9.2 & - & FP & centroid offset \\
        3345775 b & 6.2 & $0.87_{-0.04}^{+0.05}$ & $0.091_{-0.002}^{+0.001}$ & $2320_{-50}^{+55}$ & $6867.32_{-568.34}^{+678.38}$ & 7.8 & - & FP & Stellar variability \\
        3347135 b & 226.5 & $2.16_{-0.10}^{+0.12}$ & $0.765_{-0.011}^{+0.010}$ & $318_{-8}^{+8}$ & $2.42_{-0.24}^{+0.27}$ & 10.0 & - & FP & Stellar variability \\
        3662290 b & 288.2 & $2.01_{-0.13}^{+0.17}$  & $0.974_{-0.023}^{+0.018}$ & $391_{-19}^{+17}$ & $5.53_{-0.99}^{+1.05}$ & 6.5 & - & FP & Likely noise \\
        3728762 b & 6.7 & $1.35_{-0.11}^{+0.12}$ & $0.079_{-0.001}^{+0.003}$ & $1372_{-59}^{+63}$ & $845.85_{-140.52}^{+159.92}$ & 8.2 & $1.45\times10^{-5}$ & CAND & \\
        3967744 b & 57.9 & $3.10_{-0.25}^{+0.30}$ & $0.360_{-0.004}^{+0.005}$ & $801_{-34}^{+40}$ & $97.73_{-15.76}^{+21.24}$  & 7.8 & - & FP & Stellar variability \\
        4346258 b & 4.9 & $1.19_{-0.11}^{+0.12}$ & $0.054_{-0.001}^{+0.001}$ & $899_{-29}^{+31}$ & $155.10_{-19.27}^{+22.71}$ & 8.3 & - & FP & Stellar variability \\
        4551429 b & 35.4 & $0.74_{-0.05}^{+0.06}$ & $0.176_{-0.001}^{+0.001}$ & $294_{-3}^{+4}$ & $1.78_{-0.07}^{+0.10}$ & 9.0 & - & FP & Stellar variability \\
        4556565 b & 5.5 & $1.74_{-0.20}^{+0.26}$ & $0.067_{-0.001}^{+0.001}$ & $1213_{-54}^{+82}$ & $513.70_{-85.37}^{+153.38}$ & 9.2 & - & FP & Stellar variability \\
        5095499 b & 4.3 & $1.17_{-0.08}^{+0.09}$ & $0.054_{-0.001}^{+0.001}$ & $1317_{-48}^{+50}$ & $714.34_{-98.96}^{+114.32}$ & 8.9 & - & FP & Stellar variability \\
        5184017 b & 6.3 & $1.72_{-0.30}^{+0.23}$ & $0.080_{-0.002}^{+0.001}$ & $1555_{-132}^{+93}$ & $1387.16_{-415.38}^{+364.15}$ & 7.7 & - & FP & Stellar variability \\
        5342061 c & 11.5 & $1.59_{-0.13}^{+0.15}$ & $0.103_{-0.002}^{+0.002}$ & $884_{-34}^{+35}$ & $144.89_{-21.26}^{+24.70}$ & 7.8 & - & FP & Stellar variability \\
        5628770 b & 11.4 & $1.10_{-0.08}^{+0.09}$ & $0.106_{-0.002}^{+0.001}$ & $969_{-31}^{+32}$ & $208.86_{-25.80}^{+28.60}$ & 8.1 & - & FP & Stellar variability\\
        5649129 b & 2.8 & $3.30_{-0.32}^{+0.37}$ & $0.046_{-0.003}^{+0.002}$ & $2068_{-106}^{+104}$ & $4338.53_{-824.17}^{+941.03}$ & 8.2 & - & FP & Stellar variability \\
        5794479 b & 5.9 & $2.03_{-0.14}^{+0.27}$ & $0.099_{-0.006}^{+0.002}$ & $2865_{-511}^{+309}$ & $15980.38_{-8692.91}^{+8086.37}$ & 9.1 & - & FP & Stellar variability \\
        5893807 b & 7.7 & $0.85_{-0.09}^{+0.12}$ & $0.087_{-0.002}^{+0.002}$ & $1309_{-71}^{+74}$ & $695.49_{-138.60}^{+172.15}$ & 8.0 & $1.70\times10^{-6}$ & CAND & \\
        6021193 e & 26.5 & $1.60_{-0.09}^{+0.14}$ & $0.189_{-0.001}^{+0.002}$ & $763_{-13}^{+19}$ & $80.41_{-5.5}^{+8.14}$ & 8.3 & - & FP & Stellar variability\\
        6126245 b & 3.5 & $0.68_{-0.06}^{+0.06}$ & $0.052_{-0.001}^{+0.001}$ & $1739_{-86}^{+91}$ & $2166.86_{-398.49}^{+488.70}$  & 7.3 & $6.65\times10^{-3}$ & CAND & \\
        6139884 b & 4.8 & $0.51_{-0.04}^{+0.04}$ & $0.055_{-0.001}^{+0.001}$ & $1053_{-31}^{+36}$ & $291.76_{-33.13}^{+42.43}$ & 8.7 & - & FP & Stellar variability \\
        6224562 b & 2.3 & $1.08_{-0.09}^{+0.12}$ & $0.033_{-0.001}^{+0.001}$ & $1083_{-32}^{+32}$ & $326.34_{-36.61}^{+40.05}$ & 9.3 & 0.110 & CAND & \\
        6347299 d & 38.6 & $1.08_{-0.07}^{+0.08}$ & $0.229_{-0.003}^{+0.002}$ & $444_{-9}^{+10}$ & $22.48_{-1.37}^{+1.58}$ & 8.2 & - & FP & Stellar variability \\
        6380164 b  & 167.8 & $3.42_{-0.23}^{+0.21}$ & $0.707_{-0.009}^{+0.007}$ &$521_{-19}^{+17}$ & $17.51_{-2.36}^{+2.34}$ & 9.2 & - & FP & Stellar variability \\
        6440915 b & 365.4 & $5.56_{-0.52}^{+0.60}$ & $1.179_{-0.019}^{+0.017}$ & $391_{-18}^{+18}$ & $5.54_{-0.96}^{+1.11}$ & 8.8 & - & FP & Likely noise \\
        6782399 b & 34.2 & $1.65_{-0.10}^{+0.12}$& $0.237_{-0.004}^{+0.004}$ & $828_{-25}^{+21}$ & $111.59_{-12.95}^{+11.86}$ & 8.2 & $1.29\times10^{-4}$ & CAND & \\
        6837899 b & 9.0 & $1.27_{-0.11}^{+0.12}$ & $0.088_{-0.001}^{+0.001}$ & $968_{-33}^{+33}$ & $207.91_{-26.61}^{+29.44}$ & 7.7 & - & FP & Stellar variability \\
        6888194 b & 46.0 & $2.76_{-0.40}^{+0.33}$ & $0.307_{-0.004}^{+0.004}$ & $858_{-59}^{+50}$ & $128.74_{-32.03}^{+33.07}$  & 7.4 & - & FP & Stellar variability \\
        6929071 b & 61.9 & $2.45_{-0.21}^{+0.24}$ & $0.363_{-0.006}^{+0.005}$ & $692_{-28}^{+27}$ & $54.38_{-8.25}^{+8.86}$ & 7.7 & - & CAND & \texttt{vespa} failed \\
        6937870 b & 27.5 & $0.65_{-0.05}^{+0.07}$ & $0.152_{-0.002}^{+0.001}$ & $368_{-8}^{+9}$ & $4.33_{-0.36}^{+0.42}$  & 7.7 & - 
        & FP & Stellar variability \\
        7020834 b & 369.5 & $4.34_{-0.39}^{+0.92}$ & $1.218_{-0.026}^{+0.039}$ & $420_{-25}^{+35}$  & $7.34_{-1.60}^{+2.80}$  & 9.6 & - & FP & Likely noise \\
        7119412 b & 10.5 & $0.89_{-0.07}^{+0.10}$ & $0.087_{-0.001}^{+0.001}$ & $623_{-12}^{+14}$ & $36.84_{-2.65}^{+3.34}$ & 9.3 & - & FP & Stellar variability \\
        7186892 b & 17.2 & $0.56_{-0.04}^{+0.06}$ & $0.122_{-0.002}^{+0.001}$ & $543_{-13}^{+13}$ & $20.61_{-1.88}^{+2.09}$ & 10.1 & - & FP & Stellar variability \\
        7187389 b & 23.8 & $1.10_{-0.09}^{+0.10}$ & $0.159_{-0.002}^{+0.002}$ & $488_{-20}^{+21}$ & $28.28_{-3.57}^{+4.36}$ & 6.8 & - & FP & Stellar variability \\
        7269798 b & 21.4 & $0.88_{-0.07}^{+0.09}$ & $0.126_{-0.001}^{+0.001}$ & $344_{-3}^{+4}$ & $3.34_{-0.12}^{+0.14}$  & 8.2 & 0.0113 & CAND & \\
        7340288 b & 142.5 & $1.51_{-0.11}^{+0.13}$ & $0.444_{-0.004}^{+0.004}$ & $194_{-3}^{+4}$ & $0.33_{-0.02}^{+0.03}$  & 7.4 & $7.91\times10^{-4}$ & CAND & Rocky HZ\\
        7747788 b & 133.1 & $1.67_{-0.11}^{+0.13}$ & $0.601_{-0.012}^{+0.009}$ & $533_{-25}^{+29}$ & $19.19_{-3.33}^{+4.61}$ & 7.6 & $4.92\times10^{-4}$ & CAND & \\
        7974496 b & 4.0 & $1.48_{-0.16}^{+0.18}$ & $0.054_{-0.001}^{+0.001}$ & $1549_{-81}^{+75}$ & $1364.17_{-262.54}^{+283.65}$ & 7.1 & - & FP & Stellar variability \\
        8172679 b & 194.0 & $9.61_{-1.09}^{+1.23}$ & $0.784_{-0.027}^{+0.029}$ & $557_{-33}^{+34}$ & $22.80_{-5.00}^{+6.11}$ & 10.5 & - & FP & Stellar variability\\
        9274173 b & 4.4 & $1.42_{-0.11}^{+0.13}$ & $0.055_{-0.001}^{+0.001}$ & $1199_{-29}^{+30}$ & $490.71_{-46.03}^{+51.21}$ & 8.7 & - & FP & Stellar variability\\
        9716483 b & 209.4 & $2.08_{-0.11}^{+0.14}$ & $0.797_{-0.017}^{+0.015}$ & $454_{-26}^{+28}$ & $10.10_{-2.15}^{+2.68}$ & 8.6 & - & FP & Likely noise \\
        9777962 b & 367.2 & $7.46_{-0.84}^{+1.35}$ & $1.215_{-0.019}^{+0.017}$ & $422_{-21}^{+21}$ & $7.51_{-1.37}^{+1.61}$ & 8.9 & - & FP & Likely noise \\
        10018357 b & 133.8 & $7.87_{-1.72}^{+0.71}$ & $0.644_{-0.023}^{+0.075}$ & $616_{-74}^{+89}$ & $34.05_{-13.69}^{+24.49}$ & 9.1 & $1.13\times10^{-8}$ & CAND & \\
        10083396 b & 113.5 & $1.14_{-0.06}^{+0.07}$ & $0.504_{-0.007}^{+0.006}$ & $495_{-12}^{+12}$ & $14.19_{-1.28}^{+1.39}$ & 7.4 & $5.86\times10^{-5}$ & CAND & \\
        10419787 b & 122.7 & $2.06_{-0.14}^{+0.17}$ & $0.515_{-0.009}^{+0.007}$ & $429_{-15}^{+15}$ & $8.03_{-1.05}^{+1.19}$ & 7.6 & - & FP & Stellar variability  \\
        10598829 b & 67.5 & $1.96_{-0.22}^{+0.30}$ & $0.369_{-0.007}^{+0.009}$ & $607_{-32}^{+42}$ & $32.29_{-6.37}^{+9.81}$  & 7.4 & 3.59$\times10^{-3}$ & CAND & \\
        10879314 b & 49.2 & $3.92_{-0.54}^{+0.57}$ & $0.313_{-0.005}^{+0.005}$ & $773_{-51}^{+43}$ & $84.68_{-20.40}^{+20.52}$ & 7.2 & - & FP & Stellar variability \\
        11092463 b & 6.9 & $1.33_{-0.14}^{+0.24}$ & $0.070_{-0.001}^{+0.001}$ & $877_{-31}^{+30}$ & $140.06_{-18.65}^{+20.17}$ & 7.8 & - & FP & Stellar variability \\
        11139863 b & 7.2 & $0.81_{-0.05}^{+0.07}$& $0.086_{-0.002}^{+0.002}$ & $1444_{-77}^{+87}$ & $1031.85_{-202.39}^{+272.09}$  & 10.6 & - & FP & Stellar variability\\
        11350118 c & 2.7 & $0.66_{-0.05}^{+0.07}$& $0.034_{-0.001}^{+0.001}$ & $935_{-31}^{+31}$ & $181.58_{-22.96}^{+25.48}$ & 8.2 & $9.62\times10^{-4}$ & CAND & KOI-4509.02\\
        11565976 b & 24.2 & $1.40_{-0.11}^{+0.12}$& $0.188_{-0.003}^{+0.004}$ & $941_{-37}^{+39}$ & $186.19_{-27.77}^{+32.67}$  & 7.4 & - & FP & Stellar variability \\
        11805835 b & 23.5 & $0.94_{-0.07}^{+0.10}$ & $0.141_{-0.001}^{+0.001}$ & $442_{-14}^{+17}$ & $9.01_{-1.08}^{+1.49}$ & 7.2 & 2.30$\times10^{-3}$ & CAND & \\
        12023559 b & 84.6 & $1.86_{-0.11}^{+0.13}$ & $0.385_{-0.006}^{+0.005}$ & $444_{-16}^{+14}$  & $9.19_{-1.25}^{+1.24}$  & 8.1 & $4.54\times10^{-4}$ & CAND & \\
        12216301 b & 116.5 & $3.66_{-0.26}^{+0.41}$ & $0.712_{-0.023}^{+0.010}$ & $1029_{-108}^{+104}$ & $265.44_{-95.33}^{+124.64}$ & 7.6 & - & FP & Stellar variability \\
        12505309 b & 2.9 & $1.20_{-0.08}^{+0.10}$ & $0.046_{-0.002}^{+0.001}$ & $1823_{-62}^{+81}$ & $2617.39_{-336.756}^{+497.97}$ & 7.4 & - & FP & Stellar variability \\
    \end{longtable*}

\subsection{Dilution}

When a nearby star is resolved in the AO images, we must consider the effects of dilution on the estimated planet radius. We note that we do not take into account changes to the measured values of the primary star's properties due to the presence of a companion. Since most of our targets have properties inferred from photometry only, light from a companion could cause the stellar type to be misidentified. However, this is likely negligible for companions with large contrast ratios.

We consider two cases: that the PC is transiting the brighter primary star (pri), or the fainter companion (sec). Using Eqns. 3 and 4 in \citet{law14}, the corresponding radius corrections are

\begin{equation}
    R_{p,\text{pri}} = R_{p}\sqrt{\frac{F_{\text{tot}}}{F_{\text{pri}}}}
\end{equation}

\noindent and

\begin{equation}
    R_{p,\text{sec}} = R_{p}\frac{R_{\text{sec}}}{R_{\text{pri}}}\sqrt{\frac{F_{\text{tot}}}{F_{\text{sec}}}}
\end{equation}

\noindent where $F_{i}/F_{\text{tot}}$ is the fraction of total light contributed by star $i$ in the aperture and $R_{p}$ is the planet radius without dilution corrections. If we assume that the total flux is provided by the two stars, $F_{\text{tot}} = F_{\text{pri}} + F_{\text{sec}}$, we can rewrite these equations in terms of magnitudes as

\begin{equation}\label{eqn:dil1}
    R_{p,\text{pri}} = R_{p}\sqrt{1 + 10^{-0.4\Delta m}}
\end{equation}

\noindent and

\begin{equation}\label{eqn:dil2}
    R_{p,\text{sec}} = R_{p}\frac{R_{\text{sec}}}{R_{\text{pri}}}\sqrt{1 + 10^{0.4\Delta m}}.
\end{equation}

Since the $\Delta m$ values in Eqns. \ref{eqn:dil1} and \ref{eqn:dil2} are in the \textit{Kepler} band ($Kp$), they must be converted from our $K_{s}$ band AO results. \citet{how12} derived the following conversion between $Kp$ and $K_{s}$ magnitudes:

\begin{equation}
\begin{split}
Kp - K_{s} & = -643.05169 + 246.00603K_{s} - 37.136501K_{s}^{2} \\
& + 2.7802622K_{s}^{3} - 0.10349091K_{s}^{4} \\
& + 0.0015364343K_{s}^{5}
\end{split}
\end{equation}

\noindent for $10 < K_{s} < 15.4$ mag, and 

\begin{equation}
Kp - K_{s} = -2.7284 + 0.3311K_{s}
\end{equation}

\noindent for $K_{s} > 15.4$ mag, allowing us to convert $\Delta K_{s}$ to $\Delta Kp$.

Furthermore, for the case that the PC transits the secondary, we require an estimate of the ratio of stellar radii, $R_{\text{sec}}/R_{\text{pri}}$. If the companion is in the background our single-band photometry is not able to constrain $R_{\text{sec}}$. However, if we assume the primary and secondary stars are bound, we can use our knowledge of the primary star to estimate the properties of the secondary.

We follow the general strategy outlined in \citet{fur17}, which we summarize here. First, we assume that $K_{p}$ magnitudes are roughly equivalent to $R$ magnitudes. We use the primary star's known $T_{\text{eff,pri}}$ (from our \texttt{isochrones} fit) with a table of colours and effective temperatures\footnote{http://www.pas.rochester.edu/$\sim$emamajek/EEM\textunderscore dwarf\textunderscore \newline UBVIJHK\textunderscore colors\textunderscore Teff.txt} \citep{pec13} to derive $(V - R)_{\text{pri}}$ colours and absolute $V$ magnitudes ($M_{V,\text{pri}}$). Then, we assume that the bound stars are the same distance to the Sun to let $M_{V,\text{sec}} = m_{V,\text{sec}} - m_{V,\text{pri}} + M_{V,\text{pri}}$, or $M_{V,\text{sec}} =  K_{p,\text{sec}} + (V-R)_{\text{sec}} - K_{p,\text{pri}} - (V-R)_{\text{pri}} + M_{V,\text{pri}}$. We find the $(V - R)_{\text{sec}}$ colour that yields a self-consistent $M_{V,\text{sec}}$ value, which in turn gives an estimate of the radius of secondary from the table.

We determined correction factors only for companions which could physically account for the observed transit depth. If the planet needed to fully obscure the companion in order to explain the transit, we ruled out this scenario for the planet host star. For example, a 1$\%$ transit depth would rule out any companion fainter than 5 mags or more as a potential planet host.

Table \ref{tbl:dilution} shows the results of each case for the candidates with resolved stars within $4^{\prime\prime}$. For both cases, the secondary was too faint to account for the transit depth or cause significant dilution.

\begin{table}[h]
    \caption{Revised planetary radii for the two candidates with AO-resolved stars within $4^{\prime\prime}$, considering whether the planet transits the primary or secondary.}\label{tbl:dilution}
        \centering
        \begin{tabular}{c|c|c|c}
        \hline\hline
        KIC & $R_{p}$ $(R_{\bigoplus})$ & $R_{p,\text{pri}}$ $(R_{\bigoplus})$ & $R_{p,\text{sec}}$ $(R_{\bigoplus})$ \\
        \hline
        7269798 b & 0.88 & 0.88 & - \\
        7340288 b & 1.51 & 1.51 & - \\
        \end{tabular}
\end{table}

\subsection{Highlighted Discoveries}\label{sec:highlight}

We highlight select discoveries from our new PC list. One of our candidates, KIC-7340288 b, is both likely rocky and in the Habitable Zone of its star, where the planet's surface temperature could allow liquid water oceans. Finding Earth-sized planets in the Habitable Zone was one of the original goals of the \textit{Kepler} mission \citep{bor10}. Furthermore, KIC-11350118 c is a new candidate associated with a known KOI system.

\begin{figure}[h!]
\centering
\includegraphics[width=\linewidth]{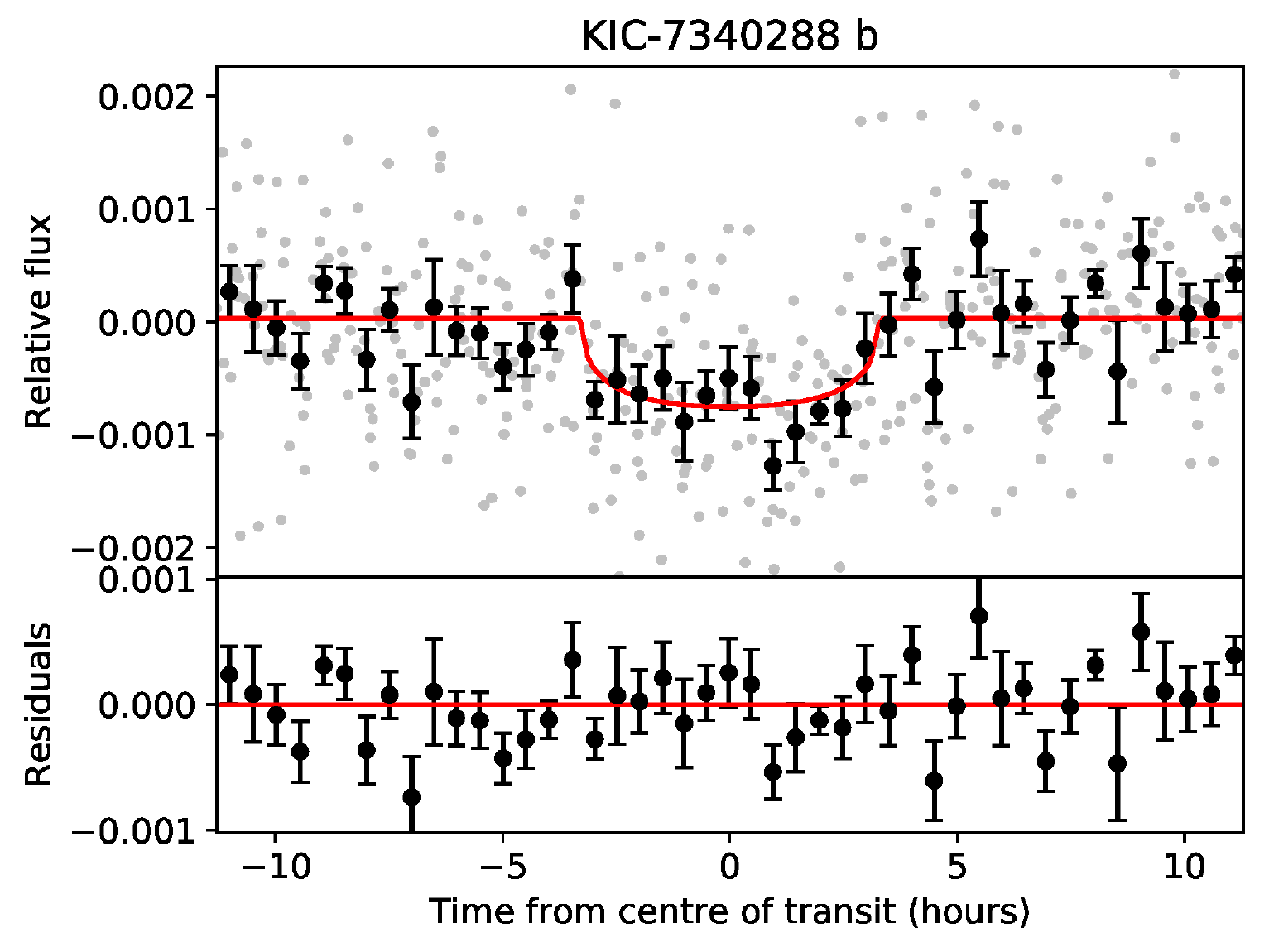}
\includegraphics[width=\linewidth]{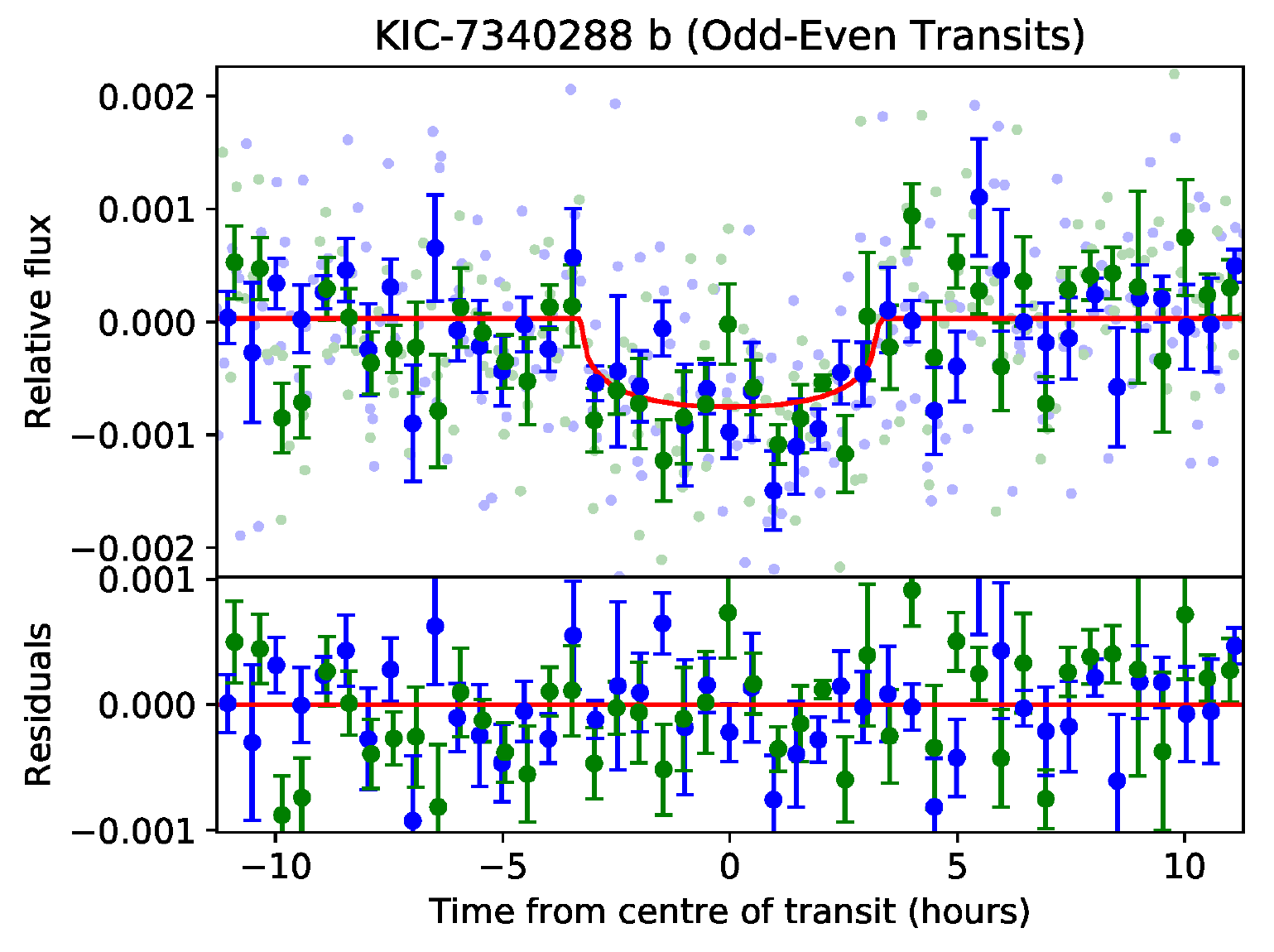}
\caption{Phase diagrams of the $1.51R_{\bigoplus}$ Habitable Zone PC KIC-7340288 b, plotting all data together (top) and indicating every odd and even transit (bottom). Odd transits are in blue and even transits are in green. Points plotted faintly in the background represents the actual data, while data binned into 30 minute bins are darker. Error bars represent the standard error of each bin. The full transit model MCMC fit to the data is plotted in red.}\label{fig:7340288}
\end{figure}

\subsubsection{KIC-7340288 b: A Candidate Super-earth in the Habitable Zone}

KIC-7340288 b is a $1.51R_{\bigoplus}$ PC orbiting a K dwarf ($T_{\text{eff}} = 3959 K$, $R_{s} = 0.547 R_{\astrosun}$, and $M_{s} = 0.574 M_{\astrosun}$) with an orbital period of 142.5 days. This candidate is in the Habitable Zone with an insolation of 0.33$ S_{\bigoplus}$, and is also likely rocky given its $< 1.6 R_{\bigoplus}$ radius.

Phase diagrams are plotted in Fig. \ref{fig:7340288}, showing the full transit as well as indicating data corresponding to odd and even transits. The odd-even plot shows consistency between their depths and durations, compared to each other as well as to the full transit's best-fit model.

\begin{figure}[h!]
\centering
\includegraphics[width=0.7\linewidth]{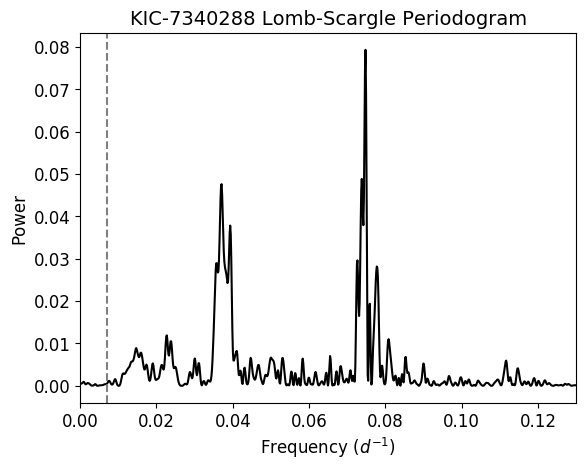}
\caption{Lomb-Scargle periodogram for the raw (un-detrended) KIC-7340288 light-curve, indicating the 142.5-day planet orbital period with a dotted line against the peaks of the periodogram. The strong peak on the right corresponds to the detected $\sim$13.4-day rotation period, while the strong peak on the left corresponds to its harmonic. No additional peaks were seen at higher frequencies (not plotted).
}\label{fig:periodogram}
\end{figure}

We also review our results from the stellar variability analysis in Section \ref{sec:variability}. Fig. \ref{fig:periodogram} shows the Lomb-Scargle periodogram, with the 142.5 day orbital period indicated by a dotted line. We find a strong rotation period at $\sim13.4$ days (half the $26.711\pm0.231$-day rotation period reported in \citet{mcq14}), but no multiples of this rotation period correspond to the orbital period.

\begin{figure}[h!]
\centering
\includegraphics[width=\linewidth]{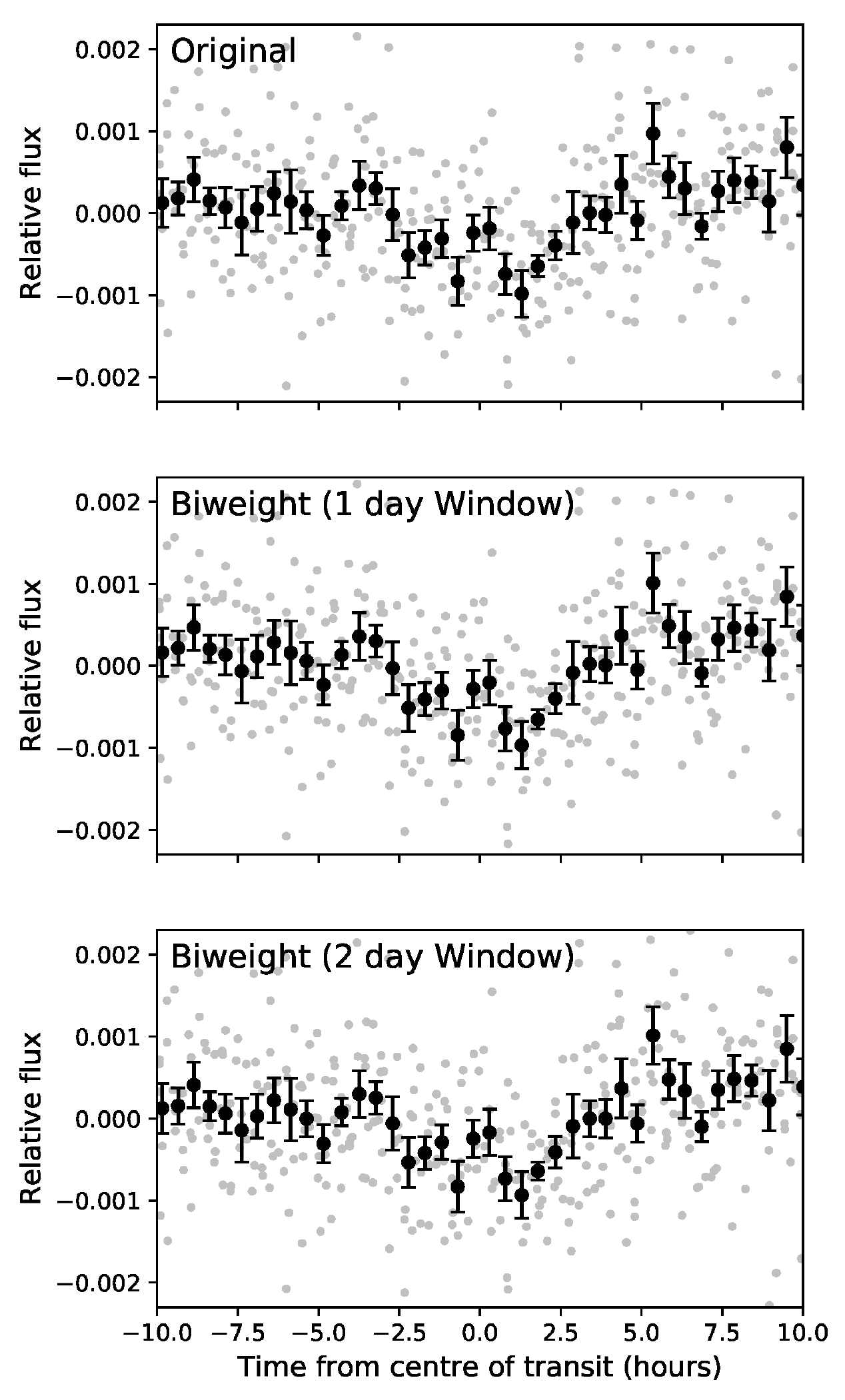}
\caption{Phase diagrams of KIC-7340288 b, plotting original data points in grey and data binned into 30 minute bins is in black. Error bars represent the standard error of each bin. Comparison can be made between the following detrending algorithms: the original detrend (top), the time-windowed slider described in \citet{hip19} with a 1 day window length (middle), and the slider with a 2 day window length (bottom).
}\label{fig:comparison}
\end{figure}

Fig. \ref{fig:comparison} also confirms that the transit remains consistent regardless of the choice of detrending algorithm used to remove the stellar variability. Plotted are phase diagrams of KIC-7340288, created by detrending the raw MAST light-curve with our original algorithm as well as the \citet{hip19} time-windowed slider with 1 and 2 day window lengths. In each case, the light-curve was folded at the BLS-detected period of the planet ($P = 142.5282$ days) and centred at the epoch ($T_{0} = 204.7231$ BKJD). Assuming the BLS-detected duration of $0.2355$ days, the transits have S/N of 7.4, 7.5, and 7.2, respectively. We also fit least-squares and MCMC transit models to each light-curve after masking the transits and re-detrending as in our standard pipeline. The best-fit parameters are given in Table \ref{tbl:comparison}, indicating good agreement within 1$\sigma$.

\begin{table*}[]
    \begin{center}
    \caption{LS + MCMC fit results for KIC-7340288 b using three different detrends.}
    \begin{tabular}{c|c|c|c|c|c}
    \hline
    \hline
        Detrend & $P$ (days) & $T_{0}$ (BKJD) & $R_{p}/R_{s}$ &  $a/R_{s}$ & $b$  \\
        \hline
        Original & $142.5324\pm0.0034$ & $204.7104\pm0.0180$ & $0.02526^{+0.00201}_{-0.00177}$ & $156.56^{+12.21}_{-32.38}$ & $0.369^{+0.311}_{-0.255}$ \\ 
        Biweight (1 day) & $142.5319\pm0.0039$ & $204.7125\pm0.0167$ & $0.02417^{+0.00183}_{-0.00198}$ & $159.98^{+13.52}_{-40.96}$ & $0.385^{+0.347}_{-0.286}$ \\
        Biweight (2 day) & $142.5319\pm0.0039$ & $204.7125\pm0.0170$ & $0.02232^{+0.00198}_{-0.00206}$ & $159.58^{+15.68}_{-44.28}$ & $0.402^{+0.351}_{-0.293}$ \\
    \end{tabular}
    \end{center}
    \label{tbl:comparison}
\end{table*}

Our AO imaging revealed one stellar companion within $4^{\prime\prime}$, with $\Delta K = 5.20$ and an angular separation of $3.9^{\prime\prime}$. Assuming this planet orbits the primary star, its radius is unchanged by dilution and it remains below the rocky limit. We are also able to rule out the scenario that the planet orbits the companion, since the faintness of the star implies it would need to be fully obscured by the planet to explain the observed 0.06$\%$ transit depth. After incorporating our AO results into \texttt{vespa}, we found an astrophysical FPP of $7.91\times10^{-4}$.

\subsubsection{KIC-11350118 b: A Small Candidate in a KOI System}

\begin{figure}[h!]
\centering
\includegraphics[width=\linewidth]{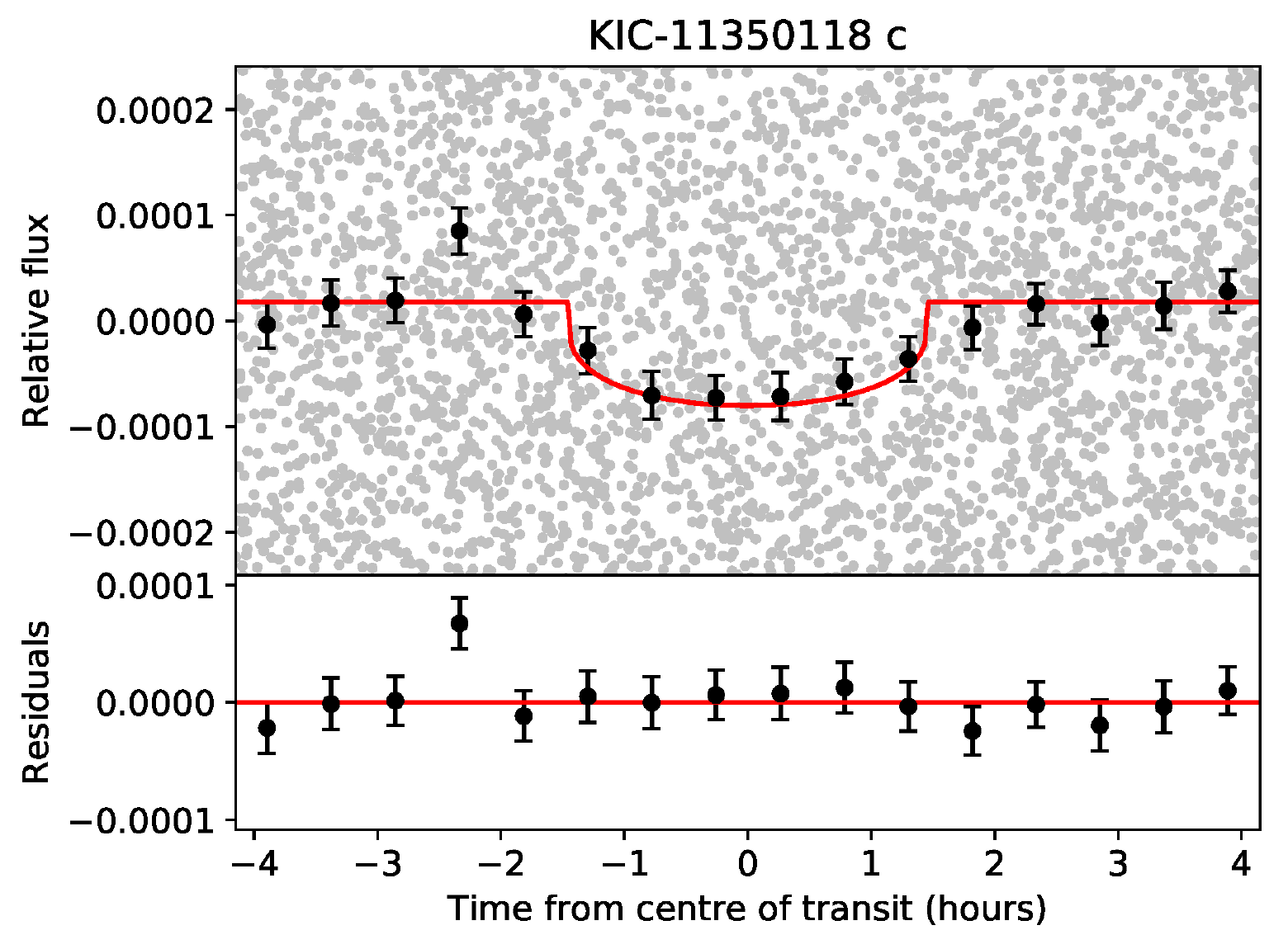}
\includegraphics[width=\linewidth]{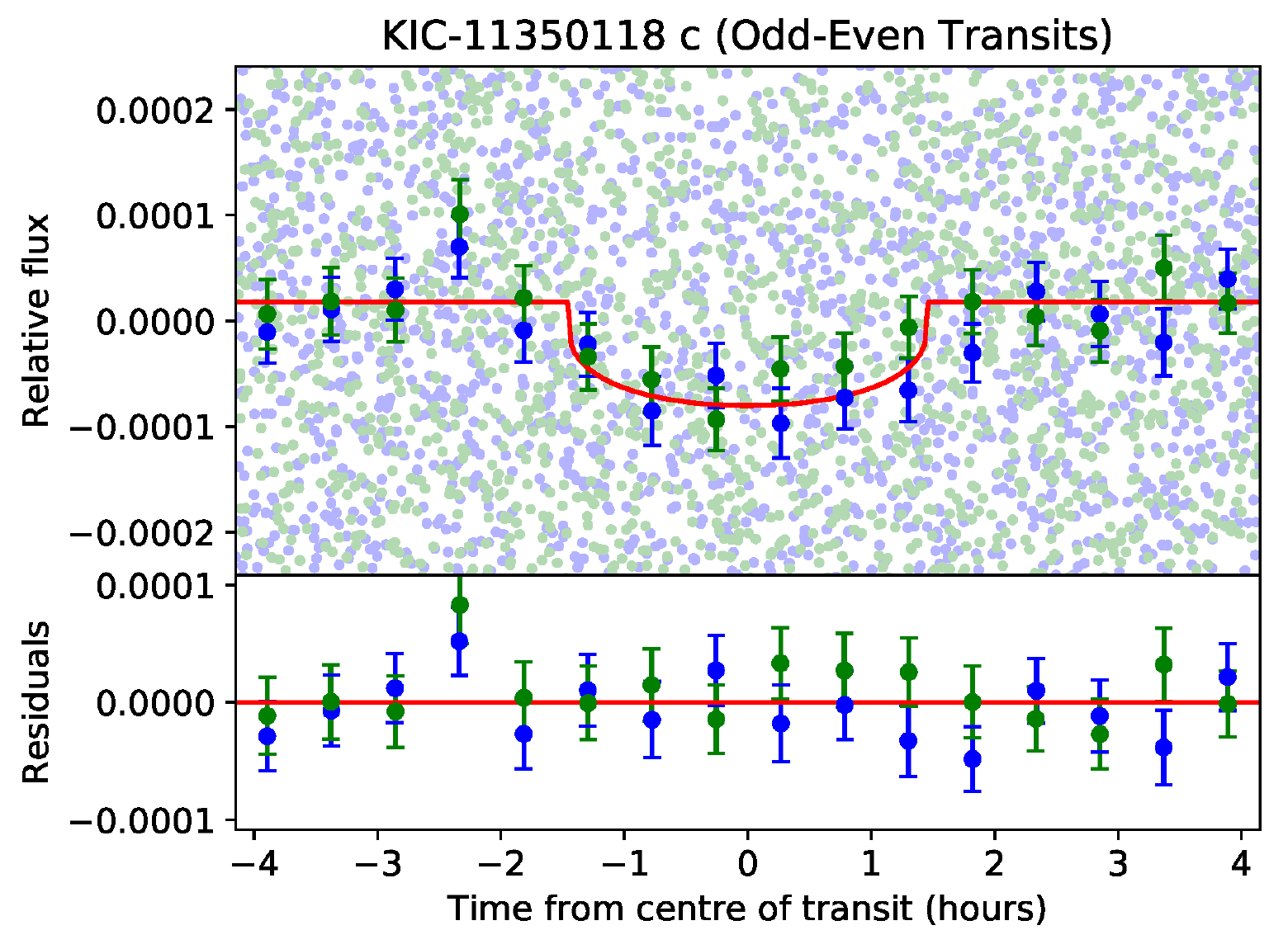}
\caption{Phase diagrams of KIC-11350118 c, otherwise known as KOI-4509.02, plotting all data together (top) and indicating every odd and even transit (bottom); (see Fig. \ref{fig:7340288}).}\label{fig:11350118}
\end{figure}

The KIC-11350118 system, also known as KOI-4509, already has a single known candidate with $P = 12.0$ days and $R_{p} = 0.97R_{\bigoplus}$. We detect an additional, smaller candidate with $P = 2.7$ days and $R_{p} = 0.66 R_{\bigoplus}$. The full transit and odd-even phase diagrams are plotted in Fig. \ref{fig:11350118}.

The Robo-AO survey observed the host star and did not detect any companions within 4$^{\prime\prime}$ \citep{zie17}. Furthermore, its membership in a multiplanet system lowers its FPP \citep{lis12}. For systems containing two planets in the \textit{Kepler} field, \citet{lis12} estimated a ``multiplicity-boost'' factor of 25 to the planet prior probability. After incorporating this into our \texttt{vespa} calculation, we found an astrophysical FPP of $9.62\times10^{-4}$.

\section{Conclusions and Future Work}

Using an independent search and vetting pipeline on all four years of \textit{Kepler} data, we report 17 new PCs. Twelve of these have astrophysical FPPs less than 1$\%$, and many of our new detections represent valuable additions to particular parameter spaces. For instance, KIC-7340288 b is both rocky and in the Habitable Zone, a part of phase space currently occupied by only 15 confirmed KOIs. The occurrence rates of such planets are of great interest to the exoplanet community, but are also poorly constrained. Furthermore, there are not many planets with radii $R_{p} < 0.7 R_{\bigoplus}$. Our survey also finds one new small candidate, KIC-11350118 c, which is larger than only 15 confirmed KOIs.

Our AO follow-up emphasizes the role that high-resolution imaging can play in validating planets. The input of contrast curves to \texttt{vespa} reduced astrophysical FPPs by 16 times on average for the six targets for which we obtained observations. The confirmation of no nearby stars bright enough to significantly dilute the transit depth of our Habitable Zone planet KIC-7340288 b is also valuable, considering how close this planet is to the rocky limit.

Using our final planet catalogue, we will be calculating occurrence rate statistics. Estimates based on an independent search of the same scope as \textit{Kepler} will be a valuable contribution to our understanding of exoplanet occurrence rates. We also have plans to apply our pipeline to other missions such as \textit{Kepler}'s follow-up K2, and the more recent Transiting Exoplanet Survey Satellite (TESS). We look forward to continue supporting the role that independent searches play as valuable and reliable tools for exoplanet detection.

\section{Acknowledgements}

We thank the referee for reviewing our paper and providing positive and constructive comments. Their insights substantially helped to improve the quality of the paper. We also thank NASA for providing the wealth of \textit{Kepler} data available to the public for download, without which this paper would not be possible. For our first uniqueness test, we acknowledge and are thankful for the use of code written by Dr. Kelsey Hoffman.

Our AO follow-up was based on observations obtained at the Gemini Observatory (Programs GN-2018B-Q-134 and GN-2019A-FT-213), which is operated by the Association of Universities for Research in Astronomy, Inc., under a cooperative agreement with the NSF on behalf of the Gemini partnership: the National Science Foundation (United States), National Research Council (Canada), CONICYT (Chile), Ministerio de Ciencia, Tecnolog\'{i}a e Innovaci\'{o}n Productiva (Argentina), Minist\'{e}rio da Ci\^{e}ncia, Tecnologia e Inova\c{c}\~{a}o (Brazil), and Korea Astronomy and Space Science Institute (Republic of Korea).

This work has made use of data from the European Space Agency (ESA) mission
{\it Gaia} (\url{https://www.cosmos.esa.int/gaia}), processed by the {\it Gaia}
Data Processing and Analysis Consortium (DPAC,
\url{https://www.cosmos.esa.int/web/gaia/dpac/consortium}). Funding for the DPAC
has been provided by national institutions, in particular the institutions
participating in the {\it Gaia} Multilateral Agreement.

This research has made use of the NASA Exoplanet Archive, which is operated by the California Institute of Technology, under contract with the National Aeronautics and Space Administration under the Exoplanet Exploration Program.

\facilities{Exoplanet Archive, \textit{Gaia}, Gemini:Gillett (NIRI), \textit{Kepler}, MAST, Sloan}

\software{\texttt{astropy}~\citep{ast13,ast18}, \texttt{emcee}~\citep{for13}, \texttt{isochrones}~\citep{mor15a} \texttt{Matplotlib}~\citep{hun07}, \texttt{numpy} \citep{vdw11}, \texttt{PyKE} \citep{sti12}, \texttt{scipy} \citep{jon01}, \texttt{vespa}~\citep{mor12,mor15b}, } 

\appendix

\section{Additional AO Observations}\label{sec:moreAO}

As discussed in Section \ref{sec:AO}, we observed an additional 56 targets across our Gemini programs GN-2018B-Q-134 (45 targets with NGS-AO) and GN-2019A-FT-213 (11 targets with LGS-AO) that did not become part of our final candidate list. We present our data here for completeness. Table \ref{tbl:rem_contrast} gives our contrast curve measurements, while Table \ref{tbl:rem_companions} gives a list of all companions within 4$^{\prime\prime}$.

Twenty-one of these targets are KOIs that were also observed by Robo-AO and had detected companions. Our motivation behind these observations was to compile multiband photometry and confirm the existence of potential companions. Another twelve of our targets are KOIs that have not been observed by Robo-AO. While the corresponding PCs we detected have since failed, these observations are still useful for follow-up analysis of known \textit{Kepler} candidates around these stars.

\begin{table}[h!]
\caption{Contrast curve data for all 56 additional targets observed with Gemini NGS-AO and LGS-AO in the $K_{s}$ band. Only a portion of this table is shown here. A machine-readable version of the full table is available.}\label{tbl:rem_contrast}
    \centering
    \begin{tabular}{c|c|c|c|c|c}
        \hline\hline
        KIC & KOI & Guide Star System & UT Obs. Date & Sep. ($^{\prime\prime}$) & $\Delta K_{s}$ \\
        \hline
        7747103 & 7847 & NGS-AO & 01 July 2018 & 0.20 & 0.38955 \\
        & & & & 0.35 & 2.23853 \\
        & & & & 0.51 & 3.83136 \\
        & & & & 0.66 & 4.6674 \\
        & & & & ... & ... \\
        11350634 & 8050 & NGS-AO & 01 July 2018 & 0.20 & 0.71592 \\
        & & & & 0.35 & 3.93214 \\
        & & & & 0.51 & 4.90273 \\
        & & & & 0.66 & 5.14522 \\
        & & & & ... & ... \\
    \end{tabular}
\end{table}

\begin{longtable*}[h!]{c|c|c|c|c|c|c|c}
\caption{AO results from our Gemini North observations, reporting all companions within 4$^{\prime\prime}$.}\label{tbl:rem_companions}\\
        \hline\hline
        KIC & KOI & Guide Star System & UT Obs. Date & Comp? & Sep. ($^{\prime\prime}$) & PA ($\degree$)  & $\Delta K_{s}$\\
        \hline
        \endfirsthead
        KIC & KOI & Guide Star System & UT Obs. Date & Comp? & Sep. ($^{\prime\prime}$) & PA ($\degree$)  & $\Delta K_{s}$ \\
        \hline
        \endhead
        7747103 & 7847 & NGS-AO & 01 July 2018 & Y & 3.0980$\pm$0.0001 & $2.2708\pm0.0002$ & $2.276\pm0.003$ \\
        11350634 & 8050 & NGS-AO & 01 July 2018 & Y & $0.881\pm0.001$ & $270.501\pm0.001$ & $6.64\pm0.02$ \\
        7134626 & 7818 & NGS-AO & 01 July 2018 & N& - & - & - \\
        5894182 & 7750 & NGS-AO & 01 July 2018 & N& - & - & - \\
        8182107 & 7870 & NGS-AO & 11 July 2018 & N& - & - & - \\
        11152511 & 5874 & NGS-AO & 11 July 2018 & N& - & - & - \\
        11360571 & 2069 & NGS-AO & 30 July 2018 &Y &  $1.257\pm0.003$ & $112.737\pm0.002$ & $2.27\pm0.02$ \\
        6938264 & 4180 & NGS-AO & 11 Oct 2018 & Y & $2.4738\pm0.0001$ & $35.0627\pm0.0001$ & $1.047\pm0.001$\\
        & & & & & $3.440\pm0.0005$ & $37.6856\pm0.0001$ & $3.763\pm0.010$ \\
        10684670 & 2317 & NGS-AO & 11 Oct 2018 &Y &  $1.5105\pm0.0007$ & $113.3984\pm0.0005$ & $4.28\pm0.01$\\
        7103919 & 4310 & NGS-AO & 11 Oct 2018 & N& - & - & - \\
        4141593 & 7685 & NGS-AO & 11 Oct 2018 & Y & $1.5573\pm0.0001$ & $221.4698\pm0.0001$ & $2.541\pm0.003$ \\
        9898447 & 2803 & NGS-AO & 17 Oct 2018 & Y & $3.8182\pm0.0002$ & $60.4070\pm0.0001$ & $2.066\pm0.006$ \\
        7749773 & 2848 & NGS-AO & 03 Nov 2018 & Y & $2.1854\pm0.0006$ & $29.8321\pm0.0003$ & $3.72\pm0.01$ \\
        7983117 & 3214 & NGS-AO & 03 Nov 2018 & Y & $0.4855\pm0.0001$ & $318.3203\pm0.0001$ & $1.362\pm0.001$\\
        & & & & & $1.3119\pm0.0001$ &  $199.7957\pm0.0001$ & $2.222\pm0.003$ \\
        6837283 & 2914 & NGS-AO & 14 Nov 2018 &Y &  $3.804\pm0.002$ & $231.2994\pm0.0004$ & $5.15\pm0.04$ \\
        7097965 & 2083 & NGS-AO & 14 Nov 2018 & Y & $0.2517\pm0.0001$ & $164.7515\pm0.0003$ & $1.646\pm0.001$ \\
        1161345 & 984 & NGS-AO & 14 Nov 2018 & Y & $1.7747\pm0.0001$ & $41.9867\pm0.0001$ & $0.1787\pm0.0008$\\
        7449136 & 1890 & NGS-AO & 14 Nov 2018 & Y & $0.4070\pm0.0001$ & $143.6840\pm0.0002$ & $2.042\pm0.002$ \\
        11869052 & 120 & NGS-AO & 14 Nov 2018 & Y & $1.5793\pm0.0001$ & $129.5042\pm0.0001$ & $0.624\pm0.001$ \\
        9469494 & 7938 & NGS-AO & 06 Dec 2018 & Y & $0.2917\pm0.0004$ & $266.04\pm0.001$ & $3.610\pm0.002$\\
        4252322 & 396 & NGS-AO & 08 Dec 2018 & Y & $1.906\pm0.003$ & $184.500\pm0.002$ & $5.76\pm0.06$ \\
        10198225 & 7991 & NGS-AO & 08 Dec 2018 & Y & $3.393\pm0.002$ & $95.5980\pm0.0006$ & $5.54\pm0.05$ \\
        7976520 & 687 & NGS-AO & 14 Dec 2018 & Y & $0.7012\pm0.0001$ & $12.3221\pm0.0002$ & $1.360\pm0.002$ \\
        10905911 & 2754 & NGS-AO & 15 Mar 2019 & Y & $0.7859\pm0.0001$ & $260.1817\pm0.0001$ & $1.564\pm0.001$ \\
        5796675& 652 & NGS-AO & 16 Mar 2019 & Y & $1.239\pm0001$ & $267.2008\pm0.0001$ & $0.5812\pm0.0006$ \\
        10199984 & 5776 & NGS-AO & 21 Mar 2019 & N & - & - & - \\
        11401253 & 4823 & NGS-AO & 21 Mar 2019 & Y & $1.3055\pm0.0001$ & $153.2864\pm0.0001$ & $0.2808\pm0.0007$ \\
        & & & & & $1.218\pm0.002$ & $336.697\pm0.001$ & $4.90\pm0.01$ \\
        7287028 & 7832 & NGS-AO & 22 Mar 2019 & N& - & - & - \\
        10932270 & 7389 & NGS-AO & 23 Mar 2019 &Y &  $1.921\pm0.003$ & $70.205\pm0.001$  & $5.53\pm0.06$ \\
        8332521 & 4567 & NGS-AO & 22 May 2019 & Y & $1.3275\pm0.0001$ & $141.8448\pm0.0001$ & $1.619\pm0.001$ \\
        8765560 & 3891 & NGS-AO & 24 May 2019 & Y & $1.973\pm0.001$ & $138.0832\pm0.0005$ & $4.40\pm0.02$ \\
        & & & & & $0.956\pm0.003$ & $241.774\pm0.004$ & $5.79\pm0.02$ \\
        4770174 & 2971 & NGS-AO & 24 May 2019 & Y & $0.2350\pm0.0004$ & $273.012\pm0.001$ & $3.681\pm0.002$ \\
        7190107 & - & NGS-AO & 31 May 2019 & N& - & - & - \\
        3662290 & - & NGS-AO & 31 May 2019 & Y & $1.632\pm0.001$ & $154.4306\pm0.0006$ & $4.78\pm0.02$ \\
        5628770 & - & NGS-AO & 31 May 2019 &Y &  1.307$\pm$0.003 & 201.39$\pm$0.15 & 5.35$\pm$0.02 \\
        6139884 & - & NGS-AO & 31 May 2019  & Y & 3.761$\pm$0.001 & 25.79$\pm$0.01 & 2.591$\pm$0.003 \\
        7020834 & - & NGS-AO & 31 May 2019 & Y & $2.4194\pm0.0006$ & $10.2823\pm0.0003$  & $4.29\pm0.01$ \\
        11565976 & - & NGS-AO & 31 May 2019 & Y & 0.8245$\pm$0.0002 & 162.8190$\pm$0.0003 & 2.827$\pm$0.004 \\
        3345775 & - & NGS-AO & 01 June 2019 & N& - & - & - \\
        7186892 & - & NGS-AO & 01 June 2019  &Y &  0.8192$\pm$0.0001 & 179.0674$\pm$0.0001 & 0.282$\pm$0.001 \\
        9823433 & - & NGS-AO & 01 June 2019 & N& - & - & -\\
        12505309 & - & NGS-AO & 02 June 2019 &   N & - & - \\
        3531436 & - & NGS-AO & 11 June 2019 & N & - & - & - \\
        6380164 & - & NGS-AO & 12 June 2019 & Y & $2.752\pm0.004$ & $257.99\pm0.08$ & $6.65\pm0.09$ \\
        8172679 & - & NGS-AO & 12 June 2019 & N &  - & - & - \\
        2985262 & - & LGS-AO & 30 June 2019 & N & - & - & - \\
        4551429 & - & LGS-AO & 30 June 2019 & N & - & - & - \\
        4569091 & - & LGS-AO & 30 June 2019 & Y & $3.7120\pm0.0001$ & $245.5372\pm0.0001$ & $1.557\pm0.002$ \\
        5803540 & - & LGS-AO & 30 June 2019 & Y & $2.5206\pm0.0001$ & $294.8547\pm0.0001$ & $2.318\pm0.002$ \\
        7119412 & - & LGS-AO & 30 June 2019 & N& - & - & - \\
        9274173 & - & LGS-AO & 30 June 2019 & N& - & - & - \\
        11092463 & - & LGS-AO & 30 June 2019 & Y &  0.7069$\pm$0.0002 & 247.1112$\pm$0.0003 & 2.936$\pm$0.004 \\
        5095499 & - & LGS-AO & 01 July 2019 & N& - & - & - \\
        12307455 & - & LGS-AO & 01 July 2019 & N& - & - & - \\
        4346258 & - & LGS-AO & 03 July 2019 & N & - & - & - \\
        6937870 & - & LGS-AO & 03 July 2019 & N & - & - & - \\
    \end{longtable*}

\end{document}